\documentclass[10pt,twocolumn]{article}
\usepackage[numbers,sort&compress]{natbib}
\usepackage[T1]{fontenc}
\usepackage[utf8]{inputenc}
\usepackage{amsmath,amssymb,amsfonts,esint}
\usepackage{xcolor}
\usepackage{bm}
\usepackage{graphicx}
\usepackage{booktabs}
\usepackage{stfloats}
\usepackage{float}
\usepackage{longtable}
\usepackage{array}
\usepackage{makecell}
\usepackage{caption}
\usepackage[hidelinks]{hyperref}
\usepackage[a4paper,top=1.55cm,bottom=1.75cm,left=1.55cm,right=1.55cm,columnsep=0.65cm,headheight=18pt]{geometry}
\usepackage{siunitx}
\usepackage{enumitem}
\usepackage{microtype}
\usepackage{placeins}
\usepackage{fancyhdr}
\usepackage{titlesec}
\usepackage{flushend}
\usepackage{doi}

\DeclareUnicodeCharacter{2013}{-}
\DeclareUnicodeCharacter{2014}{-}
\DeclareUnicodeCharacter{202F}{\,}
\DeclareUnicodeCharacter{2212}{-}
\DeclareUnicodeCharacter{03BC}{\ensuremath{\mu}}

\definecolor{atsblue}{RGB}{0,70,150}
\definecolor{atsgray}{RGB}{65,65,65}
\definecolor{wrongred}{RGB}{200,30,30}
\definecolor{fixgreen}{RGB}{0,120,55}
\newif\ifhaveulem
\IfFileExists{ulem.sty}{\haveulemtrue}{\haveulemfalse}
\ifhaveulem\usepackage[normalem]{ulem}\fi
\ifhaveulem
  
\else
  
\fi
\newcommand{\MODNEW}[1]{#1} 
\newcommand{\NEW}[1]{#1}

\definecolor{r2add}{RGB}{200,100,0}
\definecolor{r2del}{RGB}{200,0,120}
\definecolor{r2mv}{RGB}{120,40,200}

\ifhaveulem
  
\else
  
\fi

\setlist{nosep,leftmargin=*}
\titleformat{\section}{\large\bfseries\sffamily}{\thesection.}{0.45em}{}
\titleformat{\subsection}{\normalsize\bfseries\sffamily}{\thesubsection.}{0.45em}{}
\titleformat{\subsubsection}{\normalsize\itshape\sffamily}{\thesubsubsection.}{0.45em}{}
\titlespacing*{\section}{0pt}{8pt plus 2pt minus 1pt}{4pt}
\titlespacing*{\subsection}{0pt}{6pt plus 2pt minus 1pt}{3pt}
\titlespacing*{\subsubsection}{0pt}{5pt plus 1pt minus 1pt}{2pt}

\begin{document}
\twocolumn[{%
\begin{@twocolumnfalse}
\thispagestyle{fancy}
\noindent\begin{minipage}{0.48\textwidth}
{\sffamily\bfseries\small }
{\scriptsize\sffamily }
\end{minipage}\hfill
\begin{minipage}{0.45\textwidth}\raggedleft
{\sffamily\bfseries\small }
{\scriptsize\sffamily }
\end{minipage}
\vspace{1.0em}

{\LARGE\bfseries Geometric Phonon Energy Pumping in a Layer-Hybridized Moir\'e Exciton Manifold\par}
\vspace{0.65em}
{\large Bitap Raj Thakuria and Himangshu Prabal Goswami*\par}
\vspace{0.25em}
{\small QuAinT Research Group, Department of Chemistry, Gauhati University, GNB Nagar, Jalukbari, Guwahati 781014, Assam, India\par}
{\small E-mail: hpg@gauhati.ac.in;\par ORCID - B.~R.~Thakuria: 0009-0002-3336-7231, H. P. Goswami: 0000-0001-8981-9917\par}
\vspace{0.9em}

{\small\noindent\textbf{Abstract}\quad\quad
Slow cyclic modulation of an open quantum system can generate a geometric contribution to its energy transfer
statistics, separable from the  dynamic background. In this work we construct an experimentally anchored
four state open system model for a gate-tunable WSe$_2$/WS$_2$
moir\'e exciton manifold within a population level
full counting statistics framework.
We first drive the system uniformly through a closed gate and pump loop and separate the geometric contribution from the dynamic background by reversing the loop direction. We then examine how this response changes with environmental rates, exciton coupling, control conditions, and static spectral broadening. A shorter cycle period followed by nonuniform traversal of the driving loop substantially improves detectability while preserving the geometric response. Combined with enhanced phonon relaxation and reduced radiative loss, this gives about a fourteenfold fixed time signal to noise gain relative to the uniform loop. Our results show that the geometric phonon response remains robust against realistic variations in the system and its environment, and that we can improve its detectability more effectively by suppressing dynamical noise through frequency modulated driving.

\par}

\vspace{0.45em}
{\small\noindent\textbf{Keywords}\quad
geometric phase; full counting statistics; geometric pumping;
phonon emission; moir\'e excitons; non-equilibrium open quantum
systems\par}
\vspace{0.85em}
\end{@twocolumnfalse}
}]
\section{Introduction}

Geometric phases were first identified in adiabatically cycled closed
quantum systems,\cite{Berry1984} with related effects later established
in driven open and stochastic systems.\cite{Sinitsyn2009} In full
counting statistics (FCS), slow cyclic modulation can generate a
path-dependent contribution to transported charge, particle, or energy
statistics.\cite{Sinitsyn2007,Ren2010,Yuge2012} This contribution is
associated with a geometric curvature over the oriented loop
area.\cite{Sinitsyn2007,Ren2010,Sagawa2011} A parity separates
geometric pumping from a larger time-extensive background.\cite{Sinitsyn2007,Sinitsyn2009,Ren2010,Yuge2012}

The dynamic response depends on instantaneous rates and local dwell time,
whereas the adiabatic geometric response depends primarily on the
oriented contour. Loop geometry is therefore an additional
nonequilibrium control variable alongside energies, rates, biases, and
bath parameters. The first geometric cumulant contributes to the mean
transported quantity, while higher cumulants describe its fluctuations.

Geometric pumping has been developed through Berry-Pancharatnam type stochastic pump phases in chemical kinetics \cite{Sinitsyn2007}, geometric heat pumping in molecular junctions \cite{Ren2010} quantum master equation formulations of
geometric transport~\cite{Yuge2012}, and geometric contributions to transfer
statistics in adiabatically driven quantum junctions~\cite{Goswami2016}. It has also been linked to excess entropy production \cite{Sagawa2011}, interacting quantum dots \cite{Splettstoesser2006,Pluecker2017}, geometric refrigeration \cite{Monsel2022}, quantum thermal machines and heat engines \cite{Bhandari2020,Hino2021}. Energy-transfer statistics in a solid-state platform whose Hamiltonian is independently fixed by spectroscopy remain comparatively unexplored.

Such a platform requires a tunable level structure with a resolved energy
quantum, two independently controlled slow parameters enclosing a finite
loop, and a relaxation pathway linked to an identifiable terminal. A
spectroscopy-constrained Hamiltonian also limits model freedom. The level
structure and hybridization cannot be chosen solely to maximize the
response, so the geometric contribution must survive within a
experimentally supported control range and under material uncertainty.

Moir\'e excitons in transition-metal-dichalcogenide heterobilayers offer
these ingredients. Lattice mismatch or twist creates a long-period
moir\'e superlattice that can trap excitons in flat, optically active
minibands.\cite{Huang2022,Yu2017,Jin2019,Tran2019,WuLovorn2017} These
electrically tunable systems support moir\'e-trapped excitons, correlated
states, generalized Wigner crystals, and Hubbard-model
physics.\cite{Regan2020,TangHubbard2020,Seyler2019,Alexeev2019,
Kennes2021,Wilson2021,MakShan2022} Their gate-tunable energies, optically controlled populations,
layer-dependent radiative lifetimes, and intrinsic exciton-phonon
relaxation  make them possible nonequilibrium energy transfer
platforms.

We propose the use of an experimentally anchored three-level exciton Hamiltonian of a
gate-tunable WSe$_2$/WS$_2$ heterobilayer \cite{Tang2021}, in
which a two-parameter geometric protocol can be
implemented. A dressed manifold in a Markovian population-level
tilted Liouvillian with optical injection, phonon relaxation, and
radiative recombination serve as effective terminals. A counting field on the
retained phonon transition yields the mean deposited energy and its
fluctuations.\cite{Esposito2009,Campisi2011} The Hamiltonian fixes the
dressed energies and compositions, while the ground state, terminals,
phenomenological rates, counting field, and driving protocol are
theoretical extensions.

The required driving loop capable of realizing geometric contribution uses two experimentally meaningful controls. The gate voltage
\(V_G(t)\) changes the dressed energies, gaps, and bright/interlayer
character through the Hamiltonian, whereas \(I_{\mathrm{pump}}(t)\)
modulates optical injection through the Liouvillian. A finite phase lag
forms a closed oriented loop in the \((V_G(t),I_{\mathrm{pump}}(t))\) plane,
and reversal changes the sign of the geometric phonon-terminal
response.\cite{Sinitsyn2007,Ren2010,Yuge2012} This gate-pump protocol probes the combined
parameter dependence of dressed exciton structure and dissipative
injection.
The central question is whether an experimentally available,
phase-locked gate-pump cycle can generate a geometric contribution in an exciton
manifold, rather than in charge or particle counting. The challenge is
also to separate it from the dynamic response under
meaningful constraints. 

The objectives are fourfold. First, we identify a single
phonon-accessible dressed-state relaxation channel over the complete
cycle and test for nonzero orientation-odd energy cumulants using
finite-time CW and CCW tilted-Liouvillian propagation. Second, we test
robustness against spectroscopic reconstruction, branch assignment,
residual coupling, phenomenological rates, temperature, phase lag, and
static disorder. Third, we cross check loop-reversal separation from the
dynamic background and compare the first cumulant with an independent
frozen generator geometric curvature calculation. Fourth, we
redistribute traversal time along the same loop to suppress the dynamic
mean and variance while preserving the path-dependent response, and
formulate a calibrated optical proxy for the counted statistics.

These objectives test whether layer-hybridized moir\'e excitons can serve as material-specific platforms for geometric energy statistics \cite{Tang2021,Huang2022,MakShan2022} and whether loop geometry and traversal schedule can control the geometric signal and time extensive background. Such control could support geometric modulation of heat flow \cite{Ren2010,Bhandari2020,Monsel2022}, optical sensing of phonon-assisted transport \cite{Tang2021,Huang2022}, and low-noise cyclic protocols \cite{Impens2025,Wang2024} in van der Waals devices. We do not attempt a complete
microscopic description of all exciton-phonon pathways. Strict
phonon accessibility retains the cleanest channel in an experimentally
anchored open-system extension. The geometric
curvature flux is neither assumed to be quantized nor topologically
protected, and it should not be confused with the Berry-Pancharatnam curvature of
closed-system wave functions. Nor do we claim macroscopic cooling or useful
heat removal. We ask whether a statistically
identifiable geometric contribution can be defined and experimentally
sought in a solid-state system.

The paper is organized as follows. Section~2 presents the Hamiltonian, four-state model, environmental
channels, and tilted-Liouvillian FCS formulation. Section~3 describes the
spectroscopic reconstruction and finite-time propagation. Section~4
presents the cumulants, loop-reversal separation, robustness,
 optical proxy, and detectability
optimization. Section~5 summarizes the implications and experimental
detectability.

\FloatBarrier
\section{Theory and Formalism}\label{sec:framework}

\setcounter{equation}{3}

\subsection{Open-System Geometric Contributions}\label{sec:general_fcs}

For an open quantum system with multiple environments, the total Hamiltonian is partitioned into a system part, a set of effective environmental channels, and a system-bath coupling,
\begin{equation}
\hat{H}_{\mathrm{tot}}(t)
=
\hat{H}_{\mathrm{sys}}(t)
+
\hat{H}_{\mathrm{bath}}
+
\hat{H}_{\mathrm{SB}}(t).
\label{eq:Htot}
\end{equation}

Here $\hat{H}_{\mathrm{sys}}(t)$ contains the instantaneous system levels and states, $\hat{H}_{\mathrm{bath}}$ represents the external terminals, and $\hat{H}_{\mathrm{SB}}(t)$ specifies which system transitions exchange particles or energy with those terminals. This partition is the microscopic starting point of the rate model: $\hat{H}_{\mathrm{sys}}(t)$ tells us what the instantaneous eigenstates and transition energies are, while $\hat{H}_{\mathrm{SB}}(t)$ tells us which jumps are allowed and what terminal each jump belongs to. In this work, we  describe $\hat{H}_{\mathrm{sys}}(t)$ using driven moir\'e excitons, which we shall present later. 

Under the usual weak-coupling, Markov, and secular approximations, the total Hamiltonian in Eq.~\eqref{eq:Htot} is reduced to an effective population-level master equation, and coherences are decoupled.\cite{Breuer2002,Campaioli2024} In general, the system state is represented by a probability vector $\mathbf{p}(t)=[p_1(t),p_2(t),\ldots,p_N(t)]^T$, where $p_i(t)$ is the occupation probability of the i$^{th}$ instantaneous system state. Keeping the discussion general and not yet fixing the number of states, the reduced dynamics is an adiabatic master equation,\cite{Yuge2012,Paulino2024}
\begin{equation}
\frac{d}{dt}\mathbf{p}(t)=\mathcal{L}[\mathbf{R}(t)]\,\mathbf{p}(t),
\label{eq:master_general}
\end{equation}
where $\mathbf{R}(t)$ is the set of externally controlled parameters and $\mathcal{L}[\mathbf{R}(t)]$ is the ordinary population Liouvillian, or Markov generator. Its off-diagonal elements are the transition rates between system states.  $\hat{H}_{\mathrm{sys}}(t)$ contains the instantaneous energies and state characters, while $\hat{H}_{\mathrm{bath}}$ and $\hat{H}_{\mathrm{SB}}(t)$ give rise to the dissipative channels and the rates that build $\mathcal{L}[\mathbf{R}(t)]$.

To quantify the statistics of the energy exchanged by the system with a
selected terminal,\NEW{the population vector is resolved with respect to
the net exchanged energy $Q$, giving $\mathbf{p}(Q,t)$, whose entries are
the joint probabilities that the system occupies the corresponding state
at time $t$ and that a net energy $Q$ has been exchanged with the counted
terminal up to that time. Thus $Q$ is not an additional system state but
a bookkeeping variable that records how much energy has entered the
counted terminal up to time $t$.}The counting-field-resolved vector is defined by a (bilateral) Laplace transform
\begin{equation}
\mathbf{p}_{s}(t)=\int dQ\,e^{sQ}\,\mathbf{p}(Q,t),
\label{eq:laplace}
\end{equation}
with the corresponding sum over $Q$ when the exchanged energy is discrete. The auxiliary variable $s$ is the counting field.  Applying Eq.~\eqref{eq:laplace} to Eq.~\eqref{eq:master_general} gives the tilted master equation
\begin{equation}
\frac{d}{dt}\mathbf{p}_{s}(t)=\mathcal{L}_{s}[\mathbf{R}(t)]\,\mathbf{p}_{s}(t),
\qquad
\mathcal{L}_{s}[\mathbf{R}(t)]\big|_{s=0}=\mathcal{L}[\mathbf{R}(t)].
\label{eq:tilted_general}
\end{equation}
The operator $\mathcal{L}_{s}$ is the tilted Liouvillian. The statistics over one driving period $\tau$ follow from the counting field dependent propagator and the \MODNEW{finite period} cumulant-generating function
\MODNEW{\begin{align}
\mathcal{U}_{s}(\tau,0)
&=
\mathcal{T}\exp\!\left[
\int_0^{\tau}\mathcal{L}_{s}[\mathbf{R}(t)]\,dt
\right],
\\
G(s,\tau)
&=
\ln\!\left[
\mathbf{1}^{T}\mathcal{U}_{s}(\tau,0)\,\mathbf{p}(0)
\right].
\label{eq:propagator_general}
\end{align}}
where $\mathcal{T}$ is the time-ordering operator, $\mathbf{p}(0)$ is the initial population vector, and $\mathbf{1}^{T}$ is a unit vector identity matrix. For a time-independent
Liouvillian, the long-time statistics are controlled by the dominant
eigenvalue of \(\mathcal{L}_s\). In the present driven problem,
however, \(\mathcal{L}_s[\mathbf{R}(t)]\) changes during the cycle, so
the cumulants are obtained from the full time-ordered propagator rather
than from a single frozen Liouvillian eigenvalue.

\MODNEW{For slow cyclic driving, the control vector \(\mathbf{R}(t)\) varies
slowly compared with the relaxation time of \(\mathcal{L}\). At each
fixed \(s\) and each frozen control point \(\mathbf{R}\), the tilted
Liouvillian \(\mathcal{L}_{s}(\mathbf{R})\) has instantaneous
eigenvalues and left and right eigenvectors. In the adiabatic
interpretation, the dominant instantaneous eigenvalue
\(\lambda_0(s,\mathbf{R})\), together with its left and right
eigenvectors, determine the dynamic and geometric contributions to
\(G(s,\tau)\).}
Inserting the adiabatic (slow-driving) expansion of the leading eigenvalue of $\mathcal{U}_{s}(\tau)$ into Eq.~\eqref{eq:propagator_general} splits the generating function into a dynamic and a geometric part,\cite{Paulino2024}
\begin{equation}
G(s)=G_{\mathrm{dyn}}(s)+G_{\mathrm{geo}}(s).
\label{eq:Gsplit_general}
\end{equation}
The dynamic part is the time integral of the dominant eigenvalue along the path,
\begin{equation}
G_{\mathrm{dyn}}(s)=\int_{0}^{\tau}\lambda_0[s,\mathbf{R}(t)]\,dt,
\label{eq:Gdyn_general}
\end{equation}
and depends on how long the system dwells at each control point. The geometric part depends instead on the oriented area enclosed by the loop in control space. It is referred to as the geometric connection and curvature \cite{Sinitsyn2009,Sinitsyn2007,Yuge2012}. The geometric generating function is,
\begin{equation}
G_{\mathrm{geo}}(s)=\oiint_{S}dq_1\,dq_2\,B_s^{q_1q_2}.
\label{eq:Ggeo_general}
\end{equation}
with
\(B_s^{q_1q_2}=\partial_{q_1}A_s^{q_2}-\partial_{q_2}A_s^{q_1}\), where \(A_s^{q}=\langle l_0(s,\mathbf{R})|\partial_q r_0(s,\mathbf{R})\rangle
\). This is not the Berry-Pancharatnam curvature of closed-system wave functions,\cite{Sinitsyn2009} but it is the curvature of the dominant left and right eigenvectors of the tilted population Liouvillian.  Suppressing the arguments $(s,\mathbf{R})$, these vectors are defined as,
\begin{equation}
\mathcal{L}_{s}(\mathbf{R})|r_0\rangle=\lambda_0|r_0\rangle,
\qquad
\langle l_0|\mathcal{L}_{s}(\mathbf{R})=\lambda_0\langle l_0|,
\qquad
\label{eq:eigen_general}
\end{equation}
with $\langle l_0|r_0\rangle=1 .$
Here $|r_0\rangle$ and $\langle l_0|$ are the right and left dominant eigenvectors. Because $\mathcal{L}_{s}$ is a non-Hermitian population generator they are not Hermitian conjugates, and they are normalized biorthogonally as in Eq.~\eqref{eq:eigen_general}.

The cumulants of the energy transferred into the counted terminal follow by differentiation,
\begin{equation}
\begin{aligned}
C_n
&=
\left.\frac{\partial^n G(s)}{\partial s^n}\right|_{s=0}
\\
&\MODNEW{=
\left.\frac{\partial^n G_{\mathrm{dyn}}(s)}{\partial s^n}\right|_{s=0}
+
\left.\frac{\partial^n G_{\mathrm{geo}}(s)}{\partial s^n}\right|_{s=0}}
\\
&=C_n^{\mathrm{dyn}}+C_n^{\mathrm{geo}}.
\end{aligned}
\label{eq:cumulants_general}
\end{equation}
Here \(C_1\) is the first energy cumulant, i.e. the mean energy
transferred into the counted terminal during one cycle, while \(C_2\) is
the second cumulant, measuring the corresponding cycle to cycle
fluctuation.
The dynamic cumulants come from the eigenvalue term and are even under reversal of the loop orientation. The geometric cumulants come from the curvature term and are odd. This orientation parity is the basis of the loop-reversal separation which we shall employ to isolate the geometric contributions to the cumulants for the moir\'{e} system.  In the moir\'{e}-exciton manifold that we propose as a test-bed to realize these geometric effects, we shall drive two externally controllable experimental parameters, namely the gate voltage and the pump intensity, i.e $\mathbf{R}(t)=(V_G(t),I_{\mathrm{pump}}(t))$.

\subsection{Four-State System Hamiltonian}\label{sec:fourstate}

\NEW{We now apply this open-system construction to the experimentally reported layer-hybridized WSe$_2$/WS$_2$ moir\'{e}-exciton platform~\cite{Tang2021}. A spatially indirect interlayer exciton is coupled to two optically bright intralayer moir\'{e} excitons. The gate voltage $V_G$ controls the measured exciton spectrum, while the out-of-plane electric field $F_z$ mainly Stark shifts the interlayer exciton through the quantum-confined Stark effect. This tunes the dressed exciton energies and their mixed interlayer and intralayer character.}  The Hamiltonian is an effective one-exciton Hamiltonian. It describes the optical resonance structure of a single excitonic excitation, not multi-exciton or exciton-exciton interaction physics. In the bare-exciton basis
\(
B_T=\{|iX\rangle,|X_1\rangle,|X_2\rangle\},
\)
it is written as
\begin{equation}
\hat{H}_T(F_z,V_G)
=
\begin{pmatrix}
E_0(V_G)+D(V_G)F_z & W_1(V_G) & W_2(V_G)\\
W_1(V_G) & E_1(V_G) & 0\\
W_2(V_G) & 0 & E_2(V_G)
\end{pmatrix}.
\label{eq:HT}
\end{equation}
Here $|iX\rangle$ denotes the interlayer exciton, in which the electron and hole are mainly located in different layers of the WSe$_2$/WS$_2$ heterobilayer. This spatial separation gives the interlayer exciton a strong sensitivity to the out-of-plane electric field. The states $|X_1\rangle$ and $|X_2\rangle$ denote two optically bright intralayer moir\'{e}-exciton branches, in which the electron and hole are mainly associated with the same layer. The first diagonal element, $E_0(V_G)+D(V_G)F_z$, is the field-shifted interlayer-exciton energy. Here $E_0(V_G)$ is the gate-dependent interlayer-exciton energy at zero applied out-of-plane field, while $D(V_G)F_z$ is the Stark shift produced by the out of plane field. The quantities $E_1(V_G)$ and $E_2(V_G)$ are the gate-dependent energies of the two intralayer moir\'{e} exciton branches. Following the already reported minimal Hamiltonian \cite{Tang2021}, so the two intralayer branches are coupled only indirectly through their shared hybridization with the interlayer exciton.

The gate-dependent functions $E_0(V_G)$, $E_1(V_G)$, $E_2(V_G)$, $W_1(V_G)$, $W_2(V_G)$, and $D(V_G)$ are reconstructed from the fitted data in Ref.~\cite{Tang2021} and linearly interpolated over the experimental gate-voltage range. The spectrally dominant $X_1$-like fitted branch is used as the input function $E_1(V_G)$. \MODNEW{The simultaneously resolved weaker branch is also retained for the branch-choice sensitivity check reported in Sec.~S6.3 of the Supporting Information.

\MODNEW{The geometric protocol is generated by keeping the out-of-plane field
fixed at \(F_z^{(0)}\) while driving the gate coordinate as
\(V_G(t)\). Thus \(F_z=F_z^{(0)}\) and the experimental Hamiltonian
input of Eq.~\eqref{eq:HT} becomes
\(\hat{H}_T^{(F_z^{(0)})}(t)\equiv
\hat{H}_T[F_z^{(0)},V_G(t)]\). The second driven coordinate,
\(I_{\mathrm{pump}}(t)\), does not enter \(\hat{H}_T\).It enters the
optical injection rate in Eq.~\eqref{eq:an}. This separation is what
makes the loop a gate-pump loop rather than a single-parameter
Hamiltonian drive.}
Instantaneous dressed excitons are now obtainable by diagonalizing this time-dependent excitonic block:
\begin{equation}
\hat{H}_T^{(F_z^{(0)})}(t)|\psi_n(t)\rangle
=
\epsilon_n(t)|\psi_n(t)\rangle,
\qquad n=0,1,2.
\label{eq:eig_time}
\end{equation}
The eigenvalues are ordered as $\epsilon_0(t)<\epsilon_1(t)<\epsilon_2(t)$, so $|\psi_0(t)\rangle$, $|\psi_1(t)\rangle$, and $|\psi_2(t)\rangle$ denote the lower, middle, and upper dressed exciton branches, respectively. These eigenvalues and eigenstates are used to build  the required system Hamiltonian $\hat{H}_{\mathrm{sys}}(t)$.   To complete the open-system approach, we add a no-exciton ground state from which optical injection is possible to these three excitonic states. Optical injection promotes population from the no-exciton sector into the exciton manifold, while radiative recombination returns population from the dressed excitons back to the no-exciton sector. The system part of the open-system model now contains four  states,
$
\{|g\rangle,|\psi_0(t)\rangle,|\psi_1(t)\rangle,|\psi_2(t)\rangle\}.
$
Here $|g\rangle$ is the no-exciton ground state, and $|\psi_n(t)\rangle$, with $n=0,1,2$, are the instantaneous dressed exciton branches 
and the corresponding energies $E_i(t)$ are $
\{E_g,\epsilon_0(t),\epsilon_1(t),\epsilon_2(t)\}$. 
The four state system Hamiltonian is 
\begin{equation}
\hat{H}_{\mathrm{sys}}(t)
=
E_g|g\rangle\langle g|
+
\sum_{n=0}^{2}\epsilon_n(t)|\psi_n(t)\rangle\langle\psi_n(t)|,
\qquad
E_g=0.
\label{eq:Hsys}
\end{equation}
Here $E_g$ is the no-exciton ground-state energy and $\epsilon_n(t)$ is the instantaneous energy of the dressed exciton $|\psi_n(t)\rangle$. Equation~\eqref{eq:Hsys} is diagonal because the coherent interlayer-intralayer hybridization has already been included by diagonalizing the $\hat{H}_T$. Optical injection, phonon relaxation, and radiative recombination are therefore not written as coherent off-diagonal terms in $\hat{H}_{\mathrm{sys}}(t)$ and we account for these through $\hat{H}_{\mathrm{SB}}(t)$ so that these translate into transition rates in the population Liouvillian.  $\hat{H}_T$ supplies only the dressed exciton sector to  $\hat{H}_{\mathrm{sys}}(t)$ which embeds this sector into the open-system space by adding the no-exciton ground state. 
The  construction of $\hat{H}_{sys}$ fixes the scope of the experimental input. Only the static three-level Hamiltonian in Eq.~\eqref{eq:HT}, together with its gate-dependent fitted parameters, is anchored to the reported experimental setup\cite{Tang2021}. The no-exciton ground state, optical injection, environmental terminals, rate model, phonon tilted Liouvillian, and geometric cumulants are theoretically proposed extensions introduced here to convert the experimentally anchored exciton Hamiltonian into an open quantum system.
\subsection{Environmental Channels and System-Bath Coupling}\label{sec:bathcoupling}

We now address the remaining two terms of the open-system partition
Eq.~\eqref{eq:Htot}. The effective bath Hamiltonian and the system-bath
coupling. Each of the three reservoirs introduced below corresponds
to a physical subsystem that is already present in the gate-tunable
WSe$_2$/WS$_2$ device~\cite{Tang2021}. We do not postulate new environments and  only organize the existing ones
into environmental terminals. The environmental Hamiltonian is treated as a bath of harmonic oscillators,
\begin{equation}
\begin{aligned}
\hat{H}_{\mathrm{bath}}
&=\hat{H}_{\mathrm{pump}}+\hat{H}_{\mathrm{phonon}}+\hat{H}_{\mathrm{rad}} \\
&=
\sum_{\alpha\in O_p}\omega_{\alpha}a_{\alpha}^{\dagger}a_{\alpha}
+
\sum_{q\in P}\omega_q b_q^{\dagger}b_q
+
\sum_{\ell\in R}\omega_{\ell}c_{\ell}^{\dagger}c_{\ell},
\end{aligned}
\label{eq:Hbath}
\end{equation}
Here $O_p$, $P$, and $R$ label the optical-pump, phonon, and
radiative terminals, respectively. The terminal $O_p$ injects excitons,
$P$ accepts the counted phonon energy, and $R$ removes excitons
radiatively. Physically, the pump terminal $O_p$ is the photon field of the
excitation light. In the original experiment this field is a weak,
essentially stationary probe used for reflectance and
photoluminescence~\cite{Tang2021}. To introduce the second parameter's driven protocol, we propose that the pump 
intensity to be  modulated as $I_{\mathrm{pump}}(t)$, which can be
experimentally achieved by standard acousto-optic or electro-optic
modulation of the excitation laser~\cite{SalehTeich2019} with defined cycle periods. The
phonon terminal $P$ is the lattice of the WSe$_2$/WS$_2$ bilayer
and its hBN encapsulation, thermalized to the cryostat at the few-kelvin
temperatures of the measurement~\cite{Tang2021}. The interbranch
relaxation by phonon emission that happens into this reservoir is an intrinsic
channel in such samples itself ~\cite{Christiansen2017,Brem2020}. The radiative terminal $R$
is the electromagnetic vacuum into which the bright excitons
recombine~\cite{Robert2016,Rivera2015}, and it is a directly monitorable
reservoir in the experiment, since photoluminescence is precisely the
emission into this terminal. Although the pump and radiative terminals
belong to the same electromagnetic field, they involve disjoint mode
sets, the externally driven laser modes and the undriven vacuum modes,
and carry independently measurable rates, which justifies treating them
as separate reservoirs. We choose a minimal system-bath coupling
\begin{equation}
\begin{aligned}
\hat{H}_{\mathrm{SB}}(t)
=&
\sum_{n,\alpha}
\left[\eta_{n\alpha}(t)|\psi_n(t)\rangle\langle g|a_{\alpha}+\mathrm{h.c.}\right]
\\
&+
\sum_{n,\ell}
\left[\xi_{n\ell}(t)|g\rangle\langle\psi_n(t)|c_{\ell}^{\dagger}+\mathrm{h.c.}\right]
\\
&+
\sum_{i<j,q}
\left[g^c_{ij,q}(t)|\psi_i(t)\rangle\langle\psi_j(t)|+\mathrm{h.c.}\right]
\left(b_q+b_q^{\dagger}\right).
\end{aligned}
\label{eq:HSB}
\end{equation}
The first term describes optical injection from $|g\rangle$ into the
dressed manifold, the second describes radiative recombination back to
$|g\rangle$, and the third describes phonon-mediated transitions between
dressed exciton branches. The state dependence of these couplings is determined by the
instantaneous composition of the dressed exciton branches. In
particular, the optical injection amplitudes
\(\eta_{n\alpha}(t)\) and radiative emission amplitudes
\(\xi_{n\ell}(t)\) are governed by the optical dipole matrix elements
of the corresponding dressed state. To make this dependence explicit,
each dressed exciton is expanded in the bare-exciton basis as
\(
|\psi_n(t)\rangle
=
c_{iX,n}(t)|iX\rangle
+
c_{X1,n}(t)|X_1\rangle
+
c_{X2,n}(t)|X_2\rangle.
\)
The projection amplitudes are
\(
c_{iX,n}(t)=\langle iX|\psi_n(t)\rangle
\),
\(
c_{X1,n}(t)=\langle X_1|\psi_n(t)\rangle
\), and
\(
c_{X2,n}(t)=\langle X_2|\psi_n(t)\rangle
\), with the corresponding weights satisfying
\(
|c_{iX,n}(t)|^2
+
|c_{X1,n}(t)|^2
+
|c_{X2,n}(t)|^2
=
1
\).
The optically bright fraction is therefore defined by the total
intralayer weight,
\begin{equation}
f_n^{\mathrm{bright}}(t)
=
|c_{X1,n}(t)|^2
+
|c_{X2,n}(t)|^2,
\label{eq:fbright}
\end{equation}
while the interlayer fraction is
\begin{equation}
f_n^{\mathrm{iX}}(t)
=
|c_{iX,n}(t)|^2.
\label{eq:fiX}
\end{equation}
The bright fraction determines the optical matrix-element weight and
therefore controls the state dependence of the optical injection and
radiative transition rates, whereas the interlayer fraction tracks the
spatially indirect character of the dressed exciton~\cite{Wang2018}.
The same dressed-state construction is also used for the phonon
coupling in Eq.~\eqref{eq:HSB}. In this case, the microscopic
exciton-phonon interaction is projected between pairs of instantaneous
dressed states, giving the couplings \(g^c_{ij,q}(t)\). These couplings
originate from deformation-potential and Fr\"{o}hlich
exciton-phonon interactions~\cite{Kaasbjerg2012,Shree2018}.

\subsection{Geometric loop construction}
\label{sec:drivencontrols}

\MODNEW{The driving vector \(\mathbf{R}(t)\) is defined by two slow
external controls: the gate voltage \(V_G(t)\) and the dimensionless
pump-control parameter \(I_{\mathrm{pump}}(t)\).}
\MODNEW{The pump does not enter the exciton Hamiltonian directly. Instead,
it enters through the optical-injection term in the system-bath coupling
\(\hat{H}_{\mathrm{SB}}(t)\), thereby modulating the injection rates.} Equivalently, the optical-pump coupling amplitude is
\cite{SalehTeich2019}
\(
\eta_{n\alpha}(t)
=
\eta_{\alpha}^{0}
\sqrt{I_{\mathrm{pump}}(t)}\,M_n(t),
\)
where \(M_n(t)\) is the optical transition matrix element between
\(|g\rangle\) and the instantaneous dressed exciton
\(|\psi_n(t)\rangle\). Since the absorption rate is proportional to the
square of the optical field amplitude, the corresponding golden-rule
transition rate is proportional to \(I_{\mathrm{pump}}(t)\).
\MODNEW{For the dressed states used here, the optical matrix element is
dominated by the intralayer bright components.
\(|M_n(t)|^2\) is represented by the bright fraction
\(f_n^{\mathrm{bright}}(t)\). This leads to an optical-injection rate
proportional to
\(I_{\mathrm{pump}}(t)f_n^{\mathrm{bright}}(t)\), as used in the rate
model below.}

\MODNEW{The two controls are driven periodically as}
\begin{equation}
V_G(t)=V_G^{(0)}+A_G\sin(\omega t+\phi_G),
\label{eq:VG_drive}
\end{equation}
\begin{equation}
I_{\mathrm{pump}}(t)=I_{\mathrm{pump}}^{(0)}
+A_I\sin(\omega t+\phi_I),
\label{eq:Ipump_drive}
\end{equation}

\MODNEW{where \(V_G^{(0)}\) and \(I_{\mathrm{pump}}^{(0)}\) are the
loop-centre coordinates, \(A_G\) and \(A_I\) are the gate and pump
modulation amplitudes, \(\phi_G\) and \(\phi_I\) are the phases, and
\(\omega=2\pi/\tau\) is the angular frequency for a cycle of period
\(\tau\). We refer to \(V_G^{(0)}\), \(I_{\mathrm{pump}}^{(0)}\),
\(A_G\), and \(A_I\) collectively as the loop-geometry quantities.} The relative phase is
\(
\Delta\phi=\phi_I-\phi_G .
\)
\MODNEW{The out-of-plane field \(F_z^{(0)}\) is held fixed during one
loop. Thus \(V_G(t)\) changes the instantaneous Hamiltonian
\(\hat{H}_T[F_z^{(0)},V_G(t)]\), and the dressed energies, gaps, and
exciton characters. By contrast, \(I_{\mathrm{pump}}(t)\) changes only
the optical-injection rates and does not modify \(\hat{H}_T\). For nonzero
\(A_G\), nonzero \(A_I\), and nonzero \(\Delta\phi\), the trajectory
encloses an oriented area in the
\((V_G(t),I_{\mathrm{pump}}(t))\) control plane. For the sinusoidal
protocol above, this oriented area is proportional to
\(A_GA_I\sin\Delta\phi\). Changing
\(\Delta\phi\rightarrow-\Delta\phi\) reverses the loop orientation and
therefore reverses the sign of the geometric contribution, leaving
the orientation-even dynamic contribution unchanged.}

\subsection{From Microscopic Coupling to Population Rates}
\label{sec:micro_to_rates}

The second-order Born Markov secular \cite{Breuer2002,Campaioli2024} reduction
\(\hat{H}_{\mathrm{SB}}(t)\), including the bath correlation functions,
golden-rule transition structure, and projection onto the Pauli
population equation, is given in Sec.~S3.3 of the Supporting
Information. For the population vector,
\(
\mathbf{p}(t)
=
[p_g(t),p_0(t),p_1(t),p_2(t)]^{T},
\)
where \(p_g(t)\) is the no-exciton ground-state population and
\(p_n(t)\) is the population of
\(\lvert\psi_n(t)\rangle\), the ordinary population Liouvillian is
\begingroup
\setlength{\arraycolsep}{3pt}
\begin{equation}
\mathcal{L}(t)
=
\begin{pmatrix}
-\sum\limits_{n=0}^{2}a_n & b_0 & b_1 & b_2\\
a_0 & -b_0 & 0 & 0\\
a_1 & 0 & -(b_1+u) & d\\
a_2 & 0 & u & -(b_2+d)
\end{pmatrix}.
\label{eq:Lordinary}
\end{equation}
\endgroup

For compactness, the common time arguments are suppressed in
Eq.~\eqref{eq:Lordinary}, so that
\(a_n\equiv a_n(t),\ b_n\equiv b_n(t),\
d\equiv d_{ij}(t),\ u\equiv u_{ij}(t),\
\Omega\equiv\Omega_{ij}(t)\). The optical injection rates are
\(a_n(t)\) from \(|g\rangle\) to \(|\psi_n(t)\rangle\), the radiative
loss rates are \(b_n(t)\) from \(|\psi_n(t)\rangle\) to
\(|g\rangle\), and phonon rates between dressed branches. The three
quantities \(a_0,a_1,a_2\) are the optical injection rates into the
three dressed exciton branches, while \(b_0,b_1,b_2\) are the
corresponding radiative recombination rates back to the no-exciton
ground state. The optical injection rate is
\begin{equation}
a_n(t)
=
\Gamma_{\mathrm{pump}}^{0}
I_{\mathrm{pump}}(t)
f_n^{\mathrm{bright}}(t).
\label{eq:an}
\end{equation}

Here, \(\Gamma_{\mathrm{pump}}^{0}\) is the optical-injection
prefactor, \(I_{\mathrm{pump}}(t)\) is the externally controlled pump
factor, and \(f_n^{\mathrm{bright}}(t)\) is the bright state fraction
of the instantaneous dressed branch. The radiative recombination rate is
\begin{equation}
b_n(t)
=
\gamma_{\mathrm{rad}}^{\mathrm{intra}}
f_n^{\mathrm{bright}}(t)
+
\gamma_{\mathrm{rad}}^{\mathrm{inter}}
f_n^{\mathrm{iX}}(t).
\label{eq:bn}
\end{equation}
whose first term is the intralayer-bright contribution and
second term the weaker interlayer contribution.

The instantaneous dressed-state gap is
\begin{equation}
\Omega_{ij}(t)
=
\epsilon_j(t)-\epsilon_i(t),
\qquad i<j.
\label{eq:Omegaij}
\end{equation}

The downward and upward phonon rates are
\begin{align}
d_{ij}(t)
&=
\gamma_{c,ij}(t)
\left\{
n_c[\Omega_{ij}(t)]+1
\right\},
\\
u_{ij}(t)
&=
\gamma_{c,ij}(t)
n_c[\Omega_{ij}(t)],
\label{eq:d12u12}
\end{align}
with phonon occupation
\begin{equation}
n_c[\Omega_{ij}(t)]
=
\left[
\exp\!\left(
\frac{\Omega_{ij}(t)}{k_{\mathrm B}T_c}
\right)-1
\right]^{-1}.
\label{eq:bose}
\end{equation}

The corresponding overlap-weighted phonon coupling is
\begin{equation}
\begin{aligned}
\gamma_{c,ij}(t)
={}&
\gamma_c^{\mathrm{intra}}
\left|
c_{X1,i}^{*}(t)c_{X1,j}(t)
+
c_{X2,i}^{*}(t)c_{X2,j}(t)
\right|^2
\\
&+
\gamma_c^{\mathrm{inter}}
\left|
c_{iX,i}^{*}(t)c_{iX,j}(t)
\right|^2.
\end{aligned}
\label{eq:gammacij}
\end{equation}

The first term in Eq.~\eqref{eq:gammacij} is the coherent intralayer
overlap contribution, while the second is the interlayer overlap
contribution. The constants
\(\Gamma_{\mathrm{pump}}^{0}\),
\(\gamma_{\mathrm{rad}}^{\mathrm{intra}}\),
\(\gamma_{\mathrm{rad}}^{\mathrm{inter}}\),
\(\gamma_c^{\mathrm{intra}}\), and
\(\gamma_c^{\mathrm{inter}}\)
are phenomenological rate prefactors.

\subsection{Phonon Counting Statistics}\label{sec:onephonon}

The rate model allows phonon-mediated transitions between dressed exciton
branches. To define a strict phonon counting problem, we first apply
a single-phonon accessibility filter to the instantaneous dressed-state
gaps
$
\Omega_{ij}(t)$.
Only those transitions whose gaps remain within the  phonon
energy window over the full driving loop are retained as possible
phonon counting channels. \MODNEW{Tilting the retained phonon transition $|\psi_i(t)\rangle\leftrightarrow|\psi_j(t)\rangle$, with $i<j$, gives the tilted master equation~\cite{Campisi2011,Esposito2009},}
\begin{equation}
\frac{d}{dt}\mathbf{p}_s(t)
=
\mathcal{L}_{s}(t)\,\mathbf{p}_s(t),
\label{eq:tilted_master}
\end{equation}
where $\mathcal{L}_{s}^{}(t)$ is the phonon counting tilted Liouvillian given by
\begingroup
\setlength{\arraycolsep}{3pt}
\begin{equation}
\mathcal{L}_{s}^{}
=
\begin{pmatrix}
-\sum\limits_{n=0}^{2}a_n & b_0 & b_1 & b_2\\
a_0 & -b_0 & 0 & 0\\
a_1 & 0 & -(b_1+u) & d e^{s\Omega_{ij}}\\
a_2 & 0 & u e^{-s\Omega_{ij}} & -(b_2+d)
\end{pmatrix}.
\label{eq:tiltedL}
\end{equation}
\endgroup
All quantities in Eq.~\eqref{eq:tiltedL} are evaluated at the same time $t$. The factors $e^{+s\Omega_{ij}}$ and $e^{-s\Omega_{ij}}$ tag, respectively, the downward phonon emission and reverse phonon absorption through the selected phonon channel. At $s=0$, Eq.~\eqref{eq:tiltedL} reduces to the ordinary population Liouvillian. 
In the numerics we propagate Eq.~\eqref{eq:tiltedL} directly rather
than evaluating the curvature integral of
Eq.~\eqref{eq:Ggeo_general}. The curvature is computed
separately in Sec.~S9 of the Supporting Information as an independent
cross-check.

\subsection{Loop-Reversal Estimator}\label{sec:loopreversal}

\MODNEW{The dynamic part is even and the geometric part is odd\cite{Sinitsyn2007,Yuge2012}.} A one-period propagator is evaluated for both loop orientations. Let
$\eta=\mathrm{CW},\mathrm{CCW}$ label a clockwise or counter-clockwise
traversal of the control loop, corresponding to the phase-lag choices
$\Delta\phi$ and $-\Delta\phi$. The tilted generator of
Eq.~\eqref{eq:tiltedL} is propagated over one period as
\begin{equation}
\mathcal{U}_{s,\eta}(\tau)
=
\mathcal{T}
\exp\!\left[
\int_0^{\tau}\mathcal{L}_{s,\eta}(t)\,dt
\right].
\label{eq:U_orientation}
\end{equation}
The only difference between the two orientations is the time order in
which the loop visits its control points; everything else in the
generator is identical.
\MODNEW{The orientation-resolved finite-time generating function is}
\begin{equation}
G_{\eta}(s,\tau)
=
\ln\!\left[
\mathbf{1}^{T}\mathcal{U}_{s,\eta}(\tau)\,\mathbf{p}(0)
\right].
\label{eq:finite_time_G}
\end{equation}
Its derivatives at $s=0$ give the orientation-resolved cumulants,
\begin{equation}
C_n^{\eta}(\tau)
=
\left.
\frac{\partial^nG_{\eta}(s,\tau)}{\partial s^n}
\right|_{s=0},
\label{eq:Cn_orientation}
\end{equation}
\MODNEW{Writing the finite-time cumulant as orientation-even and orientation-odd parts, the clockwise value is}
\begin{equation}
C_n^{\mathrm{CW}}(\tau)
=
C_n^{\mathrm{even}}(\tau)
+
C_n^{\mathrm{odd}}(\tau).
\label{eq:CW_split}
\end{equation}
Reversing the loop leaves the even part unchanged and flips the sign
of the odd part,
\begin{equation}
C_n^{\mathrm{CCW}}(\tau)
=
C_n^{\mathrm{even}}(\tau)
-
C_n^{\mathrm{odd}}(\tau).
\label{eq:CCW_split}
\end{equation}
 
Adding and subtracting Eqs.~\eqref{eq:CW_split} and
\eqref{eq:CCW_split} separates the two parts. The even part is the
half-sum,
\begin{equation}
C_n^{\mathrm{even}}(\tau)
=
\tfrac{1}{2}\left[
C_n^{\mathrm{CW}}(\tau)+C_n^{\mathrm{CCW}}(\tau)
\right].
\label{eq:Ceven}
\end{equation}
The odd part is the half-difference,
\begin{equation}
C_n^{\mathrm{odd}}(\tau)
=
\tfrac{1}{2}\left[
C_n^{\mathrm{CW}}(\tau)-C_n^{\mathrm{CCW}}(\tau)
\right].
\label{eq:Codd}
\end{equation}
In the
slow-driving expansion, reversing the loop changes the sign of the
geometric contribution while leaving the dynamic contribution
unchanged~\cite{Sinitsyn2007,Ren2010,Yuge2012,Paulino2024}. Thus,
\begin{equation}
G_{\eta}(s,\tau)
=
G_{\mathrm{dyn}}(s,\tau)
+
\sigma_{\eta}G_{\mathrm{geo}}(s)
+
\mathcal{O}(\tau^{-1}),
\label{eq:orientation_generating_expansion}
\end{equation}
where
\(\sigma_{\mathrm{CW}}=+1\) and
\(\sigma_{\mathrm{CCW}}=-1\) are the corresponding orientation-sign
factors \(\mathcal{O}(\tau^{-1})\) denote the leading
finite-speed correction, which vanishes at least as \(1/\tau\) in the
slow-driving limit \(\tau\rightarrow\infty\). Differentiating Eq.~\eqref{eq:orientation_generating_expansion}
according to Eq.~\eqref{eq:Cn_orientation} gives
\begin{equation}
C_n^{\eta}(\tau)
=
C_n^{\mathrm{dyn}}(\tau)
+
\sigma_{\eta}C_n^{\mathrm{geo}}
+
\mathcal{O}(\tau^{-1}),
\label{eq:orientation_cumulant_expansion}
\end{equation}
where the final term represents the corresponding finite-period
correction to the \(n\)th cumulant. Substituting
Eq.~\eqref{eq:orientation_cumulant_expansion} into
Eqs.~\eqref{eq:Ceven} and \eqref{eq:Codd} yields
\begin{align}
C_n^{\mathrm{even}}(\tau)
&=
C_n^{\mathrm{dyn}}(\tau)
+
\mathcal{O}(\tau^{-1}),
\label{eq:even_dynamic_limit}
\\
C_n^{\mathrm{odd}}(\tau)
&=
C_n^{\mathrm{geo}}
+
\mathcal{O}(\tau^{-1}).
\label{eq:odd_geometric_limit}
\end{align}
Taking the slow-driving limit of
Eq.~\eqref{eq:odd_geometric_limit} gives

\begin{equation}
C_n^{\mathrm{geo}}
=
\lim_{\tau\rightarrow\infty}
C_n^{\mathrm{odd}}(\tau).
\label{eq:geo_limit}
\end{equation}
At a
finite period for which the orientation-even sector is time extensive
and the orientation-odd sector has reached its plateau, we therefore
use the operational identification

\begin{equation}
C_n^{\mathrm{dyn}}(\tau)
=
C_n^{\mathrm{even}}(\tau).
\label{eq:dyn_finite}
\end{equation}

\section{Numerical Implementation}
\label{sec:paramsurface}
The exciton Hamiltonian $\hat{H}_T$ in Eq.~\eqref{eq:HT} is reconstructed
from the available data for the experimentally studied
layer-hybridized WSe$_2$/WS$_2$ moir'e-exciton system~\cite{Tang2021}.
The reconstructed parameter values are listed in Table~S2 of the
Supporting Information. Energies and hybridization couplings are
reported in meV, the Stark slope $D$ in
$\mathrm{meV}/(\mathrm{V,nm^{-1}})$, and rates in
$\mathrm{ps^{-1}}$. The gate-voltage axis spans
$0.5$ to $6.5~\mathrm{V}$ in $0.5~\mathrm{V}$ steps, and at each
gate voltage the dressed exciton branches are ordered by increasing
energy as $\epsilon_0<\epsilon_1<\epsilon_2$.
Within the one-exciton manifold, the off-diagonal matrix elements
$W_1(V_G)$ and $W_2(V_G)$ describe hybridization of the interlayer
exciton $\vert{}iX\rangle$ with the intralayer excitons
$\vert{}X_1\rangle$ and $\vert{}X_2\rangle$, respectively~\cite{RuizTijerina2019}.
Their inclusion therefore produces dressed exciton states containing
both interlayer and intralayer character. These terms describe
single-exciton hybridization and should not be confused with
exciton-exciton interactions, which require the simultaneous presence
of two or more excitons and lie outside the reduced Hamiltonian used
here. No direct $X_1$-$X_2$ coupling is included.
A separate consideration is required for the $X_1$ energy used in
the reconstructed Hamiltonian. The experimentally reported $X_1$
resonance does not evolve as a single smooth branch over the full gate
range~\cite{Tang2021}. Across the
$\nu=0\!\to\!1$ and $\nu=1\!\to\!2$ transitions, two
$X_1$-like spectral features can be resolved simultaneously, with
different oscillator strengths. The experimental fit reports both
features together with their corresponding oscillator strengths.
For the reduced model used here, however, a single effective
$X_1$ energy is required at each gate voltage. We therefore choose
the spectrally stronger fitted branch and set
$E_1(V_G)=\varepsilon_1^{\rm dom}(V_G)$, where
$\varepsilon_1^{\rm dom}$ denotes the dominant $X_1$-like feature.
The fitted interlayer-exciton parameter $E_0$ follows the same
dominant branch assignment.
Table~\ref{tab:compact_model_parameters} summarizes the numerical
ranges, physical status, role in the model, and source or rationale for
the parameters used in the experimentally anchored phonon calculation.

\subsection{Working Driving Protocol and Rate Constants}\label{sec:baseline_protocol}
In the  \MODNEW{working driving loop used for the initial finite time tests, we use} two parameter driving protocol of \MODNEW{Eqs.~\eqref{eq:VG_drive}-\eqref{eq:Ipump_drive},} with
$V_G^{(0)}=
3.5, A_G=3.0\,\omega = 2\pi /\tau)~\mathrm{V},
I_{pump}^{(0)}=
1.3, A_I=1.2.$
The phase lag is \(\Delta\phi=\pi/2\), and the cycle period considered  \MODNEW{for the initial propagation} is
\(\tau=5000~\mathrm{ps}\). This period is later varied to check finite time convergence. The driven gate trajectory covers the full
experimentally-supported range \(V_G(t)=0.5\)-\(6.5~\mathrm{V}\), while the
dimensionless pump-control parameter lies in the interval
\(I_{\mathrm{pump}}(t)=0.1\)-\(2.5\). The reversed loop needed to extract the 
geometric contribution is obtained by changing
\(\Delta\phi\rightarrow-\Delta\phi\). The remaining numerical parameters are reported in Table ~\ref{tab:compact_model_parameters}. 
 The intralayer radiative scale
$\gamma_{\mathrm{rad}}^{\mathrm{intra}}=0.500~\mathrm{ps^{-1}}$
corresponds to a $\sim2~\mathrm{ps}$ lifetime, consistent with the
picosecond intrinsic bright-exciton radiative lifetimes reported for TMD
monolayers~\cite{Robert2016}, and the interlayer scale
$\gamma_{\mathrm{rad}}^{\mathrm{inter}}=5\times10^{-4}~\mathrm{ps^{-1}}$
corresponds to $ \sim2~\mathrm{ns}$ lifetime, close to the long-lived
interlayer-exciton lifetimes reported in TMD
heterobilayers~\cite{Rivera2015}. The optical-pump prefactor
$\Gamma_{\mathrm{pump}}^0$ and the phonon scales
$\gamma_c^{\mathrm{intra}},\gamma_c^{\mathrm{inter}}$ are not fitted to a
specific measured rate; they are phenomenological values chosen to
represent efficient optical injection and phonon-assisted exciton relaxation, motivated by established exciton-phonon scattering and
phonon-cascade processes in TMD
monolayers~\cite{Christiansen2017,Brem2020,Shree2018,Paradisanos2021}.} The sensitivity of the 
geometric response to these \MODNEW{to these phenomenological rate prefactors is tested later in } 
Section~\ref{sec:res_sensitivity} \MODNEW{after the central loop and comparison conditions have been selected through the control-centre scan}.

\subsection{Finite-Time FCS Propagation}
\label{sec:fcs_propagation}

 \MODNEW{The cumulants are obtained from Eq.~\eqref{eq:tiltedL} using Eqs.~\eqref{eq:Ceven} and \eqref{eq:Codd}, within the standard numerical procedure outlined in Sec. S5 of supporting information.}

\MODNEW{Eqs.~\eqref{eq:Ceven} and \eqref{eq:Codd} give the exact finite-time even and odd cumulants. Their dynamic/geometric identification requires two numerical conditions.} First, the
orientation-odd cumulant agree with its long-period reference value
within the considered tolerance,
\begin{equation}
\varepsilon_{n}^{\mathrm{odd}}(\tau)
=
\frac{
\left|
C_n^{\mathrm{odd}}(\tau)
-
C_n^{\mathrm{odd}}(\tau_{\mathrm{ref}})
\right|
}{
\left|
C_n^{\mathrm{odd}}(\tau_{\mathrm{ref}})
\right|
}
<
\varepsilon_{\mathrm{tol}},
\label{eq:odd_plateau_criterion}
\end{equation}
where \(\tau_{\mathrm{ref}}\) is a cycle period on the numerically
converged long-period plateau. Second, the orientation-even cumulant
must display time-extensive scaling,
\begin{equation}
\varepsilon_{n}^{\mathrm{even}}(\tau)
=
\left|
\frac{
C_n^{\mathrm{even}}(\tau)/\tau
}{
C_n^{\mathrm{even}}(\tau_{\mathrm{ref}})/\tau_{\mathrm{ref}}
}
-1
\right|
<
\varepsilon_{\mathrm{tol}}.
\label{eq:even_extensive_criterion}
\end{equation}
We ensure Eqs.~\eqref{eq:odd_plateau_criterion} and
\eqref{eq:even_extensive_criterion} are satisfied, and identify
\begin{equation}
C_n^{\mathrm{odd}}(\tau)
\simeq
C_n^{\mathrm{geo}}(\tau),
\qquad
C_n^{\mathrm{even}}(\tau)
\simeq
C_n^{\mathrm{dyn}}(\tau).
\label{eq:finite_time_geo_dyn_identification}
\end{equation}

As an additional time-scale diagnostic, the cycle period is taken to be
much longer than the slowest population-relaxation time encountered
along the loop, \(\tau\gg\tau_{\mathrm{rel}}^{\max}\), where
 \(\tau_{\mathrm{rel}}^{\max}=1/\Delta_L^{\min}\).
Here, \(\Delta_L^{\min}\) is the smallest nonzero population-relaxation
rate encountered over the complete driving loop, obtained from the
instantaneous spectrum of the ordinary Liouvillian. Its inverse,
\(\tau_{\mathrm{rel}}^{\max}=1/\Delta_L^{\min}\), is the corresponding
longest characteristic population-relaxation time. The condition
\(\tau\gg\tau_{\mathrm{rel}}^{\max}\) therefore ensures that the cycle
duration is much longer than the slowest intrinsic population-relaxation
time along the loop.
No one of these three checks is sufficient by itself. The relaxation
condition establishes
time-scale separation but does not prove convergence of the geometric
cumulant. Similarly, an apparently flat odd cumulant alone does not
verify the time-extensive dynamic sector. The dynamic/geometric
terminology is therefore used only when the odd-cumulant plateau and
even-cumulant scaling are both verified, with the Liouvillian
relaxation time providing an independent supporting diagnostic.

\begin{table*}[t]
\centering
\scriptsize
\setlength{\tabcolsep}{2.5pt}
\renewcommand{\arraystretch}{1.08}
\caption{Parameters used in the experimentally anchored
WSe$_2$/WS$_2$ phonon energy transfer model. Hamiltonian inputs are
digitized from Ref.~\cite{Tang2021}. The control protocols, operating
field, dressed gaps, and phonon selection cap are chosen or calculated
in this work. The radiative rates are guided by reported exciton
lifetimes, while the optical injection and phonon rates are
phenomenological baseline values.}
\label{tab:compact_model_parameters}

\begin{tabular}{p{2.6cm} p{2.8cm} p{2.4cm} p{3.4cm} p{4.8cm}}
\toprule
\textbf{Parameter} &
\textbf{Value or range} &
\textbf{Description} &
\textbf{Use in this work} &
\textbf{Source or rationale}
\\
\midrule

Hamiltonian input domain
&
\(0.5\leq V_G\leq6.5~\mathrm{V}\)
&
Experimental input
&
Defines the gate range used to reconstruct
\(E_i(V_G)\), \(W_i(V_G)\), and \(D(V_G)\) before diagonalizing \(\hat{H}_T\).
&
Experimentally supported gate range from Ref.~\cite{Tang2021}.
\\

\(E_0,E_1,E_2\)
&
\(1939\)-\(2040~\mathrm{meV}\)
&
Experimental input
&
Gate dependent uncoupled exciton energies. The \(E_1\) branch follows
the dominant \(X_1\) character.
&
Digitized from the fitted curves in Ref.~\cite{Tang2021}.
\\

\(W_1,W_2\)
&
\(W_1=37\)-\(48~\mathrm{meV}\);
\(W_2=18\)-\(33~\mathrm{meV}\)
&
Experimental input
&
Hybridization strengths entering \(\hat{H}_T\).
&
Digitized from the fitted curves in Ref.~\cite{Tang2021}.
\\

Stark slope \(D\)
&
\(292\)-\(348~\mathrm{meV/(V\,nm^{-1})}\)
&
Experimental input
&
Controls the field shift \(D F_z\) of the interlayer exciton.
&
Digitized from the fitted curves in Ref.~\cite{Tang2021}.
\\

Gate drive \(V_G(t)\)
&
\(0.5\leq V_G(t)\leq6.5~\mathrm{V}\)
&
External control
&
Changes the dressed energies, gaps, and state character along the loop.
&
Chosen to span the experimentally supported gate range of
Ref.~\cite{Tang2021}.
\\

Out of plane field \(F_z^{(0)}\)
&
Main value:
\(+0.17~\mathrm{V\,nm^{-1}}\)
&
Fixed external field
&
Selects the working dressed gap window.
&
Chosen from the calculated field scan using the Hamiltonian anchored to
Ref.~\cite{Tang2021}.
\\

Pump modulation \(I_{\mathrm{pump}}(t)\)
&
\(0.1\)-\(2.5\);
dimensionless
&
External control
&
Scales optical injection without entering \(\hat{H}_T\).
&
Chosen dimensionless control in this work.
\\

Optical injection prefactor
\(\Gamma_{\mathrm{pump}}^0\)
&
\(0.060~\mathrm{ps^{-1}}\)
&
Phenomenological rate
&
Sets the injection scale at \(I_{\mathrm{pump}}=1\) before applying the
pump modulation and bright state fraction.
&
Motivated by incoherent exciton pumping schemes used in quantum dot cavity and moir'e exciton models \cite{Yao2010,Song2025}.
\\

Cycle period \(\tau\)
&
\(200\)-\(50000~\mathrm{ps}\);
target \(20\)-\(50~\mathrm{ns}\)
&
Adiabatic control
&
Sets the loop frequency and finite time convergence.
&
Chosen in this work. The geometric pumping framework follows
Refs.~\cite{Sinitsyn2007,Ren2010,Paulino2024}.
\\

Phonon bath temperature \(T_c\)
&
\(5~\mathrm{K}\);
scan \(4\)-\(20~\mathrm{K}\)
&
Cryogenic bath
&
Determines the Bose factor \(n_c[\Omega_{12}(t)]\).
&
Chosen as a representative cryogenic range for TMDC spectroscopy
\cite{Tang2021,Robert2016}.
\\

Counted dressed gap \(\Omega_{12}(t)\)
&
\(31.03\)-\(58.21~\mathrm{meV}\)
&
Counted energy gap
&
Energy carried by the
\(\lvert\psi_2\rangle\rightarrow\lvert\psi_1\rangle\)
phonon transition.
&
Calculated from the anchored Hamiltonian. Related TMDC phonon energies
and phonon assisted processes are discussed in
Refs.~\cite{Zhao2013,Paradisanos2021,Brem2020}.
\\

Excluded dressed gaps
\(\Omega_{01}(t),\Omega_{02}(t)\)
&
\(67.64\)-\(82.82~\mathrm{meV}\);
\(104.99\)-\(141.04~\mathrm{meV}\)
&
Excluded channels
&
Remain above the phonon selection cap and are not counted.
&
Calculated from the anchored Hamiltonian.
\\

Phonon selection cap
\(\Omega_{\mathrm{cut}}\)
&
\(60~\mathrm{meV}\)
&
Model threshold
&
Retains the full \(\Omega_{12}(t)\) path while excluding the higher gaps.
&
Chosen relative to the WS$_2$ first order Raman modes near
\(356~\mathrm{cm^{-1}}\) and \(420~\mathrm{cm^{-1}}\), corresponding
to about \(44\) and \(52~\mathrm{meV}\), respectively
\cite{Zhao2013}.
\\

Radiative prefactors
&
\(\gamma_{\mathrm{rad}}^{\mathrm{intra}}
=0.500~\mathrm{ps^{-1}}\);
\(\gamma_{\mathrm{rad}}^{\mathrm{inter}}
=5\times10^{-4}~\mathrm{ps^{-1}}\)
&
Phenomenological rates
&
Set the intralayer and interlayer recombination strengths before
applying the dressed state fractions.
&
Their inverse times, \(2~\mathrm{ps}\) and \(2~\mathrm{ns}\), are close
to reported lifetimes of \(1.8~\mathrm{ps}\) and \(1.8~\mathrm{ns}\)
in related TMDC systems \cite{Robert2016,Rivera2015}.
\\

Phonon prefactors
&
\(\gamma_c^{\mathrm{intra}}=0.070~\mathrm{ps^{-1}}\);
\(\gamma_c^{\mathrm{inter}}=0.010~\mathrm{ps^{-1}}\)
&
Phenomenological rates
&
Set the retained \(1\leftrightarrow2\) phonon relaxation scale before
applying the overlap factors.
&
Motivated by phonon-assisted exciton relaxation in related TMD
systems~\cite{Christiansen2017,Brem2020,Shree2018,Paradisanos2021}. The rates
are phenomenological and are not extracted from these
references.
\\

Counting field \(s\)
&
\(\mathrm{meV^{-1}}\);
evaluated at \(s=0\)
&
FCS variable
&
Tags energy transferred through the \(\Omega_{12}(t)\) channel.
&
Defined within the full counting statistics framework
\cite{Esposito2009,Sinitsyn2007,Paulino2024}.
\\

Cumulants
&
\(C_1\): \(\mathrm{meV/cycle}\);
\(C_2\): \(\mathrm{meV^2/cycle}\)
&
Energy statistics
&
Give the mean and noise of phonon energy transfer per cycle.
&
Definitions and geometric interpretation follow
Refs.~\cite{Esposito2009,Ren2010,Paulino2024}.
\\

\bottomrule
\end{tabular}
\end{table*}
\FloatBarrier
\section{Results and Discussion}\label{sec:results}
 We first
establish the selection of the working operating field and the relaxation pathway which satisfies the phonon energy condition. We then test the cycle-period convergence of the
finite-time cumulants, followed by examination of how the response depends on the
range constrained loop centre
and on
phenomenological rate-prefactor sensitivity. We then check its
robustness against static energy disorder, and estimate the possible scale of
residual higher-order relaxation pathways through the higher
dressed gaps. Finally discuss an optical loop-reversal proxy
together with an enhanced-signal budget. At each instantaneous value of \(V_G(t)\), the Hamiltonian is
diagonalized to obtain the dressed energies \(\epsilon_n(t)\), the gaps
\(\Omega_{ij}(t)=\epsilon_j(t)-\epsilon_i(t)\), and the bright and
interlayer fractions defined in
Eqs.~\eqref{eq:fbright}-\eqref{eq:fiX}.
The corresponding input checks are shown in Fig.~S1.

\MODNEW{We adopt \(60~\mathrm{meV}\) as a practical phonon energy selection
threshold. This value is chosen as a modest extension of the first order
WS$_2$ Raman modes near
\(420~\mathrm{cm^{-1}}\), corresponding to approximately \(52~\mathrm{meV}\) \cite{Zhao2013}. We select
\(F_z^{(0)}=+0.17~\mathrm{V\,nm^{-1}}\)based on a gap window scan in Fig.~S2 for which the complete
\(\Omega_{12}(t)\) trajectory remains below 60 meV, whereas
\(\Omega_{01}(t)\) and \(\Omega_{02}(t)\) remain above it.} \MODNEW{The threshold retains the counted dressed transition and
excludes the two higher dressed gaps.} \MODNEW{The dressed state bright and interlayer fractions in
Figs.~S2(c,d) and S3 determine the rates in
Eqs.~\eqref{eq:an}-\eqref{eq:gammacij}.} The consistency of retaining \(\Omega_{12}(t)\) as the only counted
phonon assisted channel is checked in Sec. S6 of the Supporting Information. \MODNEW{We then evaluate the orientation even and orientation odd
cumulants using Eqs.~\eqref{eq:Ceven} and \eqref{eq:Codd},
respectively. We next test the convergence with the cycle period.}

\FloatBarrier
\subsection{Cycle-period validation and finite-time cumulants}
\label{sec:res_tau}
The working gate–pump loop fixes the closed path in the (\(V_G(t),I_{\mathrm{pump}}(t))\) control plane, while the cycle period \(\tau \) determines the rate at which this path is traversed. Thus, varying \(\tau \) leaves the loop shape and enclosed control space region unchanged but rescales the driving speed, thereby controlling the magnitude of finite driving corrections and the convergence of the orientation odd cumulants toward their slow driving geometric limit.

\begin{figure}[htbp]
\centering
\includegraphics[width=\columnwidth]
{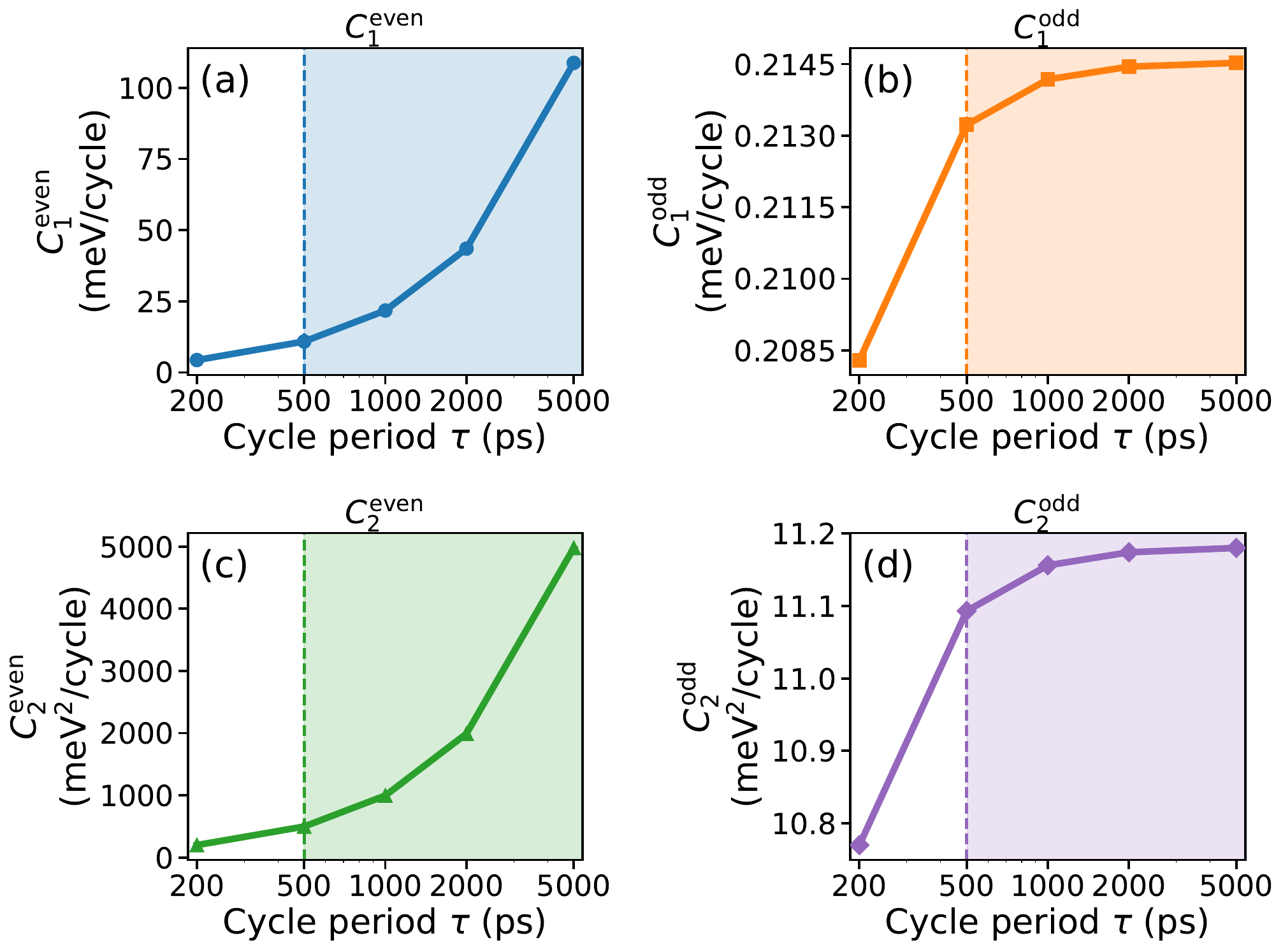}
\caption{Calculated cycle-period dependence of the exact
finite-time orientation-even and orientation-odd cumulants per completed
loop. Panels~(a) and (c) show \(C_1^{\mathrm{even}}\) and
\(C_2^{\mathrm{even}}\), while panels~(b) and (d) show
\(C_1^{\mathrm{odd}}\) and \(C_2^{\mathrm{odd}}\). The shaded region,
beginning at \(500~\mathrm{ps}\), marks the sampled periods for which
both the orientation-odd plateau criterion and the orientation-even
time-extensive-scaling criterion are satisfied within
\(1\%\) tolerance for \(n=1,2\), and remain satisfied at all longer
sampled periods. Within this validated region,
\(C_n^{\mathrm{even}}\) and \(C_n^{\mathrm{odd}}\) are identified with
\(C_n^{\mathrm{dyn}}\) and \(C_n^{\mathrm{geo}}\), respectively.}
\label{fig:tau_cumulants}
\end{figure}

\begin{figure}[htbp]
\centering
\includegraphics[width=\columnwidth]
{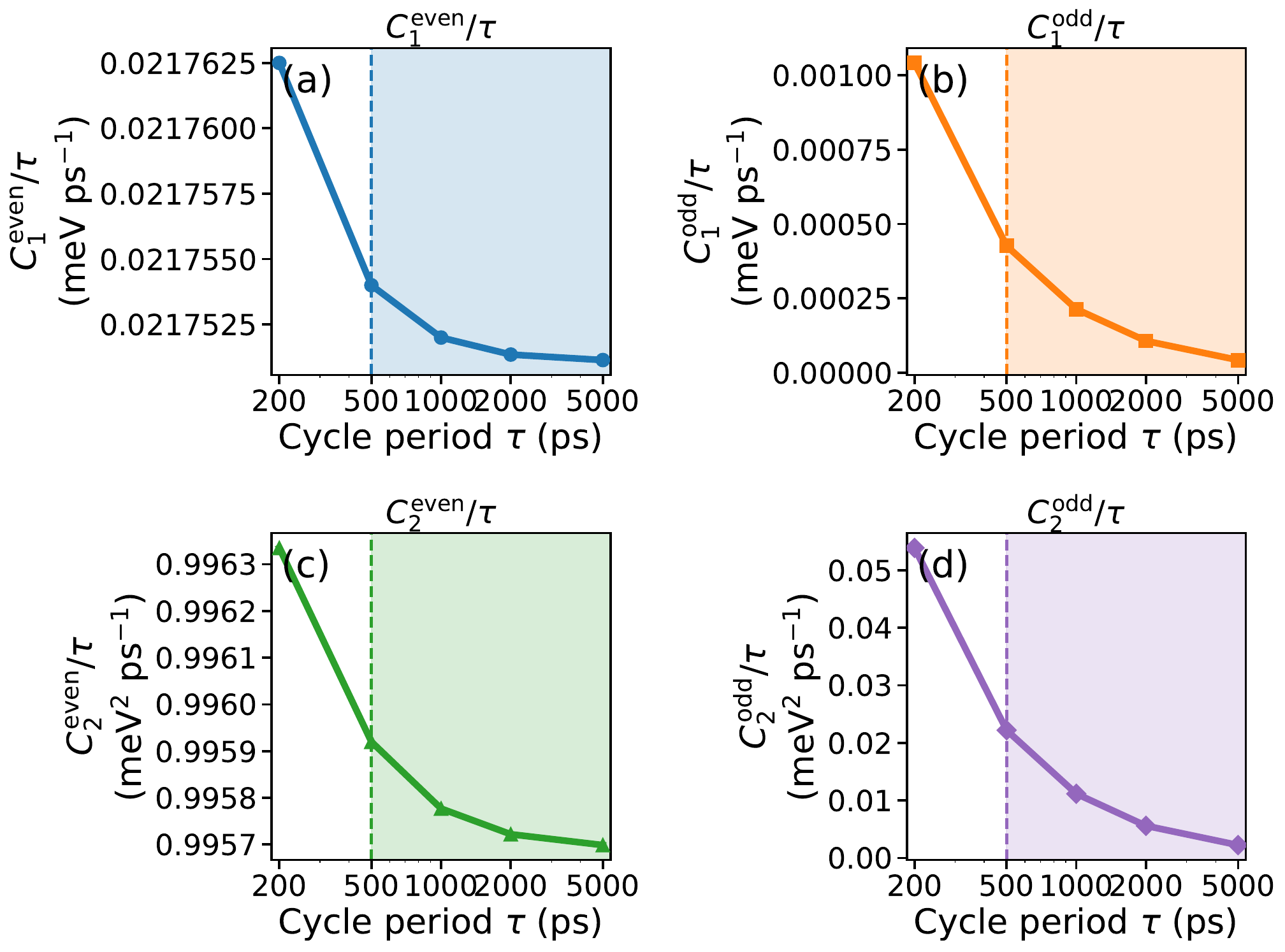}
\caption{Calculated cycle-period dependence of the finite-time
orientation-even and orientation-odd cumulants per unit time. Panels~(a) and (c) show \(C_1^{\mathrm{even}}/\tau\) and
\(C_2^{\mathrm{even}}/\tau\), while panels~(b) and (d) show
\(C_1^{\mathrm{odd}}/\tau\) and
\(C_2^{\mathrm{odd}}/\tau\). The nearly constant orientation-even
rates verify the time-extensive condition in
Eq.~\eqref{eq:even_extensive_criterion}. The orientation-odd rates
decrease approximately as \(1/\tau\) because their unnormalized amounts
approach constants per completed loop. The shaded region has the same
meaning as in Fig.~\ref{fig:tau_cumulants} and marks the sampled periods
satisfying the combined \(1\%\) criterion}
\label{fig:tau_cumulants_normalized}
\end{figure}

We evaluate the cumulants for five cycle periods
\(\tau=200, 500, 1000, 2000,\) and \(5000~\mathrm{ps}\).
The cycle-period dependence of the four finite-time cumulants is plotted
in Fig.~\ref{fig:tau_cumulants}, while their corresponding values per
unit time are plotted in
Fig.~\ref{fig:tau_cumulants_normalized}.

\MODNEW{In Fig.~\ref{fig:tau_cumulants}, the first and second orientation-even cumulants,
\(C_1^{\mathrm{even}}\) and \(C_2^{\mathrm{even}}\), increase strongly
with \(\tau\). A longer cycle permits more optical injection, counted
phonon relaxation, and radiative recycling.} \MODNEW{ \(C_1^{\mathrm{odd}}\) and
\(C_2^{\mathrm{odd}}\) attain long period values
\(0.2145~\mathrm{meV/cycle}\) and
\(11.18~\mathrm{meV^2/cycle}\), respectively. Their short-period
variation is a finite-speed correction to the orientation-odd
contribution per completed loop and is not expected to scale linearly
with \(\tau\) \cite{Sinitsyn2009,Ren2010}. Thus, the plateau of the
unnormalized orientation-odd cumulants, together with the
time-extensive orientation-even sector provides the two criteria Eq.~\eqref{eq:odd_plateau_criterion} and Eq.~\eqref{eq:even_extensive_criterion}, required to identify the finite-time half-difference and
half-sum with the geometric and dynamic contributions, respectively.}

\MODNEW{A period is accepted only when the plateau and
time-extensive criteria are simultaneously satisfied within a
\(1\%\) tolerance for both \(n=1\) and \(n=2\). Over this validated
period range, the orientation-odd and orientation-even cumulants are
identified with the geometric and dynamic cumulants, respectively.}

\MODNEW{Here, for Eqs.~\eqref{eq:odd_plateau_criterion} and
\eqref{eq:even_extensive_criterion}  
\(\tau_{\mathrm{ref}}\) is fixed at \(5000~\mathrm{ps}\). In Fig.~\ref{fig:tau_cumulants}(b) and (d) at \(\tau=200~\mathrm{ps}\), the first and second
orientation-odd cumulants differ from their long-period reference
values by approximately \(2.9\%\) and \(3.7\%\), respectively, and
therefore remain outside the  \(1\%\) plateau tolerance. At
\(\tau=500~\mathrm{ps}\), these deviations decrease to approximately
\(0.61\%\) and \(0.78\%\), respectively. At the same period in Fig.~\ref{fig:tau_cumulants_normalized}(a) and (c),
\(C_1^{\mathrm{even}}/\tau\) and
\(C_2^{\mathrm{even}}/\tau\) also satisfy the time-extensive-scaling
criterion. Both conditions remain satisfied at every larger sampled
period. The shaded region in Fig.~\ref{fig:tau_cumulants} and Fig.~\ref{fig:tau_cumulants_normalized} therefore
begins at \(\tau=500~\mathrm{ps}\) and marks the sampled range over
which \(C_n^{\mathrm{even}}\) and \(C_n^{\mathrm{odd}}\) are identified
with \(C_n^{\mathrm{dyn}}\) and \(C_n^{\mathrm{geo}}\), respectively,
for \(n=1,2\).}

To connect this cumulant based validation with the intrinsic relaxation
time of the open system, we examine the instantaneous eigenspectrum of
the ordinary Liouvillian,
\(\mathcal{L}(t)\equiv\mathcal{L}_{s=0}(t)\).
Probability conservation gives one stationary eigenvalue,
\(\lambda_0(t)=0\), while the remaining eigenvalues
\(\lambda_m(t)\), with \(m\neq0\), describe the decay of population
perturbations toward the instantaneous stationary state. We define the
instantaneous Liouvillian relaxation gap as
\(
\Delta_L(t)
=
\min_{m\neq0}
\left[-\operatorname{Re}\lambda_m(t)\right].\)

Thus, \(\Delta_L(t)\) is the slowest nonzero population-decay rate at a
given point on the gate-pump loop. Its minimum over one complete loop
and the corresponding longest characteristic relaxation time are
\(
\Delta_L^{\min}
=
\min_{0\leq t\leq\tau}\Delta_L(t)
\simeq
0.160~\mathrm{ps^{-1}}
\),
\(\tau_{\mathrm{rel}}^{\max}
\simeq
\frac{1}{\Delta_L^{\min}}
=
6.25~\mathrm{ps}\) respectively. \(\Delta_L^{\min}\) is the smallest relaxation gap encountered
along the loop, while \(\tau_{\mathrm{rel}}^{\max}\) is the associated
longest population-relaxation time. A cycle period much longer than
\(\tau_{\mathrm{rel}}^{\max}\) allows the populations to adjust to the
changing controls and supports the slow-driving interpretation.

The frequency to gap
indicator (
\(
\epsilon_{\mathrm{ad}}
=
\frac{2\pi/\tau}{\Delta_L^{\min}}
\)
) decreases from \(0.196\) at \(\tau=200~\mathrm{ps}\) to \(0.0079\) at
\(\tau=5000~\mathrm{ps}\).This decrease shows that the driving
frequency becomes progressively smaller than the slowest Liouvillian
relaxation rate. The system therefore has more time to follow the
instantaneous steady state along the loop, reducing finite-speed
corrections and bringing the dynamics closer to the slow-driving
regime. The complete time-scale classification, including an
extended \(50~\mathrm{ns}\) calculation, is summarized in Table~S6 and
Sec.~S5.3 of the Supporting Information.

\MODNEW{A strict open-system adiabatic
condition also involves derivatives of the instantaneous left and right
eigenvectors. The simultaneous orientation-odd plateau and
orientation-even time-extensive scaling therefore provide the
operational basis for the dynamic/geometric identification. The tested
periods constitute a broad scan within the slow-driving sector rather
than a complete crossover from sudden or strongly nonadiabatic
dynamics.}

At \(\tau_{\mathrm{ref}}\)
\(=5~\mathrm{ns}\), we obtain
\(C_1^{\mathrm{dyn}}=108.76~\mathrm{meV/cycle}\),
\(C_1^{\mathrm{geo}}=0.2145~\mathrm{meV/cycle}\),
\(C_2^{\mathrm{dyn}}=4.9785\times10^3~\mathrm{meV^2/cycle}\), and
\(C_2^{\mathrm{geo}}=11.18~\mathrm{meV^2/cycle}\).
\MODNEW{The geometric cumulants remain stable from
\(5\) to \(50~\mathrm{ns}\) as shown in Sec.~S5.3 of the Supporting Information. Together with
\(\epsilon_{\mathrm{ad}}(5000~\mathrm{ps})=0.0079\), this confirms that
the selected reference period lies safely on the converged geometric
plateau.} Thus, we retain the cycle period of \(5~\mathrm{ns}\) as the conservative long-period
reference for the subsequent analysis.

We next test whether the geometric response depends on the
loop-geometry quantities of the gate-pump trajectory. The working driving
loop spans the full experimentally-supported gate range used to construct
\(\hat{H}_T[F_z^{(0)},V_G(t)]\), giving
\(0.5\leq V_G(t)\leq6.5~\mathrm{V}\). Thus, the gate trajectory remains
inside the Hamiltonian input surface. The pump-control trajectory spans
\(0.1\leq I_{\mathrm{pump}}(t)\leq2.5\) and modulates only the
optical-injection rates.

\begin{figure}[htbp]
\centering
\includegraphics[width=1\columnwidth]{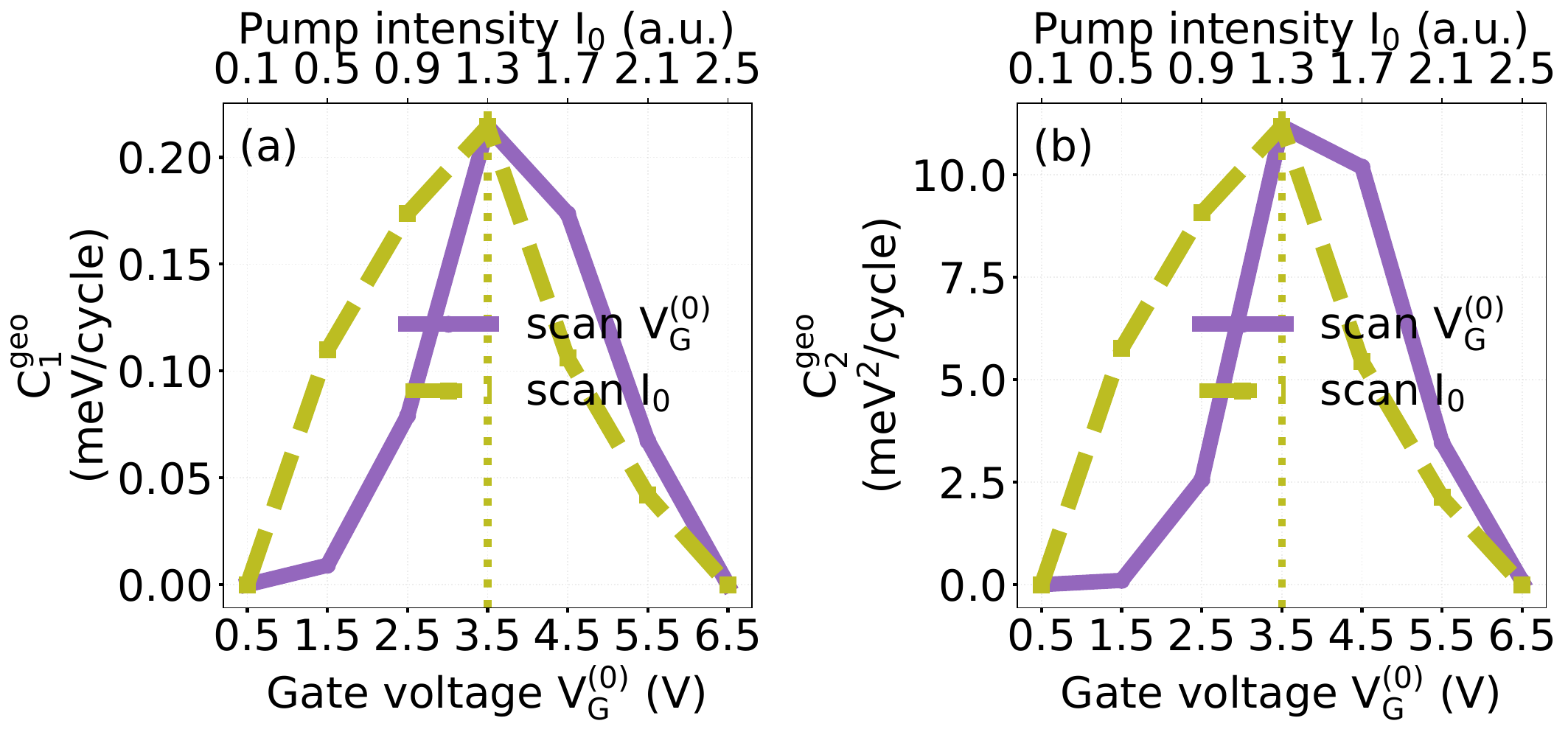}
\caption{(Calculated.) One-parameter control-centre scan of the geometric
cumulants \MODNEW{under baseline conditions}. The loop centre in the driving equations
Eqs.~\eqref{eq:VG_drive}-\eqref{eq:Ipump_drive} is displaced while the
trajectory is kept inside the validated gate and pump ranges. Panel (a)
contains \(C_1^{\mathrm{geo}}\), and panel (b) contains
\(C_2^{\mathrm{geo}}\). The solid curves denote the gate-centre scan in
\(V_G^{(0)}\), and the dashed curves denote the pump-centre scan in
\(I_{\mathrm{pump}}^{(0)}\). Dotted guides mark the selected centre value centre
\((V_G^{(0)}=3.5~\mathrm{V},I_{\mathrm{pump}}^{(0)}=1.3)\). The
corresponding dynamic-cumulant scan is reported in Fig.~S4 and Sec.~S4.1}
\label{fig:image3_controls}
\end{figure}

We perform two one-parameter control-centre scans. In the first scan,
\(V_G^{(0)}\) is varied while \(I_{\mathrm{pump}}^{(0)}=1.3\) is kept
fixed. In the second scan, \(I_{\mathrm{pump}}^{(0)}\) is varied while
\(V_G^{(0)}=3.5~\mathrm{V}\) is kept fixed. For each varied centre,
the corresponding modulation amplitude is adjusted so that the complete
trajectory remains inside the supported gate-pump control window. The
corresponding geometric cumulants are plotted in Figure~\ref{fig:image3_controls}. In panel (a) 
\(C_1^{\mathrm{geo}}\), and in panel (b)
\(C_2^{\mathrm{geo}}\), both were extracted using Eq.~\eqref{eq:Codd}.

The geometric cumulants are largest near the central full-range
loop. Displacing the loop centre reduces the admissible modulation
amplitude in at least one control direction and therefore reduces the
effective enclosed area in the \((V_G(t),I_{\mathrm{pump}}(t))\) plane.
Since the geometric contribution is the orientation-odd part of the
tilted-Liouvillian response, this area reduction suppresses
\(C_n^{\mathrm{geo}}\). The corresponding dynamic-cumulant scan is
reported in Fig.~S4 and Sec.~S4.1 to document the orientation-even
background. The dynamic cumulants \(C_1^{\mathrm{dyn}}\) and
\(C_2^{\mathrm{dyn}}\) in Fig.~S4 generally increase as either
\(V_G^{(0)}\) or \(I_{\mathrm{pump}}^{(0)}\) is increased, with the
pump-centre scan showing an almost monotonic rise, while the gate-centre
scan rises up to the higher-gate region and then slightly decreases at
the largest \(V_G^{(0)}\).

We therefore select the central loop-geometry quantities as the baseline
loop for the remaining calculations because this loop gives the largest
geometric response within the supported control window while keeping the
gate trajectory inside the Hamiltonian input surface.  \MODNEW{This defines the baseline conditions against}
which the subsequent sensitivity scans are compared. The chosen
baseline geometric cumulants are
\(C_1^{\mathrm{geo,base}}=0.2145~\mathrm{meV/cycle}\) and
\(C_2^{\mathrm{geo,base}}=11.18~\mathrm{meV^2/cycle}\).

\NEW{\FloatBarrier}

\subsection{One-parameter sensitivity scan}
\label{sec:res_sensitivity}

Each one-parameter variation was carried out about the baseline
conditions listed in Table~\ref{tab:compact_model_parameters} , with the remaining
quantities fixed at their baseline values. The corresponding geometric cumulants were obtained according to
Eq.~\eqref{eq:finite_time_geo_dyn_identification}.

\begin{figure}[htbp]
\centering
\includegraphics[width=0.9\columnwidth]{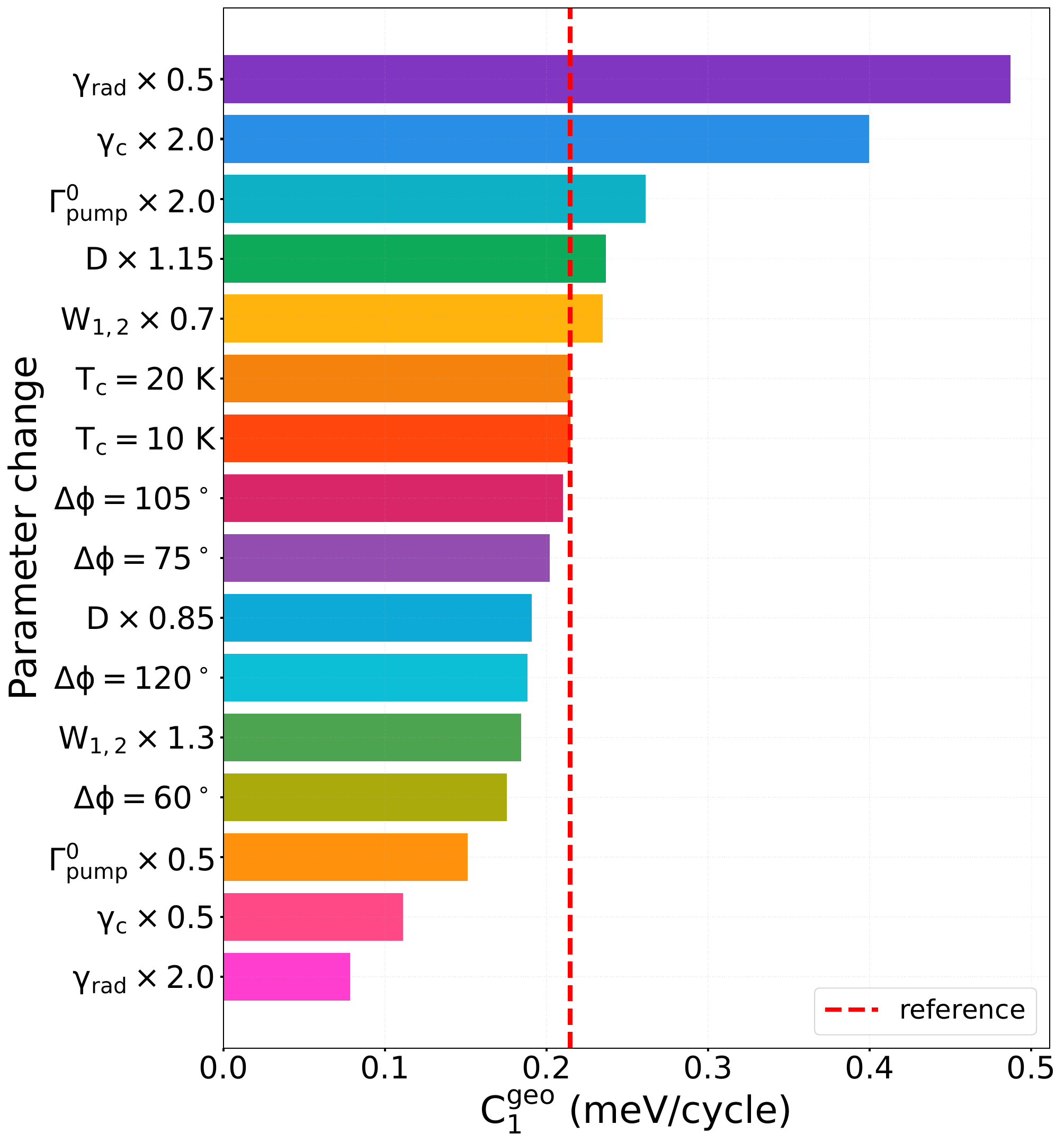}
\caption{(Calculated.) One-parameter sensitivity scan of the first geometric cumulant
\(C_1^{\mathrm{geo}}\). All unvaried parameters retain the baseline values in Table ~\ref{tab:compact_model_parameters}. In each bar, only the quantity listed
on the vertical axis is changed relative to its baseline value, while
all other quantities are kept fixed. The labels
\(\gamma_c\times0.5\) and \(\gamma_c\times2.0\) mean that both
\(\gamma_c^{\mathrm{intra}}\) and \(\gamma_c^{\mathrm{inter}}\) are
multiplied by 0.5 and 2.0, respectively. Similarly,
\(\gamma_{\mathrm{rad}}\times0.5\) and
\(\gamma_{\mathrm{rad}}\times2.0\) mean that both radiative prefactors
\(\gamma_{\mathrm{rad}}^{\mathrm{intra}}\) and
\(\gamma_{\mathrm{rad}}^{\mathrm{inter}}\) are multiplied by 0.5 and
2.0, respectively. The labels
\(\Gamma_{\mathrm{pump}}^0\times0.5\) and
\(\Gamma_{\mathrm{pump}}^0\times2.0\) scale the optical-injection
prefactor. The labels \(D\times0.85\) and \(D\times1.15\) scale the
Stark-slope input, while \((W_1,W_2)\times0.7\) and
\((W_1,W_2)\times1.3\) scale both interlayer-intralayer hybridization
inputs \(W_1\) and \(W_2\). The vertical dashed
line marks the baseline value
\(C_1^{\mathrm{geo,base}}=0.2145~\mathrm{meV/cycle}\). Bars to the
right of this line enhance the geometric response, whereas bars to the
left suppress it.}
\label{fig:one_parameter_sensitivity}
\end{figure}

For each bar in Fig.~\ref{fig:one_parameter_sensitivity}, only the
parameter listed on the vertical axis is changed, while all other
parameters are kept fixed at their baseline values. We scan 
selected parameters from Table~\ref{tab:compact_model_parameters}: the
optical-injection prefactor, the phonon prefactors, the radiative
prefactors, the Stark slope, the interlayer-intralayer hybridization
inputs \(W_1\) and \(W_2\), the phase lag, and the phonon-bath temperature.

In Fig~\ref{fig:one_parameter_sensitivity}, we show the resulting first geometric
 cumulant \(C_1^{\mathrm{geo}}\). The dashed vertical marker denotes
the baseline value
\(C_1^{\mathrm{geo,base}}=0.2145~\mathrm{meV/cycle}\). Bars to the right
of this marker correspond to parameter changes that enhance the
geometric response, whereas bars to the left correspond to parameter
changes that suppress it.

\begin{figure}[htbp]
\centering
\includegraphics[width=0.90\columnwidth]{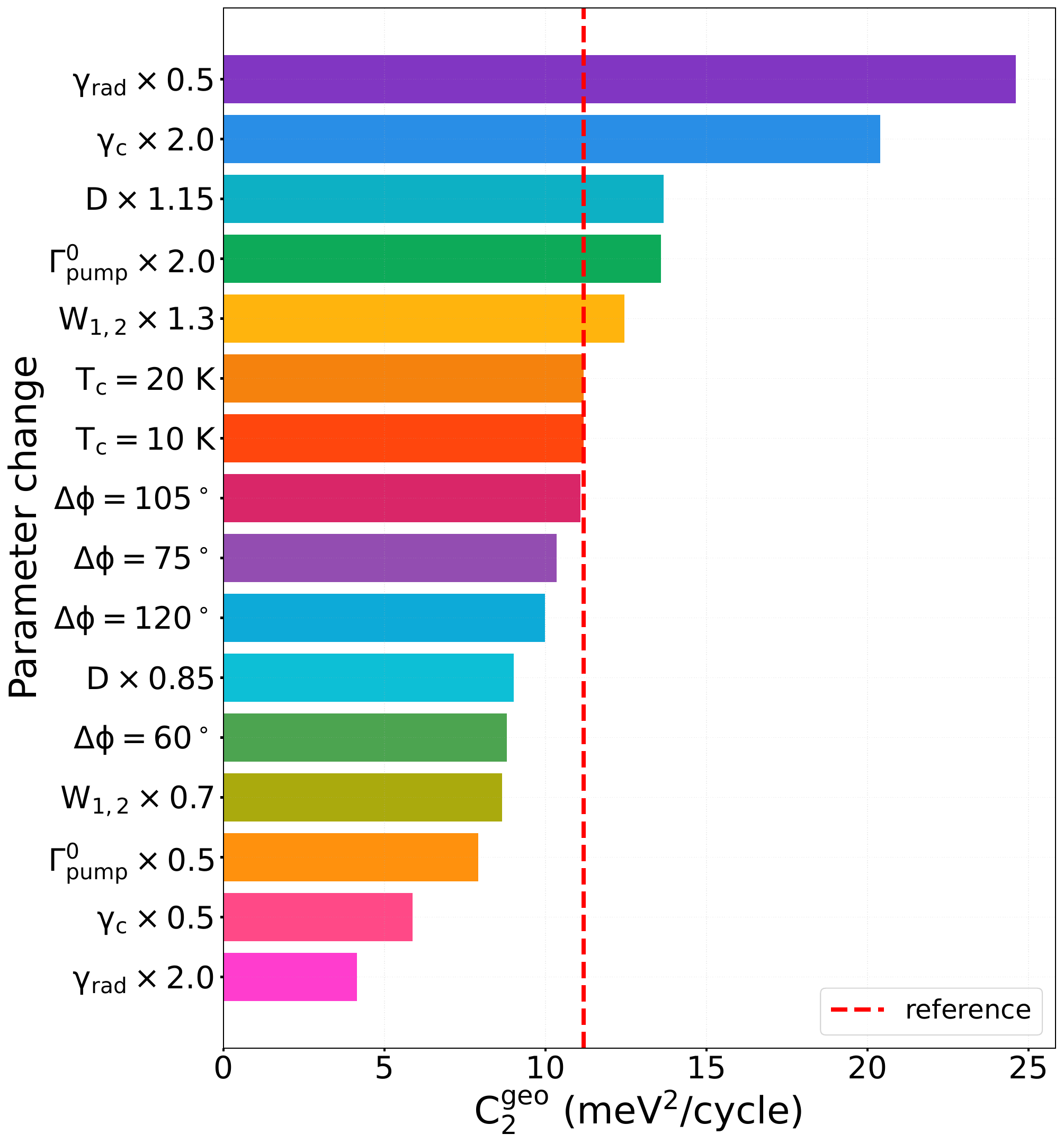}
\caption{(Calculated.) One-parameter sensitivity scan of the second geometric cumulant
\(C_2^{\mathrm{geo}}\). All unvaried parameters retain the baseline values in Table ~\ref{tab:compact_model_parameters}. In each bar, only the quantity listed
on the vertical axis is changed relative to its baseline value, while
all other quantities are kept fixed. The labels
\(\gamma_c\times0.5\) and \(\gamma_c\times2.0\) mean that both
\(\gamma_c^{\mathrm{intra}}\) and \(\gamma_c^{\mathrm{inter}}\) are
multiplied by 0.5 and 2.0, respectively. Similarly,
\(\gamma_{\mathrm{rad}}\times0.5\) and
\(\gamma_{\mathrm{rad}}\times2.0\) mean that both radiative prefactors
\(\gamma_{\mathrm{rad}}^{\mathrm{intra}}\) and
\(\gamma_{\mathrm{rad}}^{\mathrm{inter}}\) are multiplied by 0.5 and
2.0, respectively. The labels
\(\Gamma_{\mathrm{pump}}^0\times0.5\) and
\(\Gamma_{\mathrm{pump}}^0\times2.0\) scale the optical-injection
prefactor. The labels \(D\times0.85\) and \(D\times1.15\) scale the
Stark-slope input, while \((W_1,W_2)\times0.7\) and
\((W_1,W_2)\times1.3\) scale both interlayer-intralayer hybridization
inputs \(W_1\) and \(W_2\). The vertical dashed
line marks the baseline value
\(C_2^{\mathrm{geo,base}}=11.18~\mathrm{meV^2/cycle}\). Bars to the
right of this line enhance the geometric response, whereas bars to the
left suppress it.}
\label{fig:c2geo_sensitivity}
\end{figure}

The largest enhancement occurs when the radiative loss prefactors are
reduced. Increasing the phonon
prefactors also enhances \(C_1^{\mathrm{geo}}\). Increasing the
optical-injection prefactor gives a smaller enhancement by feeding more
population into the dressed exciton manifold. In contrast, increasing
the radiative loss prefactors, reducing the phonon prefactors, or
reducing the optical-injection prefactor suppresses the geometric
response. Changes in the Stark slope, the hybridization inputs \(W_1\)
and \(W_2\), the phase lag, and the phonon-bath temperature modify the
magnitude but do not remove the positive orientation-odd signal. Thus,
the geometric first cumulant is not a fine-tuned feature of one isolated parameter choice. It survives across the
tested one-parameter variations.
For the geometric second cumulant \(C_2^{\mathrm{geo}}\), the same
one-parameter procedure is used and the results are shown in
Fig.~\ref{fig:c2geo_sensitivity}. The baseline value is
\(C_2^{\mathrm{geo,base}}=11.18~\mathrm{meV^2/cycle}\). The ordering of
the parameter changes closely follows the trend found for
\(C_1^{\mathrm{geo}}\). Reducing radiative loss and increasing
phonon relaxation enhance the geometric noise, whereas the opposite
variations suppress it.

\FloatBarrier

\subsection{Energy-disorder robustness}
\label{sec:res_disorder}

Through the parameter sensitivity scans, we established that the geometric response is
not tied to a single choice of rate or control parameter. We next test a
different source of uncertainty: static inhomogeneous broadening of the
dressed exciton spectrum. This check is relevant because local strain,
twist-angle variations, electrostatic disorder, and spatial variations in
the moir\'e potential can shift the dressed energies in real devices.
Since the geometric signal is extracted from the CW-CCW half-difference
in Eq.~\eqref{eq:Codd}, it is important to verify that the positive
orientation-odd response is not a fine-tuned consequence of one exactly
clean Hamiltonian surface.

We perform the disorder test around the baseline conditions defined
by Eqs.~\eqref{eq:VG_drive}-\eqref{eq:Ipump_drive}. For each disorder realization, independent Gaussian
energy shifts with zero mean and standard deviation \(\sigma\) are added
to the two intralayer-derived dressed energies \(\epsilon_1\) and
\(\epsilon_2\). These shifts are kept fixed during one loop, so the test
represents sample-to-sample or spatial inhomogeneous broadening rather
than fast temporal noise. For each realization, the shifted dressed energies modify the
instantaneous gaps, the associated rate weights, and therefore the
tilted Liouvillian in Eq.~\eqref{eq:tiltedL}. We then repeat the
finite-time propagation for both
loop orientations and extract \(C_1^{\mathrm{geo}}\) using
Eq.~\eqref{eq:Codd}. For each value of \(\sigma\), the calculation is
averaged over independent disorder realizations.

\begin{figure}[htbp]
\centering
\includegraphics[width=1\columnwidth]{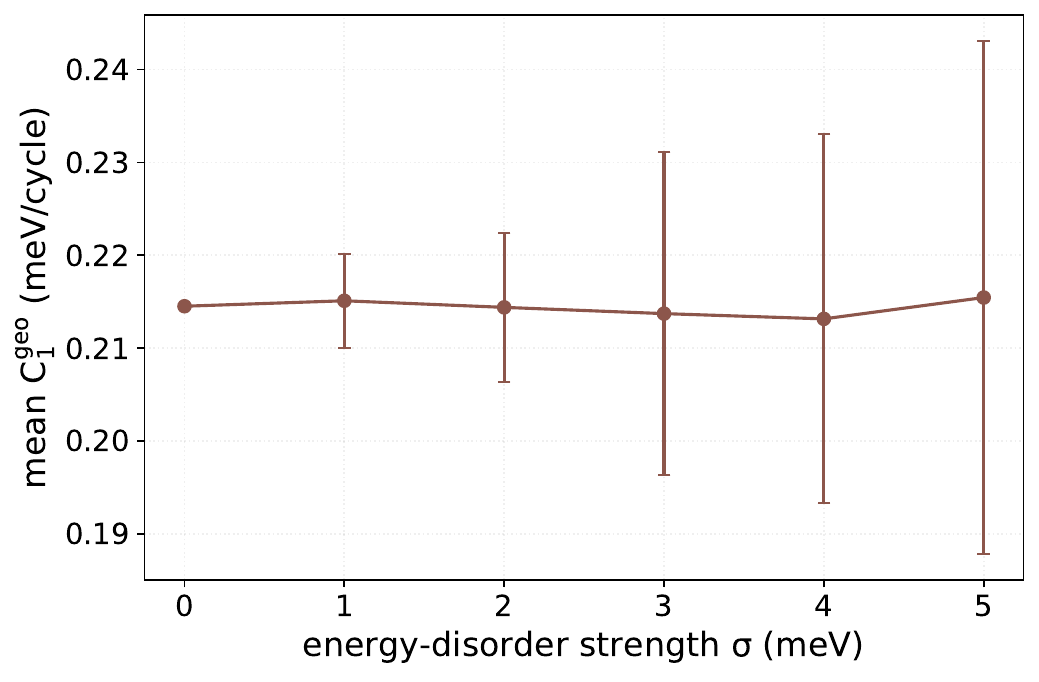}
\caption{(Calculated.) Energy-disorder robustness of the geometric first
cumulant. Independent Gaussian shifts with standard deviation
\(\sigma\) are added to the intralayer-derived dressed energies
\(\epsilon_1\) and \(\epsilon_2\). For each \(\sigma\), the finite-time
tilted-Liouvillian calculation is repeated for both loop orientations
over \(20\)-\(30\) disorder realizations. Points denote the mean
\(C_1^{\mathrm{geo}}\), and error bars indicate the standard deviation over
the realizations.}
\label{fig:dis}
\end{figure}

The disorder test results shown in Figure~\ref{fig:dis} indicate that the mean geometric cumulant remains
close to the clean baseline value throughout the tested disorder range.
At \(\sigma=5~\mathrm{meV}\), the averaged value is
\(C_1^{\mathrm{geo}}=0.211\pm0.029~\mathrm{meV/cycle}\), compared with
the clean baseline value
\(C_1^{\mathrm{geo,base}}=0.2145~\mathrm{meV/cycle}\). The sign of
\(C_1^{\mathrm{geo}}\) is preserved in every realization tested. Thus,
static energy disorder changes the magnitude slightly from realization
to realization, but it does not wash out or reverse the loop-reversal-odd
response. This result complements the parameter sensitivity scan. The
geometric signal is robust not only to controlled changes in rates and
couplings, but also to realistic static broadening of the exciton
spectrum. The response should therefore be interpreted as a property of
the oriented gate-pump trajectory and the retained \(\Omega_{12}\)
phonon channel, rather than as an accidental feature of an exactly
clean spectrum. Having checked this static-disorder robustness, we next
estimate the possible scale of higher-order phonon channels outside the
retained phonon window.

\subsection{Residual-coupling robustness}
\label{sec:res_w12}

We next test whether the orientation-odd geometric response depends
on the idealized choice \(W_{12}=0\) in the Hamiltonian of
Eq.~\eqref{eq:HT}. This check is important because a small residual
direct coupling between the two intralayer branches \(X_1\) and \(X_2\)
could be present in a real device, even if it is not resolved in the
experimental Hamiltonian input used to construct \(\hat{H}_T\). We therefore replace
\(\hat{H}_T\) by
\(\hat{H}_T+W_{12}(|X_1\rangle\langle X_2|+|X_2\rangle\langle X_1|)\)
and repeat the finite-time tilted-Liouvillian propagation under
the baseline conditions for each chosen value of \(W_{12}\).

\begin{figure}[htbp]
\centering
\includegraphics[width=0.85\columnwidth]{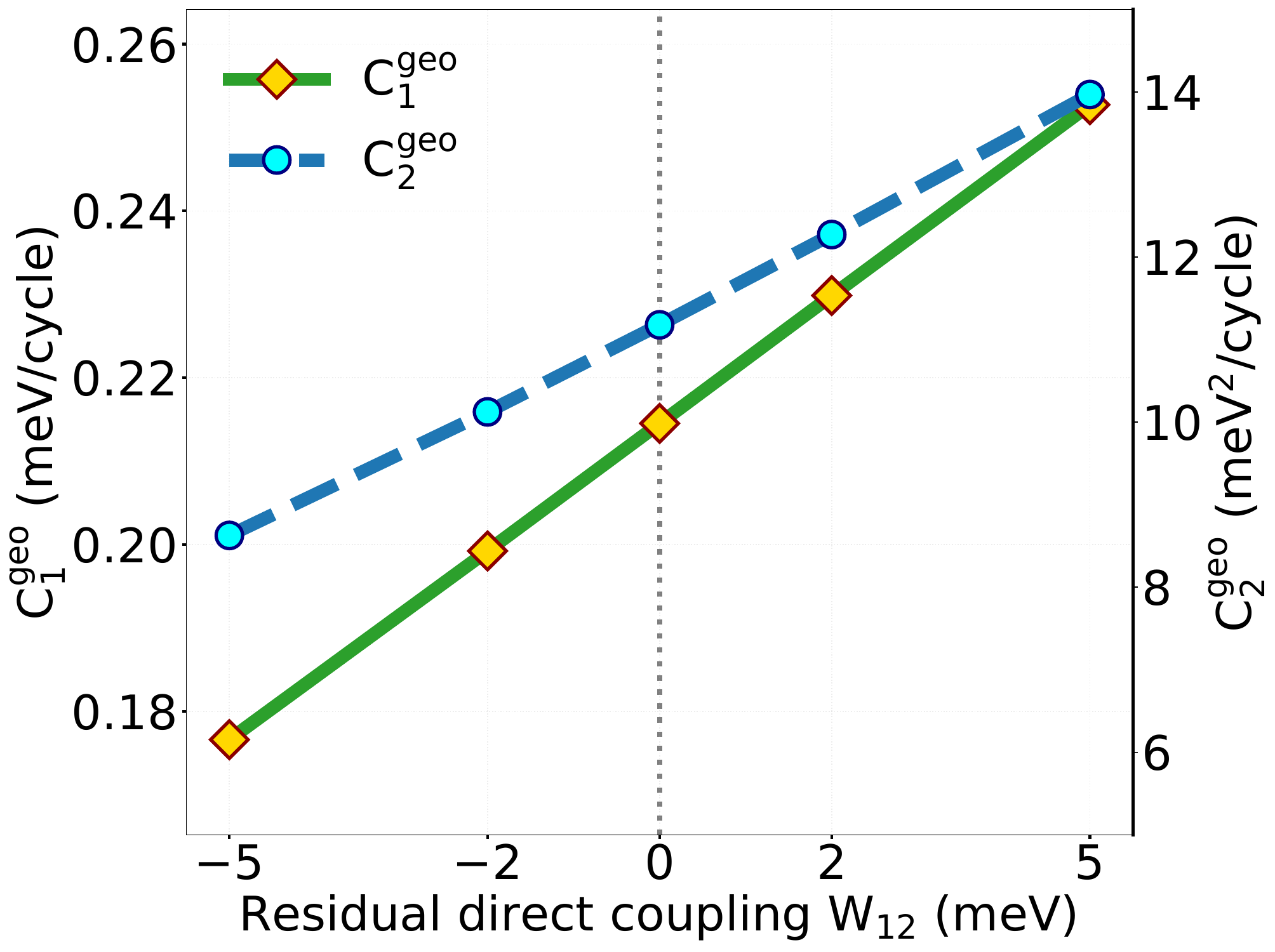}
\caption{(Calculated.) Residual-coupling robustness of the geometric cumulants under
the selected baseline conditions. A residual direct \(X_1\)-\(X_2\) coupling
\(W_{12}\) is added to the Hamiltonian in Eq.~\eqref{eq:HT}, while all other
quantities are kept fixed at their baseline values. The geometric first
cumulant \(C_1^{\mathrm{geo}}\) and second geometric cumulant
\(C_2^{\mathrm{geo}}\) are plotted as functions of \(W_{12}\), with
\(C_1^{\mathrm{geo}}\) shown on the left axis and \(C_2^{\mathrm{geo}}\)
shown on the right axis. The vertical dotted line marks the zero-coupling
case used in Eq.~\eqref{eq:HT} for the baseline model. Both cumulants vary
smoothly over the tested range, indicating that the geometric response remains
robust against moderate residual \(X_1\)-\(X_2\) coupling.
}
\label{fig:stokes_proxy}
\end{figure}

The residual-coupling results in Fig.~\ref{fig:stokes_proxy}  indicate how
\(C_1^{\mathrm{geo}}\) and \(C_2^{\mathrm{geo}}\) changes with \(W_{12}\). The magnitude of \MODNEW{both the} geometric cumulants
changes linearly as \(W_{12}\) is varied, but the sign remains positive over the
tested range. Thus, the loop-reversal-odd response is not an artifact of
the exact \(W_{12}=0\) idealization It survives a residual direct
\(X_1\)-\(X_2\) coupling. Positive coupling between $X_1$ abd $X_2$ increases the magnitude while negative values decreases the geometric cumulants. 

\FloatBarrier

\subsection{Enhanced-signal and detectability optimization}
\label{sec:res_enhanced_signal}
\begin{table*}[!t]
\centering
\scriptsize
\setlength{\tabcolsep}{4.5pt}
\renewcommand{\arraystretch}{1.12}
\caption{
Calculated cumulants and intrinsic single pair SNR for the cumulative
parameter enhancement sequence at the uniform
\(\tau=5~\mathrm{ns}\) baseline. Each row adds one favourable change
to the preceding configuration.
}
\label{tab:enhanced_uniform_budget}
\begin{tabular}{p{4.2cm}cccc}
\toprule
\textbf{Configuration}
&
\(\boldsymbol{C_1^{\mathrm{geo}}}\)
&
\(\boldsymbol{C_1^{\mathrm{dyn}}}\)
&
\(\boldsymbol{C_2^{\mathrm{dyn}}}\)
&
\(\boldsymbol{\mathrm{SNR}_{N=1}}\)
\\
&
(meV/cycle)
&
(meV/cycle)
&
(meV\(^2\)/cycle)
&
\\
\midrule

Baseline
& 0.2145 & 108.8 & 4978.5 & 0.004299 \\

\(I_{\mathrm{pump}}=0.1\text{-}4.9\)
& 0.2763 & 154.8 & 7093.2 & 0.004640 \\

\(+\,\Gamma_{\mathrm{pump}}^0\times2\)
& 0.2958 & 206.7 & 9480.3 & 0.004296 \\

\(+\,\gamma_c\times2\)
& 0.5524 & 402.1 & 18190.5 & 0.005793 \\

\(+\,\gamma_{\mathrm{rad}}\times0.5\)
& 0.8896 & 468.9 & 20704.8 & 0.008743 \\

\bottomrule
\end{tabular}
\end{table*}
We now test whether widening the pump range, doubling the optical
injection and phonon rate prefactors, and halving the radiative loss
prefactors relative to their baseline values in
Table~\ref{tab:compact_model_parameters} can improve statistical
detectability. Specifically, we examine whether these changes preserve
or enhance \(C_1^{\mathrm{geo}}\) while reducing the dominant dynamic
variance \(C_2^{\mathrm{dyn}}\), and whether the dynamic mean
\(C_1^{\mathrm{dyn}}\) can be suppressed without changing the
established driving loop.

For the numerically sampled period range beginning at
\(\tau=500~\mathrm{ps}\), the intrinsic CW-CCW signal to noise ratio
and the time normalized detectability, defined as the signal to noise
ratio per square root of total acquisition time, are
\begin{equation}
\operatorname{SNR}_{N}
=
\sqrt{2N}\,
\frac{|C_1^{\mathrm{geo}}|}{\sqrt{C_2^{\mathrm{dyn}}}},
\qquad
\mathcal{D}_{T}
=
\frac{|C_1^{\mathrm{geo}}|}{\sqrt{\tau C_2^{\mathrm{dyn}}}}.
\label{eq:main_snr_metrics}
\end{equation}
The first quantity measures the intrinsic SNR for a fixed number of cycles, whereas \(\mathcal{D}_{T}\) compares protocols for
the same total acquisition time. These quantities provide an intrinsic FCS estimate,
meaning that the signal and noise are determined solely by the mean and
variance of the transferred energy predicted by the model. Here, \(N\) denotes the number of statistically independent cycles
performed for each loop orientation. Thus, \(N=1\) corresponds to one
CW cycle and one CCW cycle, constituting a single CW-CCW pair. Detector
noise, drift, correlations between cycles, imperfect transduction, and
other technical noise are not included. The derivation is given in
Supporting Information Sec.~S5.5
\cite{Blanter2000,Bagrets2003,Esposito2009}.

The favourable changes identified in Fig.~\ref{fig:c2geo_sensitivity}
are a wider pump range, doubled injection and phonon prefactors, and
halved radiative loss prefactors. Starting from the uniform
\(5~\mathrm{ns}\) baseline, we apply these changes sequentially and
recalculate the cumulants and the single pair CW--CCW SNR after each
step.  The cumulative sequence
increases \(C_1^{\mathrm{geo}}\) by \(4.15\times\), from \(0.2145\) to
\(0.8896~\mathrm{meV/cycle}\), but the simultaneous rise of
 \(C_2^{\mathrm{dyn}}\) from \(4978.5\) to
\(20704.8~\mathrm{meV^2/cycle}\) limits the single-pair SNR gain to
\(2.03\times\).  In particular, doubling the injection prefactor raises
the signal and  variance by comparable proportions and gives almost no
SNR improvement. This signal variance competition motivates  the next two steps: optimizing the cycle period and redistributing time along the same loop.

\begin{figure}[htbp]
\centering
\includegraphics[
width=\columnwidth,
keepaspectratio
]{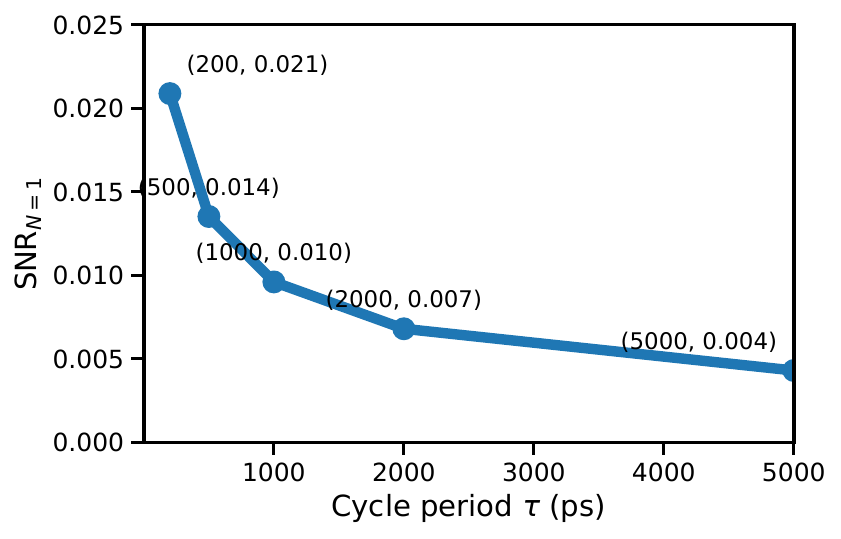}
\caption{Calculated intrinsic CW-CCW SNR versus cycle period for the
uniform baseline loop. As \(\tau\) increases, \(C_1^{\mathrm{geo}}\) remains near
its plateau while \(C_2^{\mathrm{dyn}}\) grows approximately with the
cycle duration, so the SNR decreases.
}
\label{fig:snr_tau_baseline}
\end{figure}

The period validated cumulants in Sec.~\ref{sec:res_tau} and
Fig.~\ref{fig:tau_cumulants}, combined through
Eq.~\eqref{eq:main_snr_metrics}, give the SNR dependence shown in
Fig.~\ref{fig:snr_tau_baseline}. The full period convergence and
Liouvillian time scale analyses are provided in Sec.~S5.3 of the Supporting
Information.

The uniform \(5~\mathrm{ns}\) protocol remains the converged baseline.
At \(1~\mathrm{ns}\), \(C_1^{\mathrm{geo}}\) changes only from
\(0.2145\) to \(0.2142~\mathrm{meV/cycle}\), a \(0.16\%\) loss,
whereas \(C_2^{\mathrm{dyn}}\) falls from \(4978.494\) to
\(995.777~\mathrm{meV^2/cycle}\). The SNR rises from about \(0.0043\)
to \(0.010\), giving \(2.23\times\) single-pair and \(4.99\times\)
fixed-time gains with \(99.84\%\) geometric retention. We therefore
retain \(5~\mathrm{ns}\) as the plateau reference and use
\(1~\mathrm{ns}\) as the common operating period for the subsequent
detectability optimization.

Although the uniform \(500~\mathrm{ps}\) protocol gives a larger
intrinsic SNR, it is the earliest sampled period satisfying the
\(1\%\) dynamic/geometric classification criterion and therefore lies
closest to the finite-speed boundary of the validated region. At this
period, the first and second orientation-odd cumulants remain
approximately \(0.61\%\) and \(0.78\%\) below their long-period
references. The \(1~\mathrm{ns}\) period provides a more conservative
compromise between geometric retention, reduced dynamic variance, and
the control-speed constraints of the nonuniform traversal. 
\begin{table*}[!htbp]
\centering
\scriptsize
\setlength{\tabcolsep}{5pt}
\renewcommand{\arraystretch}{1.12}
\caption{
Calculated first geometric cumulant, second dynamic cumulant, and
intrinsic SNR gains for the uniform and \(\mathrm{NUFM}\) protocols.
All calculations use \(N_t=400\). The single-pair and fixed-total-time
gains are measured relative to the uniform \(5~\mathrm{ns}\)
baseline. The \(\mathrm{NUFM}\) rows use
\(\tau=1~\mathrm{ns}\) and
\((\alpha,\beta,r,\epsilon_{\max})=(6,0,5,0.08)\). The unmodified
\(\mathrm{NUFM}\) row is the intrinsic protocol-only result; rows
with modified rate prefactors are conditional device-engineering
scenarios.
}
\label{tab:nufm_rate_combinations}
\begin{tabular}{p{5.2cm}cccc}
\toprule
\textbf{Protocol or rate modification}
&
\(\boldsymbol{C_1^{\mathrm{geo}}}\)
&
\(\boldsymbol{C_2^{\mathrm{dyn}}}\)
&
\shortstack{\textbf{Single-pair gain}\\
\textbf{vs uniform \(5~\mathrm{ns}\)}}
&
\shortstack{\textbf{Fixed-time gain}\\
\textbf{vs uniform \(5~\mathrm{ns}\)}}
\\
&
(meV/cycle)
&
(meV\(^2\)/cycle)
&
&
\\
\midrule

Uniform, \(5~\mathrm{ns}\)
& 0.214525 & 4978.494 & \(1.00\times\) & \(1.00\times\) \\

Uniform, \(1~\mathrm{ns}\)
& 0.214179 & 995.777 & \(2.23\times\) & \(4.99\times\) \\

\(\mathrm{NUFM}\), \(1~\mathrm{ns}\)
& 0.213027 & 461.557
& \(3.26\times\)
& \(7.29\times\) \\

\(\mathrm{NUFM}+\Gamma_{\mathrm{pump}}^0\times2\)
& 0.258943 & 696.977
& \(3.23\times\)
& \(7.21\times\) \\

\(\mathrm{NUFM}+\gamma_c\times2\)
& 0.396871 & 868.718
& \(4.43\times\)
& \(9.90\times\) \\

\(\mathrm{NUFM}+\gamma_{\mathrm{rad}}\times0.5\)
& 0.484361 & 1101.309
& \(4.80\times\)
& \(10.73\times\) \\

\(\mathrm{NUFM}+\Gamma_{\mathrm{pump}}^0\times2
+\gamma_c\times2\)
& 0.482964 & 1314.634
& \(4.38\times\)
& \(9.80\times\) \\

\(\mathrm{NUFM}+\Gamma_{\mathrm{pump}}^0\times2
+\gamma_{\mathrm{rad}}\times0.5\)
& 0.500485 & 1471.701
& \(4.29\times\)
& \(9.59\times\) \\

\(\mathrm{NUFM}+\gamma_c\times2
+\gamma_{\mathrm{rad}}\times0.5\)
& 0.848614 & 1953.918
& \(\mathbf{6.31\times}\)
& \(\mathbf{14.12\times}\) \\

\(\mathrm{NUFM}+\Gamma_{\mathrm{pump}}^0\times2
+\gamma_c\times2+\gamma_{\mathrm{rad}}\times0.5\)
& 0.878929 & 2621.921
& \(5.65\times\)
& \(12.62\times\) \\

\bottomrule
\end{tabular}
\end{table*}

Having selected the operating period, we redistribute the traversal time
along the same control loop using a nonuniform frequency modulated
(NUFM) protocol. The gate and pump modulations retain the same loop
geometry and phase lag as Eqs.~\eqref{eq:VG_drive} and
\eqref{eq:Ipump_drive}, while the uniform phase \(\omega t\) is replaced
by a nonuniform loop coordinate \(\theta(t)\). The complete construction,
including the speed profile, period constraint, convergence tests, and
waveform analysis, is given in Sec.~S10 of the Supporting Information.

 At
\(\tau=1~\mathrm{ns}\), the balanced parameter set explained in Sec. S10  reduces
\(C_1^{\mathrm{dyn}}\) from \(21.752\) to
\(10.274~\mathrm{meV/cycle}\) and \(C_2^{\mathrm{dyn}}\) from
\(995.777\) to \(461.557~\mathrm{meV^2/cycle}\), corresponding to
reductions of \(52.8\%\) and \(53.6\%\), respectively. In contrast,
\(C_1^{\mathrm{geo}}\) changes only from \(0.214179\) to
\(0.213027~\mathrm{meV/cycle}\), retaining \(99.46\%\) of the uniform
\(1~\mathrm{ns}\) geometric signal. This parameter set is therefore used
for the remaining NUFM calculations as a compromise between strong
suppression of the dynamic background and high retention of the
geometric contribution.

We next examine whether the performance of the selected NUFM protocol
can be further improved by modifying the rate parameters identified in
Sec.~\ref{sec:res_sensitivity} and
Fig.~\ref{fig:c2geo_sensitivity}. The considered changes are enhanced
optical injection and phonon relaxation together with reduced radiative
loss. The wider pump range is excluded because it changes the
control-space path and enclosed area, whereas all cases in
Table~\ref{tab:nufm_rate_combinations} retain the established loop.
Table~\ref{tab:nufm_rate_combinations} shows that increasing the
injection strength enhances the geometric signal but also increases the
dynamic variance, leaving the fixed-time gain close to
\(7.21\times\). Increasing the phonon relaxation strength or reducing
the radiative loss raises this gain to \(9.90\times\) and
\(10.73\times\), respectively. Since these rate modifications also
change \(j_1(\theta)\), \(\Delta_L(\theta)\), and consequently the NUFM
waveform , the cumulants are
recalculated self-consistently for each case rather than obtained by
simple rescaling.

The largest fixed total time gain of \(14.12\times\) is obtained when
enhanced phonon relaxation is combined with reduced radiative loss
while the injection strength is kept at its baseline value. This case
gives \(C_1^{\mathrm{geo}}=0.848614~\mathrm{meV/cycle}\),
\(C_2^{\mathrm{dyn}}=1953.918~\mathrm{meV^2/cycle}\), and a
\(6.31\times\) single-pair gain. Applying all three rate modifications
produces the largest absolute \(C_1^{\mathrm{geo}}\), but the associated
increase in dynamic variance lowers the fixed-time gain to
\(12.62\times\). Thus, maximizing the geometric contribution alone
does not maximize statistical detectability.

For experimental detection, the counted phonon-energy statistics may be
accessed through a spectrally resolved optical feature associated with
the same phonon-assisted exciton relaxation
\cite{Li2019DarkExcitonReplica,Liu2019ChiralPhononReplica,
Ripin2023TunablePhononic}. Repeated CW and CCW measurements would allow
orientation-odd optical proxy cumulants to be constructed from
cycle-resolved photon statistics. Such a proxy requires system-specific
calibration and is therefore proposed here as a possible readout strategy
rather than a direct quantitative conversion of the phonon cumulants.
Further details are given in Sec. S5.6 of the Supporting Information.
\FloatBarrier

\section{Conclusion}\label{sec:conclusion}

In this work, we developed an experimentally anchored four state open system model to investigate geometric phonon energy statistics in a layer hybridized WSe$_2$/WS$_2$ moir\'e exciton manifold. The model extends the experimentally reported three level exciton Hamiltonian by including the no exciton ground state together with optical injection, phonon relaxation, radiative loss, and a counting field for the selected phonon relaxation channel. The gate and pump coordinates form a phase lagged closed loop in control space, whose reversal separates the orientation odd geometric contribution from the much larger orientation even dynamic background.

The geometric energy cumulants were evaluated from the finite time moment generating function using the full time ordered propagator of the driven tilted Liouvillian. We found that the dynamic mean and variance increase strongly with cycle duration, whereas the geometric first and second cumulants approach nearly period independent values in the slow driving regime. The geometric response also remains robust against variations in the environmental and system parameters. Increasing phonon relaxation enhances the geometric mean and geometric noise, whereas radiative loss produces the opposite trend. Static inhomogeneous broadening does not wash out the geometric signal. We further introduced a residual direct coupling (W$_{12}$) between the two intralayer exciton branches (X$_1$) and (X$_2$). Positive (W$_{12}$) enhances both geometric cumulants, while negative (W$_{12}$) suppresses them. Thus, the geometric response survives controlled variations in rates and couplings as well as realistic broadening of the exciton spectrum.

An important result is that enhancement of the geometric signal does not necessarily imply improved detectability. Under uniform driving, cumulative parameter enhancement increases the first geometric cumulant by more than fourfold, but the simultaneous increase in the dynamic variance limits the SNR improvement to about twofold. In particular, increasing optical injection raises the geometric signal and the dynamic variance by comparable proportions and therefore produces almost no SNR improvement. This shows that the conditions that maximize geometric transfer are different from those that maximize its statistical detection.

We therefore used the driving period and the time distribution along the loop as additional control variables. Reducing the uniform cycle period lowers the accumulated dynamic variance while preserving almost the same geometric first cumulant. A subsequent nonuniform frequency modulated traversal of exactly the same control space loop further suppresses the dynamic mean and variance while retaining almost all of the geometric response. When combined with enhanced phonon relaxation and reduced radiative loss, this produces about a fourteenfold fixed total time SNR improvement relative to the uniform baseline. The resulting enhancement therefore originates mainly from suppressing dynamical noise rather than from increasing the geometric signal itself.

Since direct cycle resolved phonon calorimetry remains challenging in a cryogenic moir\'e heterobilayer, a clockwise and counter clockwise optical proxy based on phonon assisted spectral features provides a possible experimental route to the same orientation odd response. Overall, our results show that geometric phonon energy statistics can remain robust in an experimentally anchored moir\'e exciton platform and that environmental rates, driving period, and time distribution along the loop can serve as complementary control variables for improving their detectability. The finite time loop reversal framework developed here also provides a transferable route for investigating geometric energy statistics in other driven open quantum material systems.

\medskip
\textbf{Supporting Information}\par
Supporting Information is available from the Wiley Online Library or from the author upon reasonable request.

\medskip
\textbf{Acknowledgements}\par
B.~R.~T. and H.~P.~G. acknowledge the Department of Chemistry, Gauhati University, and the QuAinT Research Group for research support and discussion.

\medskip
\textbf{Conflict of Interest}\par
The authors declare no conflict of interest.

\medskip

\bibliographystyle{unsrtnat}
\bibliography{Reference}

\end{document}


\maketitle
\vspace{-1.4cm}

\section*{Scope and organization of the Supporting Information}

Section~\ref{sec:si_notation} fixes the notation and conventions used
throughout the Supporting Information.
Section~\ref{sec:si_inputs_protocol} records the experimentally anchored
Hamiltonian inputs and the imposed gate-pump protocol, while
Sec.~\ref{sec:si_operating_field} establishes the operating field, counted
transition, microscopic-to-rate construction, and scope of the population
model. Section~\ref{sec:si_baseline_selection} then uses the control-centre
and enclosed-area comparisons to identify the working full-range trajectory
as the baseline loop. Section~\ref{sec:si_numerical_detectability} documents
the finite-time propagation, periodic-limit-cycle initialization,
cycle-period validation, counting-field convergence, and intrinsic CW-CCW
detectability criteria for this baseline protocol.
Section~\ref{sec:si_baseline_robustness} collects the physical-robustness
and model-dependence checks. Sections~\ref{sec:si_enhanced_signal} and
\ref{sec:si_geometric_checks} present, respectively, the uniform-protocol
signal-amplitude and detectability diagnostics and the independent
Berry-Sinitsyn first-cumulant cross-check.
Section~\ref{sec:si_alternative_protocols} compares alternative loop shapes,
whereas Sec.~\ref{sec:si_current_shaped} gives the construction, constrained
optimization, response mechanism, convergence, feasibility, and
rate-model robustness of the selected same-loop
nonuniform-frequency-modulated (NUFM) traversal. The principal SI findings
are summarized in Sec.~\ref{sec:si_conclusion}.

\setcounter{section}{0}
\renewcommand{\thesection}{S\arabic{section}}

\section{Notation and conventions}
\label{sec:si_notation}

\begingroup
\scriptsize

\setlength{\tabcolsep}{3pt}
\renewcommand{\arraystretch}{1.08}
\setlength{\LTleft}{0pt}
\setlength{\LTright}{0pt}
\setlength{\LTpre}{4pt}
\setlength{\LTpost}{4pt}

\begin{longtable}{
    >{\raggedright\arraybackslash}p{0.13\textwidth}
    >{\raggedright\arraybackslash}p{0.22\textwidth}
    >{\raggedright\arraybackslash}p{0.18\textwidth}
    >{\raggedright\arraybackslash}p{0.38\textwidth}
}

\caption{
Notation and conventions used for the experimentally anchored Hamiltonian
input \(H_T\), the tilted-Liouvillian full-counting-statistics calculation,
the original uniform gate-pump protocol, and the additional
nonuniform-frequency traversal of the same parameter-space loop.
}
\label{tab:notation_conventions}
\\

\toprule
\textbf{Symbol}
&
\textbf{Meaning}
&
\textbf{Units/status}
&
\textbf{Convention}
\\
\midrule
\endfirsthead

\multicolumn{4}{c}{
\textbf{Table~\ref{tab:notation_conventions} continued from the previous page}
}
\\[2pt]

\toprule
\textbf{Symbol}
&
\textbf{Meaning}
&
\textbf{Units/status}
&
\textbf{Convention}
\\
\midrule
\endhead

\midrule
\multicolumn{4}{r}{
\textit{Continued on the next page}
}
\\
\endfoot

\bottomrule
\endlastfoot

\(\lvert iX\rangle,\lvert X_1\rangle,\lvert X_2\rangle\)
&
Bare-basis interlayer and intralayer excitons
&
Basis states
&
Used as the excitonic basis before diagonalization of \(H_T\).
\\

\(\lvert g\rangle\)
&
No-exciton ground state
&
Theoretical state
&
Added to close the optical-injection and relaxation cycle in the
open-system population model.
\\

\(\lvert\psi_n(t)\rangle\)
&
Instantaneous dressed exciton branch
&
Instantaneous eigenstate
&
Obtained by diagonalizing \(H_T[F_z^{(0)},V_G(t)]\) and ordered according to
\[
\epsilon_0(t)<\epsilon_1(t)<\epsilon_2(t).
\]
\\

\(V_G(t)\)
&
Gate-voltage control
&
V
&
For the original uniform protocol,
\[
V_G(t)
=
V_G^{(0)}
+
A_G\sin(\omega t+\phi_G).
\]
For the additional nonuniform-frequency protocol, the uniformly advancing
factor \(\omega t\) is replaced by \(\theta(t)\):
\[
V_G(t)
=
V_G^{(0)}
+
A_G\sin[\theta(t)+\phi_G].
\]
The gate control enters \(H_T\) and changes the instantaneous dressed
energies, gaps, and eigenstates.
\\

\(I_{\mathrm{pump}}(t)\)
&
Optical pump-control parameter
&
Dimensionless; a.u. in figures
&
For the original uniform protocol,
\[
I_{\mathrm{pump}}(t)
=
I_{\mathrm{pump}}^{(0)}
+
A_I\sin(\omega t+\phi_I).
\]
For the additional nonuniform-frequency protocol,
\[
I_{\mathrm{pump}}(t)
=
I_{\mathrm{pump}}^{(0)}
+
A_I\sin[\theta(t)+\phi_I].
\]
The pump control enters the Liouvillian through the optical-injection
rates but does not enter \(H_T\).
\\

\(\phi_G,\phi_I\)
&
Constant phases of the gate and pump controls
&
rad
&
These phases specify the relative placement of the gate and pump
oscillations. They are retained unchanged when the uniform factor
\(\omega t\) is replaced by the nonuniform coordinate \(\theta(t)\).
A common change in both phases only shifts the chosen starting point of
the loop.
\\

\(\Delta\phi\)
&
Relative gate-pump phase lag
&
rad
&
Defined by
\[
\Delta\phi
=
\phi_I-\phi_G.
\]
It determines the loop shape and orientation. The working loop uses
\[
\Delta\phi=\frac{\pi}{2}.
\]
Changing
\[
\Delta\phi\rightarrow-\Delta\phi
\]
reverses the orientation of the parameter-space loop.
\\

\(\omega\)
&
Angular frequency for traversal of the driving loop
&
\(\mathrm{ps^{-1}}\)
&
Defined by

\[\omega=\frac{2\pi}{\tau},\]

where \(\tau\) is the total time required to complete one traversal of
the driving loop. For the uniform protocol, \(\omega\) is constant and
the common running phase is \(\omega t\). For the nonuniform protocol,
the instantaneous angular speed is \(\dot{\theta}(t)\), whereas
\(\omega=2\pi/\tau\) remains the reference angular speed corresponding
to one complete traversal in the fixed cycle period \(\tau\). It is used
to define the imposed angular-speed bounds.
\\

\(\theta(t)\)
z&
Accumulated loop coordinate of the nonuniform protocol
&
rad
&
It replaces only the uniformly advancing factor \(\omega t\). It does not
replace either \(\phi_G\) or \(\phi_I\). It satisfies
\[
\theta(0)=0,
\qquad
\theta(\tau)=2\pi,
\]
and
\[
\frac{d\theta}{dt}
=
\dot{\theta}[\theta(t)].
\]
Thus, \(\theta(t)\) specifies the instantaneous position along the fixed
gate-pump loop.
\\

\(\dot{\theta}(\theta)\)
&
Programmed angular speed of the nonuniform traversal
&
\(\mathrm{ps^{-1}}\)
&
Chosen as
\[
\dot{\theta}(\theta)
=
\operatorname{clip}
\!\left[
K
\left(
\frac{j_1(\theta)}{j_{1,*}}
\right)^\alpha
\left(
\frac{\Delta_L(\theta)}
{\Delta_{L,*}}
\right)^\beta,
\frac{\omega}{r},
\min\!\left(
r\omega,
\epsilon_{\max}\Delta_L(\theta)
\right)
\right].
\]
It determines how quickly each part of the fixed parameter-space loop is
traversed and therefore defines a different physical driving protocol,
not a different numerical evaluation of the uniform protocol.
\\

\(F_z^{(0)}\)
&
Fixed out of plane electric field
&
\(\mathrm{V\,nm^{-1}}\)
&
Held fixed during each driven loop.
\\

\(\Omega_{ij}(t)\)
&
Instantaneous dressed-state energy gap
&
meV
&
Defined by
\[
\Omega_{ij}(t)
=
\epsilon_j(t)-\epsilon_i(t).
\]
It determines the energy transferred by a transition and is used to impose
the phonon accessibility condition.
\\

\(\tau\)
&
Total duration of one gate-pump cycle
&
ps or ns
&
For the original uniform protocol,
\[
\tau=\frac{2\pi}{\omega}.
\]
For the nonuniform protocol,
\[
\int_0^\tau
\dot{\theta}(t)\,dt
=
2\pi.
\]
The total cycle duration is fixed while the local dwell times are
redistributed around the same loop.
\\
\(j_1(\theta)\)
&
Loop-resolved frozen dynamic first-cumulant rate
&
\(\mathrm{meV\,ps^{-1}}\)
&
At a given loop position \(\theta\), the gate voltage and pump intensity
are held fixed at their instantaneous values and the stationary dynamic
first-cumulant rate is calculated from the dominant eigenvalue of the
tilted Liouvillian:
\[
j_1(\theta)
=
\left.
\partial_s\lambda_0(s,\theta)
\right|_{s=0}.
\]
Repeating this calculation around the selected loop gives
\(j_1(\theta)\) as a function of loop position. It identifies the regions
of that loop in which the local dynamic energy-transfer background is
large. It must be recalculated when the loop shape or the underlying model
parameters are changed.
\\

\(\Delta_L(\theta)\)
&
Ordinary-Liouvillian relaxation gap
&
\(\mathrm{ps^{-1}}\)
&
Defined from the nonstationary eigenvalues of the ordinary
probability-conserving Liouvillian as
\[
\Delta_L(\theta)
=
\min_{m\neq0}
\left[
-\operatorname{Re}\lambda_m(0,\theta)
\right].
\]
For stable population dynamics,
\[
\Delta_L(\theta)>0,
\]
and
\[
\tau_{\mathrm{rel}}(\theta)
=
\frac{1}{\Delta_L(\theta)}
\]
is the slowest instantaneous population-relaxation time at that loop
position. The gap contains the complete population-transfer network,
including pump-assisted, phonon-assisted, and radiative processes, and is
therefore not a phonon-only relaxation rate.
\\

\(j_{1,*}\)
&
Reference value of the loop-resolved dynamic current
&
\(\mathrm{meV\,ps^{-1}}\)
&
Chosen as the maximum value of \(j_1(\theta)\) along the selected loop,
\[
j_{1,*}
=
max_{\theta}j_1(\theta).
\]
Because \(j_1(\theta)\) is positive throughout the selected loop, the
normalized quantity satisfies
\[
0<
\frac{j_1(\theta)}{j_{1,*}}
\leq
1.
\]
The reference value makes the current-dependent factor dimensionless. It
is a normalization constant rather than an additional physical parameter,
and its overall scale is absorbed into the period-fixing factor \(K\).
\\

\(\Delta_{L,*}\)
&
Reference value of the Liouvillian gap
&
\(\mathrm{ps^{-1}}\)
&
Chosen as the maximum value of \(\Delta_L(\theta)\) along the selected
loop,
\[
\Delta_{L,*}
=
\max_{\theta}\Delta_L(\theta).
\]
The normalized gap therefore satisfies
\[
0<
\frac{\Delta_L(\theta)}{\Delta_{L,*}}
\leq
1.
\]
The reference value makes the direct gap-weighting factor dimensionless
and is not an independently varied physical parameter. Its scale is
absorbed into \(K\). For the selected protocol,
\[
\beta=0,
\]
so
\[
\left[
\frac{\Delta_L(\theta)}
{\Delta_{L,*}}
\right]^{\beta}
=
1.
\]
Consequently, \(\Delta_{L,*}\) does not affect the selected unconstrained
angular-speed profile.
\\

\(\alpha\)
&
Exponent controlling dynamic current-based angular-speed shaping
&
Dimensionless
&
Controls how strongly the loop-resolved current \(j_1(\theta)\) changes
the time assigned to different positions of the selected loop. Increasing
\(\alpha\) increases the relative angular speed where
\(j_1(\theta)\) is large. Because the total cycle period remains fixed,
more time is then assigned to positions where \(j_1(\theta)\) is smaller.
For
\[
\alpha=0,
\]
the local dynamic current does not directly shape the unconstrained
angular speed.
\\

\(\beta\)
&
Exponent controlling direct gap weighting
&
Dimensionless
&
For
\[
\beta>0,
\]
the unconstrained angular-speed profile contains the factor
\[
\left[
\frac{\Delta_L(\theta)}
{\Delta_{L,*}}
\right]^{\beta},
\]
so the relaxation gap directly influences the proposed angular speed.
\\

\(r\)
&
Loop-wide angular-speed bound factor
&
Dimensionless
&
Sets the allowed angular-speed range relative to the reference angular
frequency
\[
\omega
=
\frac{2\pi}{\tau}.
\]
The same lower and upper bounds are applied at every position around the
loop:
\[
\frac{\omega}{r}
\leq
\dot{\theta}(\theta)
\leq
r\omega.
\]

The actual upper speed at each loop position is
\[
\dot{\theta}(\theta)
\leq
\min\!\left[
r\omega,
\epsilon_{\max}\Delta_L(\theta)
\right].
\]
Therefore, the local relaxation-gap cap may reduce the allowed speed below
\(r\omega\).
\\

\(\epsilon_{\max}\)
&
Local Liouvillian-gap constraint parameter
&
Dimensionless
&
Imposes
\[
\frac{\dot{\theta}(\theta)}
{\Delta_L(\theta)}
\leq
\epsilon_{\max}.
\]

\\

\(K\)
&
Normalization factor in the nonuniform angular-speed profile
&
\(\mathrm{ps^{-1}}\)
&
Determined numerically from
\[
\int_0^{2\pi}
\frac{d\theta}
{\dot{\theta}(\theta)}
=
\tau.
\]
It ensures that the nonuniform traversal has the prescribed total cycle
duration.
\\

\(\Delta\theta\)
&
Width of one numerical loop cell
&
rad
&
For \(N_t\) equally spaced loop-coordinate points,
\[
\Delta\theta
=
\frac{2\pi}{N_t}.
\]
The same parameter-space grid is used for the uniform and nonuniform
protocols.
\\

\(\Delta t_k\)
&
Physical dwell time assigned to loop cell \(k\)
&
ps
&
For the original uniform protocol,
\[
\Delta t_k
=
\frac{\tau}{N_t}.
\]
For the nonuniform protocol,
\[
\Delta t_k
=
\frac{\Delta\theta}
{\dot{\theta}(\theta_k)}.
\]
Thus, the nonuniform protocol changes the time assigned to each loop point.
\\

\(\mathcal{T}\)
&
Time-ordering operator
&
Operator
&
Orders the generally noncommuting tilted Liouvillians evaluated at
different laboratory times or loop positions.
\\

\(\mathcal{U}_s(\tau)\)
&
One-cycle tilted propagator
&
Matrix operator
&
Defined by
\[
\mathcal{U}_s(\tau)
=
\mathcal{T}
\exp\!\left[
\int_0^\tau
\mathcal{L}_s(t)\,dt
\right].
\]
Numerically,
\[
\mathcal{U}_s(\tau)
=
\prod_k
\exp\!\left[
\mathcal{L}_s(\theta_k)\Delta t_k
\right].
\]
\\

\(s\)
&
FCS counting field
&
\(\mathrm{meV^{-1}}\)
&
The factors
\[
e^{\pm s\Omega_{12}}
\]
tag signed energy exchange with the counted phonon bath terminal.
Derivatives with respect to \(s\) generate the transferred-energy
cumulants.
\\

\(Q(t)\)
&
Net energy accumulated in the counted phonon bath terminal up to time \(t\)
&
meV
&
A downward \(2\rightarrow1\) transition at time \(t_k\) adds
\[
+\Omega_{12}(t_k),
\]
whereas the reverse transition subtracts it. Therefore, \(Q(t)\) is the
signed sum over all counted transitions and is not equal to one
instantaneous gap \(\Omega_{12}(t)\).
\\

\(G(s,\tau)\)
&
One-cycle cumulant-generating function
&
Dimensionless
&
Defined by
\[
G(s,\tau)
=
\ln\!\left[
\langle\mathbf{1}|
\mathcal{U}_s(\tau)
|\boldsymbol{p}_{\mathrm{LC}}\rangle
\right],
\]
where \(\boldsymbol{p}_{\mathrm{LC}}\) is the periodic-limit-cycle
population state. The cumulants are
\[
C_n
=
\left.
\partial_s^nG(s,\tau)
\right|_{s=0}.
\]
\\

\(\boldsymbol{p}_{\mathrm{LC}}\)
&
Periodic-limit-cycle population state
&
Probability vector
&
Chosen as the normalized fixed point of the ordinary one-cycle propagator:
\[
\mathcal{U}_0(\tau)
\boldsymbol{p}_{\mathrm{LC}}
=
\boldsymbol{p}_{\mathrm{LC}}.
\]
It removes dependence on an arbitrary transient initial population.
\\

\shortstack[l]{
\(C_n^{\mathrm{even}}(\tau)\),\\
\(C_n^{\mathrm{odd}}(\tau)\)
}
&
Finite-period orientation-even and orientation-odd combinations
&
\(C_1\): meV/cycle;
\(C_2\): \(\mathrm{meV^2/cycle}\)
&
Defined by
\[
C_n^{\mathrm{even}}
=
\frac{
C_{n,\mathrm{CW}}
+
C_{n,\mathrm{CCW}}
}{2},
\]
and
\[
C_n^{\mathrm{odd}}
=
\frac{
C_{n,\mathrm{CW}}
-
C_{n,\mathrm{CCW}}
}{2}.
\]
These names are used at arbitrary finite cycle period.
\\

\(C_n^{\mathrm{dyn}}\)
&
Dynamic cumulant
&
\(C_1\): meV/cycle;
\(C_2\): \(\mathrm{meV^2/cycle}\)
&
Identified with the slow-driving orientation-even contribution. It depends
on the physical time spent in different regions of the loop and therefore
changes when the local dwell-time distribution is modified.
\\

\(C_n^{\mathrm{geo}}\)
&
Geometric cumulant
&
\(C_1\): meV/cycle;
\(C_2\): \(\mathrm{meV^2/cycle}\)
&
Identified with the converged orientation-odd contribution. In the
adiabatic limit, it depends primarily on the oriented parameter-space loop
rather than on the local traversal speed.
\\

\(B_s^{V_G,I_{\mathrm{pump}}}\)
&
Berry-Sinitsyn curvature of the tilted Liouvillian
&
\(\mathrm{V^{-1}}\) per pump-control unit
&
Its oriented integral over the enclosed
\((V_G,I_{\mathrm{pump}})\) control-space area gives the adiabatic
geometric contribution to the cumulant-generating function.
\\

\end{longtable}

\endgroup

\FloatBarrier
\newpage  
\section{Experimental inputs and driven protocol}
\label{sec:si_inputs_protocol}

This section presents the experimentally anchored Hamiltonian input
surface reconstructed from the digitized experimental data together with
the imposed gate-pump loop. The gate coordinate changes the instantaneous
Hamiltonian, whereas the pump coordinate acts only through the
optical-injection part of the Liouvillian.

\subsection{Experimentally anchored parameter surface}
\label{sec:si_parameter_surface}

\begin{table}[H]
\centering
\scriptsize
\setlength{\tabcolsep}{4.0pt}
\renewcommand{\arraystretch}{1.10}

\caption{Experimentally anchored parameter surface obtained by
pixel-level digitization of Tang \textit{et al.}\
Fig.~4(a)-(c)~\cite{Tang2021}. Energies and couplings are in meV, and
\(D\) is in \(\mathrm{meV}/(\mathrm{V\,nm^{-1}})\).
The column \(\varepsilon_1^{\mathrm{dom}}\) gives the spectrally dominant
\(X_1\)-like fitted branch used as \(E_1(V_G)\) in the Hamiltonian,
whereas \(\varepsilon_1^{\mathrm{sub}}\) records the weaker simultaneously
resolved branch. Values marked by \(\dagger\) are linearly interpolated at
\(V_G=4.5~\mathrm{V}\), where no fitted \(E_2\) or
\(\lvert W_2\rvert\) value is reported.}
\label{tab:params}

\resizebox{\textwidth}{!}{%
\begin{tabular}{cccccccccc}
\toprule
\(V_G\)
&
\(E_0\)
&
\(\varepsilon_1^{\mathrm{dom}}\)
&
dom.
&
\(\varepsilon_1^{\mathrm{sub}}\)
&
sub.
&
\(E_2\)
&
\(\lvert W_1\rvert\)
&
\(\lvert W_2\rvert\)
&
\(D\)
\\

(V)
&
(meV)
&
(meV)
&
label
&
(meV)
&
label
&
(meV)
&
(meV)
&
(meV)
&
\(\mathrm{meV}/(\mathrm{V\,nm^{-1}})\)
\\
\midrule

0.5 & 1941 & 2004 & A & -   & - & 2040 & 37 & 33 & 299 \\
1.0 & 1943 & 2005 & A & -   & - & 2039 & 39 & 33 & 314 \\
1.5 & 1942 & 2005 & A & -   & - & 2039 & 39 & 32 & 303 \\
2.0 & 1939 & 2004 & A & 1976 & B  & 2039 & 38 & 28 & 292 \\
2.5 & 1955 & 1998 & A & 1976 & B  & 2037 & 44 & 24 & 305 \\
3.0 & 1959 & 1976 & B & 2001 & A  & 2036 & 46 & 22 & 317 \\
3.5 & 1962 & 1976 & B & -   & - & 2035 & 46 & 21 & 325 \\
4.0 & 1960 & 1976 & B & 1950 & C  & 2036 & 46 & 18 & 323 \\
4.5 & 1958 & 1978 & B & 1952 & C  & \(2029^\dagger\) & 45 &
      \(22^\dagger\) & 322 \\
5.0 & 1978 & 1951 & C & 1974 & B  & 2021 & 48 & 25 & 348 \\
5.5 & 1977 & 1951 & C & -   & - & 2019 & 48 & 24 & 345 \\
6.0 & 1978 & 1951 & C & -   & - & 2020 & 47 & 24 & 344 \\
6.5 & 1969 & 1952 & C & -   & - & 2018 & 44 & 20 & 314 \\

\bottomrule
\end{tabular}%
}
\end{table}

\FloatBarrier

\subsection{Driven gate-pump loop and Hamiltonian input functions}
\label{si:loopinputs}
\begin{figure}[htbp]
\centering
\safeincludegraphics[width=0.7\columnwidth]{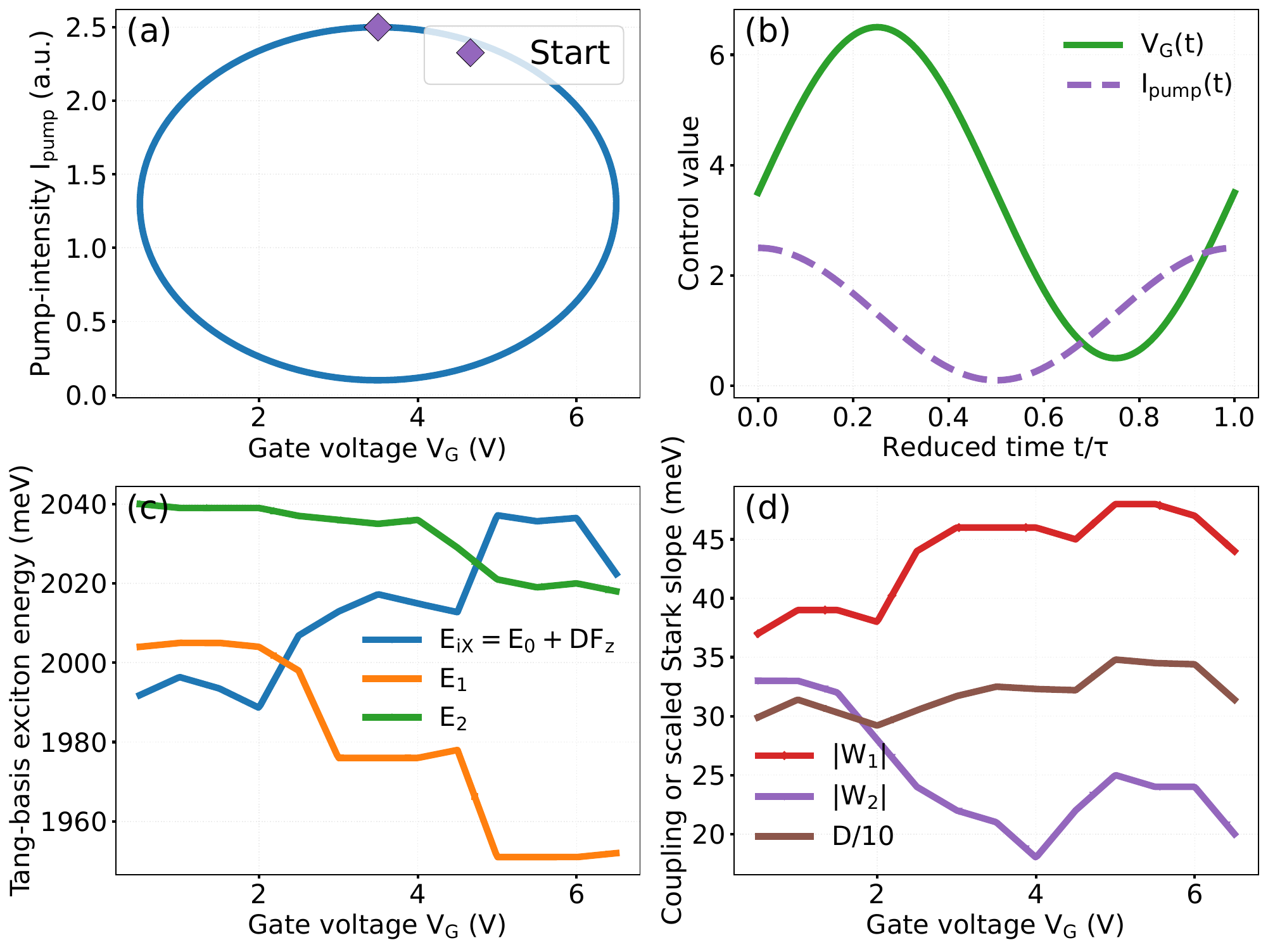}
\caption{(Calculated.) Gate-pump loop and experimentally-anchored Hamiltonian
input functions. (a) Closed trajectory in the
\((V_G(t),I_{\mathrm{pump}}(t))\) control plane. (b) Time-domain gate and pump
waveforms over one cycle. (c) Experimentally-anchored energy nodes used before
diagonalization, including the field-shifted interlayer branch at the selected
operating field. (d) Hybridization couplings \(W_1\), \(W_2\), and the scaled
Stark slope \(D/10\) across the digitized gate range.}
\label{fig:singleloopinputs}
\end{figure}

Figure~\ref{fig:singleloopinputs} summarizes the externally imposed
gate-pump controls and the experimentally anchored Hamiltonian inputs used
in the finite-time propagation. The controls are
\(
V_G(t)=3.5~\mathrm{V}+3.0~\mathrm{V}\sin(2\pi t/\tau)
\)
and
\(
I_{\mathrm{pump}}(t)
=
1.3+1.2\sin(2\pi t/\tau+\Delta\phi),
\qquad
\Delta\phi=\pi/2.
\)
The resulting trajectory spans
\(0.5\leq V_G(t)\leq6.5~\mathrm{V}\) and
\(0.1\leq I_{\mathrm{pump}}(t)\leq2.5\), enclosing a finite oriented area
in the gate-pump control plane. The gate coordinate enters
\(H_T[F_z^{(0)},V_G(t)]\), whereas the pump coordinate enters the
Liouvillian through the optical-injection rates and does not modify
\(H_T\).
The upper panels show the externally imposed controls. Panel (a) gives the
closed trajectory in the \((V_G,I_{\mathrm{pump}})\) plane, while panel (b)
gives the corresponding waveforms over the dimensionless cycle coordinate
\(t/\tau\). The gate coordinate changes the instantaneous Hamiltonian input,
whereas the pump coordinate changes the optical injection rates.

The lower panels show the gate-dependent Hamiltonian inputs used to construct
\(H_T[F_z^{(0)},V_G(t)]\). The field-shifted interlayer branch is
\(E_{iX}(F_z,V_G)=E_0(V_G)+D(V_G)F_z\), and the two intralayer inputs are
\(E_1(V_G)\) and \(E_2(V_G)\). Panel (c) shows these three energy inputs, with
\(E_1(V_G)\) chosen as the spectrally dominant \(X_1\)-like fitted branch listed
in Table~\ref{tab:params}. Panel (d) shows \(|W_1(V_G)|\), \(|W_2(V_G)|\), and
the scaled Stark slope \(D(V_G)/10\). The scaling of \(D\) is used only for
plotting. At each instantaneous \(V_G(t)\), the interpolated inputs are
diagonalized to obtain the dressed energies, gaps, and state compositions that
enter the population rates and the tilted-Liouvillian propagation.
\FloatBarrier
\clearpage
\section{Operating field, counted transition, and model scope}
\label{sec:si_operating_field}

This section first identifies the fixed out of plane field for which one
dressed state transition remains within the phonon energy
window throughout the working gate-pump loop. The gate-dependent
dressed spectrum and state composition are then examined to specify the
energy quantum and state-character factors entering the population
rates. Finally, the population-model scope and the trajectory level
counted-energy convention are stated explicitly, thereby fixing the
assumptions and sign convention used in the subsequent
tilted-Liouvillian calculation.

\subsection{Operating-field and counted-transition selection}
\label{sec:si_field_counted_selection}

\begin{figure}[htbp]
\centering
\safeincludegraphics[
width=0.7\columnwidth,
keepaspectratio
]{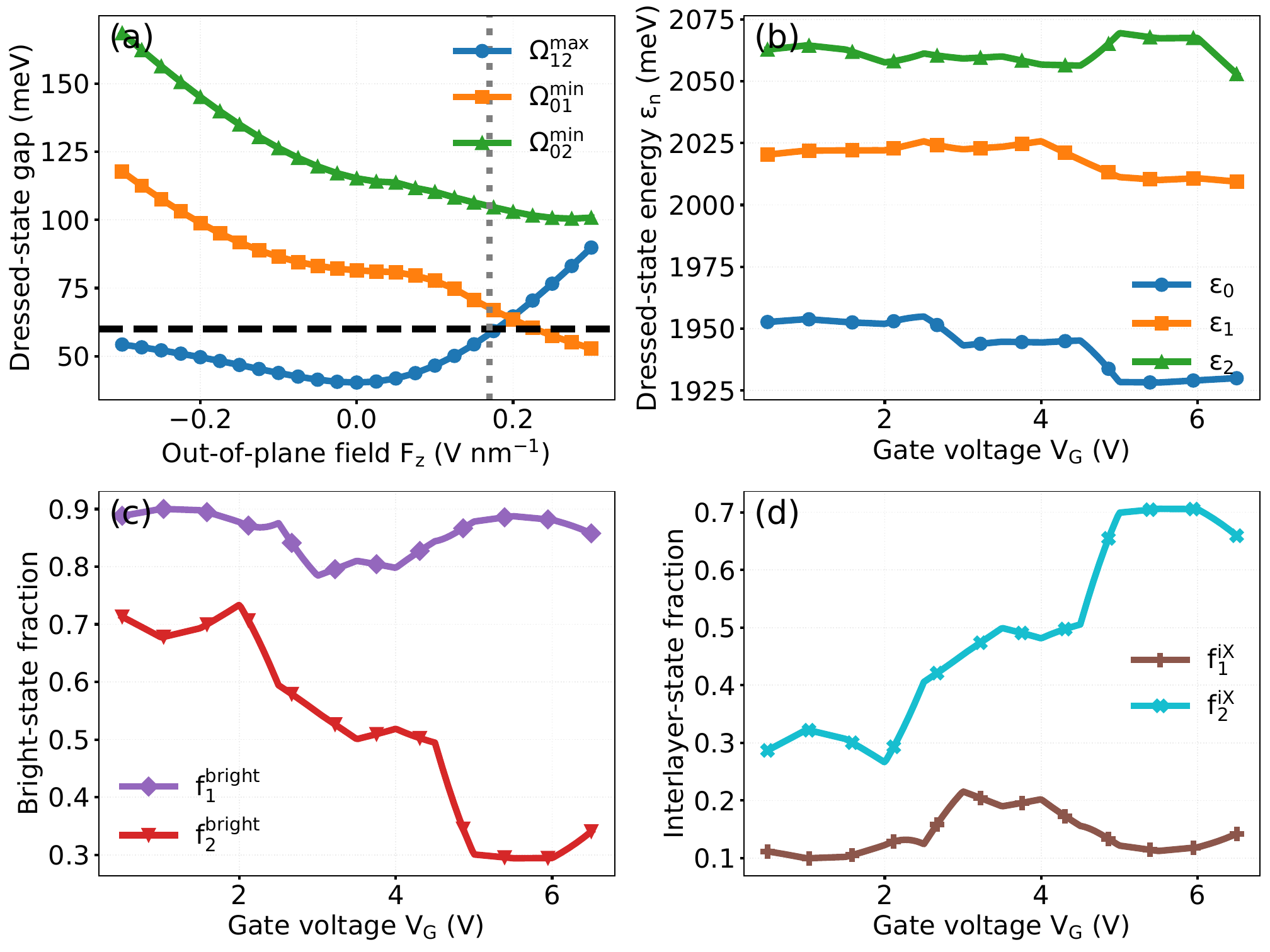}
\caption{
(Calculated.) Operating-field selection and dressed-state composition.
(a) Minimum and maximum dressed-state gaps over the gate-pump loop,
plotted as functions of the fixed out of plane field \(F_z\). The
horizontal dashed line marks the \(60~\mathrm{meV}\) phonon cutoff,
whereas the vertical dotted line marks the selected field
\(F_z^{(0)}=+0.17~\mathrm{V\,nm^{-1}}\). The gap windows at this field
indicate that \(\Omega_{12}\) remains below the cutoff throughout the
loop, whereas the minimum values of \(\Omega_{01}\) and
\(\Omega_{02}\) remain above it.
(b) Dressed eigenvalues \(\epsilon_n(t)\) plotted against \(V_G(t)\) at
\(F_z^{(0)}\).
(c,d) Bright and interlayer fractions, respectively, of the retained
dressed branches \(\lvert\psi_1(t)\rangle\) and
\(\lvert\psi_2(t)\rangle\).
}
\label{fig:fieldselection}
\end{figure}

The extrema of the dressed-state gaps plotted in
Fig.~\ref{fig:fieldselection}(a) are used to select the fixed operating
field \(F_z^{(0)}\).
For each trial value of \(F_z\), the field is held fixed while
\(V_G(t)\) traverses one complete cycle over \(0\leq t\leq\tau\). The
Hamiltonian \(H_T[F_z,V_G(t)]\) is diagonalized at each time \(t\) to
obtain the instantaneous dressed eigenvalues \(\epsilon_n(t)\), and the
corresponding dressed-state gaps are calculated as
\(\Omega_{ij}(t)=\epsilon_j(t)-\epsilon_i(t)\), with \(i<j\). The
dependence of these quantities on the selected trial value of \(F_z\)
is left implicit because \(F_z\) remains fixed during each complete
cycle. For that trial field, the minimum and maximum values attained by
each gap are
\(\Omega_{ij}^{\min}=\min_{0\leq t\leq\tau}\Omega_{ij}(t)\) and
\(\Omega_{ij}^{\max}=\max_{0\leq t\leq\tau}\Omega_{ij}(t)\),
respectively. Repeating this calculation for all trial values of \(F_z\)
gives the minimum and maximum gap curves plotted as functions of \(F_z\)
in Fig.~\ref{fig:fieldselection}(a).

The resulting gap windows are compared with the
\(60~\mathrm{meV}\) phonon cutoff. This comparison determines which
dressed-state transition can be retained as a phonon channel in the
tilted-Liouvillian calculation. Among the three dressed-state gaps, only
\(\Omega_{12}(t)\) satisfies this condition over the complete gate-voltage
range. Its maximum value remains below the cutoff, whereas the minimum
values of \(\Omega_{01}(t)\) and \(\Omega_{02}(t)\) remain above it. We
therefore choose
\(F_z^{(0)}=+0.17~\mathrm{V\,nm^{-1}}\). At this field, the counted
transition is the upper dressed-state relaxation \(\lvert\psi_2(t)\rangle
\rightarrow\lvert\psi_1(t)\rangle\),
and the counting field is attached only to the corresponding retained
\(\Omega_{12}(t)\) phonon channel.

\subsection{Dressed-state spectrum and state-character variation}
\label{sec:si_dressed_character}

The gate-dependent eigenvalues plotted in
Fig.~\ref{fig:fieldselection}(b) specify the dressed spectrum used in
the rate calculation.

At each instantaneous value of \(V_G(t)\), the fixed-field Hamiltonian
is diagonalized to obtain the three dressed eigenvalues
\(\epsilon_0(t)\), \(\epsilon_1(t)\), and \(\epsilon_2(t)\), and the
corresponding instantaneous eigenstates
\(\lvert\psi_n(t)\rangle\), with \(n=0,1,2\). In the bare exciton basis,
each dressed state is expanded as
\(
\lvert\psi_n(t)\rangle
=
c_{iX,n}(t)\lvert iX\rangle
+
c_{X1,n}(t)\lvert X_1\rangle
+
c_{X2,n}(t)\lvert X_2\rangle .
\)
The variation of the dressed eigenvalues with \(V_G(t)\) reflects the
changing detuning between the field-shifted interlayer branch and the
two intralayer branches and determines the instantaneous counted energy
quantum
\(
\Omega_{12}(t)
=
\epsilon_2(t)-\epsilon_1(t).
\)

The bright and interlayer fractions of the retained branches
\(\lvert\psi_1(t)\rangle\) and
\(\lvert\psi_2(t)\rangle\) are plotted in
Figs.~\ref{fig:fieldselection}(c) and
\ref{fig:fieldselection}(d), respectively. The bright fraction is
\(
f_n^{\mathrm{bright}}(t)
=
\left|c_{X1,n}(t)\right|^2
+
\left|c_{X2,n}(t)\right|^2,
\)
whereas the interlayer fraction is
\(
f_n^{\mathrm{iX}}(t)
=
\left|c_{iX,n}(t)\right|^2.
\)
The bright fraction determines the optical-injection and radiative
weights, whereas the interlayer fraction tracks the spatially indirect
component of each dressed state. These state-character factors enter
the population rates defined in
Sec.~\ref{sec:si_micro_to_rates}.

\begin{figure}[htbp]
\centering
\safeincludegraphics[
width=0.7\columnwidth,
keepaspectratio
]{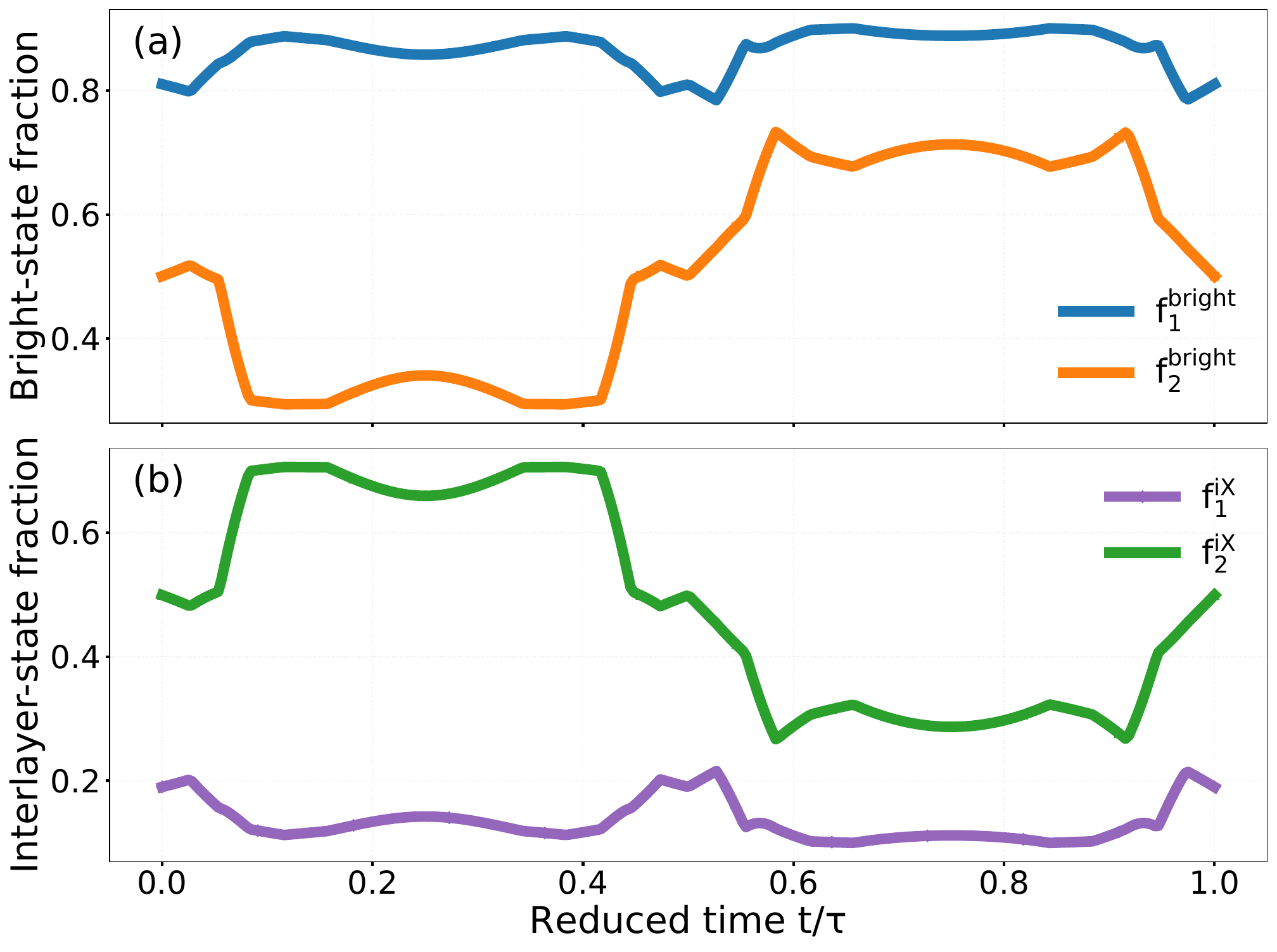}
\caption{
(a) Bright fractions
\(f_1^{\mathrm{bright}}(t)\) and
\(f_2^{\mathrm{bright}}(t)\) of the retained dressed branches
\(\lvert\psi_1(t)\rangle\) and
\(\lvert\psi_2(t)\rangle\), respectively.
(b) Interlayer fractions
\(f_1^{\mathrm{iX}}(t)\) and
\(f_2^{\mathrm{iX}}(t)\) of the same two branches, respectively.
}
\label{fig:loop_character_trajectory}
\end{figure}

The corresponding state-character variation along one complete
gate-pump cycle is plotted in
Fig.~\ref{fig:loop_character_trajectory}. The two controls have
distinct roles: \(I_{\mathrm{pump}}(t)\) sets the instantaneous
optical-injection strength, whereas \(V_G(t)\) determines the dressed
energies and state compositions.

\subsection{From microscopic coupling to population rates}
\label{sec:si_micro_to_rates}

The four-state open-system Hamiltonian is written as
\begin{equation}
H_{\mathrm{tot}}(t)
=
H_{\mathrm{sys}}(t)
+
H_{\mathrm{bath}}
+
H_{\mathrm{SB}}(t).
\label{eq:si_total_hamiltonian}
\end{equation}
In the instantaneous dressed basis, the system Hamiltonian is
\begin{equation}
H_{\mathrm{sys}}(t)
=
E_g\lvert g\rangle\langle g\rvert
+
\sum_{n=0}^{2}
\epsilon_n(t)
\lvert\psi_n(t)\rangle
\langle\psi_n(t)\rvert .
\label{eq:si_system_hamiltonian}
\end{equation}
The coherent interlayer-intralayer hybridization is already contained
in the dressed energies and states obtained by diagonalizing
\(H_T[F_z^{(0)},V_G(t)]\). Optical injection, radiative recombination,
and phonon-assisted interbranch transitions therefore enter through
the system-bath coupling rather than as off-diagonal terms in
\(H_{\mathrm{sys}}(t)\).

The three environmental terminals are represented by
\begin{equation}
\begin{aligned}
H_{\mathrm{bath}}
&=
H_{\mathrm{pump}}
+
H_{\mathrm{phonon}}
+
H_{\mathrm{rad}}
\\
&=
\sum_{\alpha\in h}
\omega_{\alpha}
a_{\alpha}^{\dagger}a_{\alpha}
+
\sum_{q\in c}
\omega_q
b_q^{\dagger}b_q
+
\sum_{\ell\in r}
\omega_{\ell}
c_{\ell}^{\dagger}c_{\ell},
\end{aligned}
\label{eq:si_bath_hamiltonian}
\end{equation}
where \(h\), \(c\), and \(r\) label the optical-pump, phonon, and
radiative terminals, respectively. A minimal coupling Hamiltonian is
\begin{equation}
\begin{aligned}
H_{\mathrm{SB}}(t)
={}&
\sum_{n,\alpha}
\left[
\eta_{n\alpha}(t)
\lvert\psi_n(t)\rangle\langle g\rvert
a_{\alpha}
+
\mathrm{h.c.}
\right]
\\
&+
\sum_{n,\ell}
\left[
\xi_{n\ell}(t)
\lvert g\rangle\langle\psi_n(t)\rvert
c_{\ell}^{\dagger}
+
\mathrm{h.c.}
\right]
\\
&+
\sum_{i<j,q}
\left[
g^{c}_{ij,q}(t)
\lvert\psi_i(t)\rangle
\langle\psi_j(t)\rvert
+
\mathrm{h.c.}
\right]
\left(
b_q+b_q^{\dagger}
\right).
\end{aligned}
\label{eq:si_system_bath_coupling}
\end{equation}
The first line produces optical injection
\(\lvert g\rangle\rightarrow\lvert\psi_n(t)\rangle\), the second
produces radiative recombination
\(\lvert\psi_n(t)\rangle\rightarrow\lvert g\rangle\), and the third
produces phonon-assisted transitions between instantaneous dressed
branches.

The pump-dependent optical matrix element is represented as
\begin{equation}
\eta_{n\alpha}(t)
=
\eta_{\alpha}^{0}
\sqrt{I_{\mathrm{pump}}(t)}\,
M_n(t),
\label{eq:si_pump_coupling_amplitude}
\end{equation}
with
\begin{equation}
\left|M_n(t)\right|^2
=
f_n^{\mathrm{bright}}(t)
=
\left|c_{X1,n}(t)\right|^2
+
\left|c_{X2,n}(t)\right|^2.
\label{eq:si_optical_matrix_weight}
\end{equation}
Thus, the square of the pump-coupling amplitude is proportional to
\(I_{\mathrm{pump}}(t)f_n^{\mathrm{bright}}(t)\).

To second order in \(H_{\mathrm{SB}}(t)\), the reduced density operator
in the interaction picture obeys the Born-Markov equation
\begin{equation}
\frac{d\widetilde{\rho}_{\mathrm S}(t)}{dt}
=
-\frac{1}{\hbar^2}
\int_{0}^{\infty}
d\tau_{\mathrm B}\,
\operatorname{Tr}_{\mathrm B}
\left[
\widetilde{H}_{\mathrm{SB}}(t),
\left[
\widetilde{H}_{\mathrm{SB}}(t-\tau_{\mathrm B}),
\widetilde{\rho}_{\mathrm S}(t)
\otimes
\rho_{\mathrm B}
\right]
\right],
\label{eq:si_born_markov}
\end{equation}
where \(\tau_{\mathrm B}\) is the bath-correlation-time variable and
\(\rho_{\mathrm B}\) is the stationary state of the environmental
terminals. Evaluation of the bath correlation functions gives
instantaneous transition rates of the golden-rule form
\begin{equation}
k_{\mu\leftarrow\nu}^{(x)}(t)
=
\frac{2\pi}{\hbar}
\sum_{\lambda\in x}
\left|
\left\langle
\mu;f_{\lambda}
\left|
H_{\mathrm{SB}}(t)
\right|
\nu;i_{\lambda}
\right\rangle
\right|^2
P_{i_{\lambda}}\,
\delta(E_f-E_i),
\label{eq:si_golden_rule}
\end{equation}
where \(x\in\{h,c,r\}\) labels the terminal,
\(P_{i_{\lambda}}\) is the occupation probability of the initial bath
state, and
\begin{equation}
k_{\mu\leftarrow\nu}(t)
=
\sum_{x\in\{h,c,r\}}
k_{\mu\leftarrow\nu}^{(x)}(t).
\label{eq:si_total_rate}
\end{equation}

With the instantaneous jump operator
\begin{equation}
A_{\mu\nu}(t)
=
\lvert\mu(t)\rangle
\langle\nu(t)\rvert ,
\label{eq:si_jump_operator}
\end{equation}
the Born-Markov-secular generator takes the form
\begin{align}
\dot{\rho}_{\mathrm S}(t)
={}&
-\frac{i}{\hbar}
\left[
H_{\mathrm{sys}}(t),
\rho_{\mathrm S}(t)
\right]
\nonumber\\
&+
\sum_{\mu\neq\nu}
k_{\mu\leftarrow\nu}(t)
\Bigg[
A_{\mu\nu}(t)
\rho_{\mathrm S}(t)
A_{\mu\nu}^{\dagger}(t)
\nonumber\\
&\hspace{3.8cm}
-\frac{1}{2}
\left\{
A_{\mu\nu}^{\dagger}(t)
A_{\mu\nu}(t),
\rho_{\mathrm S}(t)
\right\}
\Bigg].
\label{eq:si_secular_generator}
\end{align}

Defining the instantaneous populations as
\begin{equation}
p_{\mu}(t)
=
\langle\mu(t)|
\rho_{\mathrm S}(t)
|\mu(t)\rangle,
\label{eq:si_population_definition}
\end{equation}
and neglecting the derivative couplings
\(\langle\mu(t)|\dot{\nu}(t)\rangle\), the diagonal part of
Eq.~\eqref{eq:si_secular_generator} becomes the Pauli population
equation
\cite{Breuer2002,Esposito2009,Campaioli2024},
\begin{equation}
\dot p_{\mu}(t)
=
\sum_{\nu\neq\mu}
\left[
k_{\mu\leftarrow\nu}(t)p_{\nu}(t)
-
k_{\nu\leftarrow\mu}(t)p_{\mu}(t)
\right].
\label{eq:si_pauli_equation}
\end{equation}

For the present model, the optical-injection rates are
\begin{equation}
a_n(t)
\equiv
k_{n\leftarrow g}(t)
=
\Gamma_{\mathrm{pump}}^{0}
I_{\mathrm{pump}}(t)
f_n^{\mathrm{bright}}(t),
\qquad n=0,1,2.
\label{eq:si_injection_rates}
\end{equation}
The radiative recombination rates are
\begin{equation}
\begin{aligned}
b_n(t)
\equiv
k_{g\leftarrow n}(t)
={}&
\gamma_{\mathrm{rad}}^{\mathrm{intra}}
f_n^{\mathrm{bright}}(t)
\\
&+
\gamma_{\mathrm{rad}}^{\mathrm{inter}}
f_n^{\mathrm{iX}}(t),
\qquad n=0,1,2.
\end{aligned}
\label{eq:si_radiative_rates}
\end{equation}

For two dressed branches \(i<j\), the instantaneous transition energy
is
\begin{equation}
\Omega_{ij}(t)
=
\epsilon_j(t)-\epsilon_i(t).
\label{eq:si_transition_gap}
\end{equation}
The Bose occupation of the phonon terminal is
\begin{equation}
n_c[\Omega_{ij}(t)]
=
\left[
\exp\!\left(
\frac{\Omega_{ij}(t)}
{k_{\mathrm B}T_c}
\right)
-1
\right]^{-1}.
\label{eq:si_phonon_occupation}
\end{equation}
The overlap-weighted phonon coupling scale is
\begin{equation}
\begin{aligned}
\gamma_{c,ij}(t)
={}&
\gamma_c^{\mathrm{intra}}
\left|
c_{X1,i}^{*}(t)c_{X1,j}(t)
+
c_{X2,i}^{*}(t)c_{X2,j}(t)
\right|^2
\\
&+
\gamma_c^{\mathrm{inter}}
\left|
c_{iX,i}^{*}(t)c_{iX,j}(t)
\right|^2.
\end{aligned}
\label{eq:si_phonon_coupling_prefactor}
\end{equation}
The downward phonon-emission and upward phonon-absorption rates are
therefore
\begin{align}
d_{ij}(t)
&\equiv
k_{i\leftarrow j}(t)
=
\gamma_{c,ij}(t)
\left\{
n_c[\Omega_{ij}(t)]+1
\right\},
\label{eq:si_downward_rate}
\\
u_{ij}(t)
&\equiv
k_{j\leftarrow i}(t)
=
\gamma_{c,ij}(t)
n_c[\Omega_{ij}(t)].
\label{eq:si_upward_rate}
\end{align}

After applying the phonon accessibility condition, only the
\(1\leftrightarrow2\) interbranch channel is retained in the population
generator. We consequently define
\begin{equation}
d(t)\equiv d_{12}(t),
\qquad
u(t)\equiv u_{12}(t),
\qquad
\Omega(t)\equiv\Omega_{12}(t),
\label{eq:si_retained_rate_abbreviations}
\end{equation}
while the other interbranch rates are set to zero in the retained
single-channel model.

Using Eq.~\eqref{eq:si_pauli_equation}, the four population equations
are
\begin{align}
\dot p_g(t)
={}&
-\left[
a_0(t)+a_1(t)+a_2(t)
\right]p_g(t)
\nonumber\\
&+
b_0(t)p_0(t)
+
b_1(t)p_1(t)
+
b_2(t)p_2(t),
\label{eq:si_pg_equation}
\\
\dot p_0(t)
={}&
a_0(t)p_g(t)
-
b_0(t)p_0(t),
\label{eq:si_p0_equation}
\\
\dot p_1(t)
={}&
a_1(t)p_g(t)
-
\left[
b_1(t)+u(t)
\right]p_1(t)
+
d(t)p_2(t),
\label{eq:si_p1_equation}
\\
\dot p_2(t)
={}&
a_2(t)p_g(t)
+
u(t)p_1(t)
-
\left[
b_2(t)+d(t)
\right]p_2(t).
\label{eq:si_p2_equation}
\end{align}

For
\begin{equation}
\mathbf p(t)
=
\begin{pmatrix}
p_g(t)&p_0(t)&p_1(t)&p_2(t)
\end{pmatrix}^{T},
\label{eq:si_population_vector}
\end{equation}
Eqs.~\eqref{eq:si_pg_equation}-\eqref{eq:si_p2_equation} can be
written as
\begin{equation}
\dot{\mathbf p}(t)
=
\mathcal L(t)\mathbf p(t).
\label{eq:si_matrix_population_equation}
\end{equation}
Suppressing the common time arguments only inside the matrix,
\begin{equation}
a_n\equiv a_n(t),
\qquad
b_n\equiv b_n(t),
\qquad
d\equiv d(t),
\qquad
u\equiv u(t),
\label{eq:si_compact_rate_notation}
\end{equation}
the ordinary population Liouvillian is
\begingroup
\setlength{\arraycolsep}{3pt}
\begin{equation}
\mathcal L(t)
=
\begin{pmatrix}
-\displaystyle\sum_{n=0}^{2}a_n
&
b_0
&
b_1
&
b_2
\\
a_0
&
-b_0
&
0
&
0
\\
a_1
&
0
&
-(b_1+u)
&
d
\\
a_2
&
0
&
u
&
-(b_2+d)
\end{pmatrix}.
\label{eq:si_ordinary_liouvillian}
\end{equation}
\endgroup

Each column of \(\mathcal L(t)\) sums to zero, which ensures
conservation of the total probability,
\begin{equation}
p_g(t)+p_0(t)+p_1(t)+p_2(t)=1.
\label{eq:si_probability_conservation}
\end{equation}
The gate control changes the rates through the instantaneous dressed
energies and state-character factors, whereas
\(I_{\mathrm{pump}}(t)\) enters directly through the optical-injection
rates in Eq.~\eqref{eq:si_injection_rates}.

\subsection{Population-model scope and accumulated-energy convention}
\label{sec:si_population_scope}

The Liouvillian derived in
Sec.~\ref{sec:si_micro_to_rates} propagates only the populations of
the four instantaneous states
\(
\left\{
\lvert g\rangle,
\lvert\psi_0(t)\rangle,
\lvert\psi_1(t)\rangle,
\lvert\psi_2(t)\rangle
\right\}.
\)
Derivative couplings of the form
\(\langle\psi_i(t)\vert\dot{\psi}_j(t)\rangle\), together with the
coherences that they may generate, are not included. The model therefore
relies on the weak-coupling, Markovian, secular, and slow-control
assumptions under which the time-local population equation in
Eq.~\eqref{eq:si_pauli_equation} is applicable
\cite{Breuer2002,Paulino2024}. The Liouvillian-gap and cycle-period
scans test the separation between the population-relaxation and driving
time scales within this reduced model. They do not replace a coherent
open-system adiabatic analysis.

The optical-injection, radiative-loss, and phonon rates in
Eqs.~\eqref{eq:si_injection_rates}-\eqref{eq:si_upward_rate} are
phenomenological overlap-weighted rate laws. In particular, the
coherent sum
\(
c_{X1,i}^{*}(t)c_{X1,j}(t)
+
c_{X2,i}^{*}(t)c_{X2,j}(t)
\)
in Eq.~\eqref{eq:si_phonon_coupling_prefactor} represents coupling to
one common effective intralayer phonon channel. Independent
branch-resolved phonon reservoirs would instead give separate
contributions,
\(
\left|
c_{X1,i}^{*}(t)c_{X1,j}(t)
\right|^2
+
\left|
c_{X2,i}^{*}(t)c_{X2,j}(t)
\right|^2,
\)
without the interference cross term. This alternative would therefore
represent a different microscopic coupling assumption.

For an individual stochastic trajectory, the net energy transferred to
the counted phonon terminal during one cycle is
\begin{equation}
Q(\tau)
=
\sum_{k\in 2\rightarrow1}
\Omega_{12}(t_k)
-
\sum_{\ell\in 1\rightarrow2}
\Omega_{12}(t_\ell),
\label{eq:si_accumulated_energy_trajectory}
\end{equation}
where \(t_k\) is the time of the \(k\)th counted downward transition
\(
\lvert\psi_2(t_k)\rangle
\rightarrow
\lvert\psi_1(t_k)\rangle,
\)
and \(t_\ell\) is the time of the \(\ell\)th counted upward transition
\(
\lvert\psi_1(t_\ell)\rangle
\rightarrow
\lvert\psi_2(t_\ell)\rangle
\)
during \(0\leq t\leq\tau\). The indices \(k\) and \(\ell\) enumerate
all downward and upward counted jump events occurring during the
cycle. If no transition of one type occurs, the corresponding sum is
zero.

Because
\(
\Omega_{12}(t)
=
\epsilon_2(t)-\epsilon_1(t)>0,
\)
each downward transition adds the instantaneous transition energy to
\(Q(\tau)\), whereas each upward transition subtracts it. Positive
\(Q(\tau)\) therefore denotes net phonon emission into the counted
terminal, while negative \(Q(\tau)\) denotes net phonon absorption
from that terminal.

Although the system states and the number of jump events are discrete,
the jump times vary continuously, and the instantaneous gap
\(\Omega_{12}(t)\) changes continuously along the driven loop.
Consequently, the possible trajectory values of \(Q(\tau)\) are
effectively continuous. The finite-time-step tilted-Liouvillian
propagation used in the calculation is a numerical approximation to
this underlying continuous-time stochastic process
\cite{Esposito2009,Campisi2011}.

\newpage
\section{Selection of the baseline driving loop}
\label{sec:si_baseline_selection}

The full-range path is treated first as the initial candidate driving
loop. Its loop-geometry quantities are assessed through the
control-centre scans, in which the gate and pump centres are varied and
the corresponding modulation amplitudes are adjusted so that the
complete trajectory remains within the experimentally supported control
window. The reduced-range trajectories generated during these scans are
therefore part of the same loop-geometry assessment rather than separate
subsidiary loops. A nested reduced-area loop is subsequently included as
an additional comparison to isolate the effect of restricting the
enclosed control-space region. The full-range path is designated as the
baseline driving loop only after the control-centre, loop-geometry, and
enclosed-area results are considered together.

\subsection{Control-centre scan}
\label{sec:si_control_centre_scan}

\begin{figure}[htbp]
\centering
\safeincludegraphics[
width=0.8\columnwidth,
keepaspectratio
]{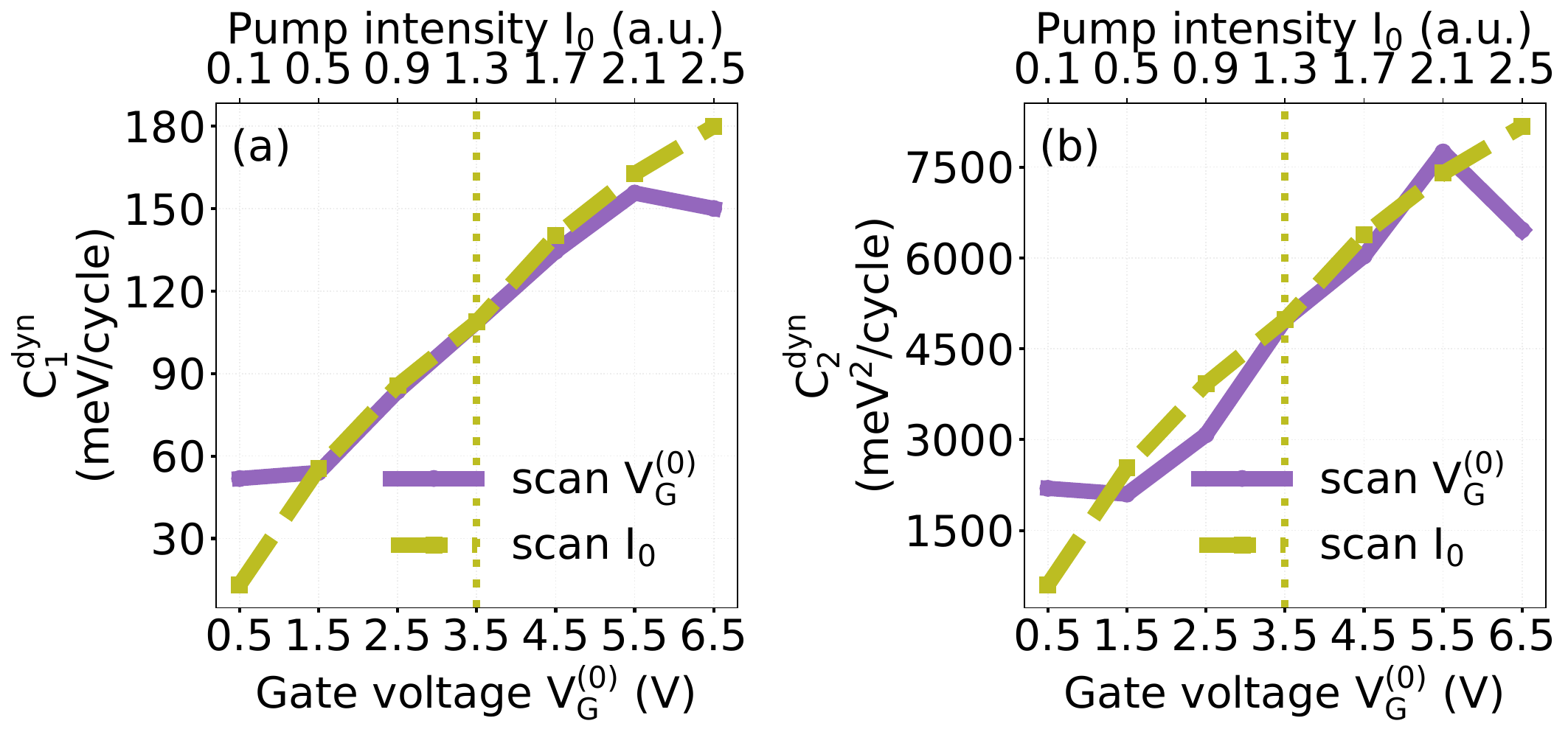}
\caption{
(Calculated.) One-parameter control-centre scan of the dynamic
cumulants at fixed
\(F_z^{(0)}=+0.17~\mathrm{V\,nm^{-1}}\) and
\(\tau=5000~\mathrm{ps}\).
(a) Dynamic first cumulant \(C_1^{\mathrm{dyn}}\).
(b) Dynamic second cumulant \(C_2^{\mathrm{dyn}}\).
In each panel, the solid curve with circular markers represents the
response obtained by displacing the gate centre \(V_G^{(0)}\) across
the experimentally supported gate window
\(0.5\leq V_G^{(0)}\leq6.5~\mathrm{V}\), whereas the dashed curve with
square markers represents the response obtained by displacing the pump
centre \(I_{\mathrm{pump}}^{(0)}\) across the pump-control window
\(0.1\leq I_{\mathrm{pump}}^{(0)}\leq2.5\).
The lower horizontal axis gives \(V_G^{(0)}\), and the upper horizontal
axis gives \(I_{\mathrm{pump}}^{(0)}\). The dotted vertical guides mark
the working centre values \(V_G^{(0)}=3.5~\mathrm{V}\) and
\(I_{\mathrm{pump}}^{(0)}=1.3\).
}
\label{fig:dynamic}
\end{figure}

The orientation-even dynamic cumulants obtained from the varied
loop centres are plotted in Fig.~\ref{fig:dynamic}.
Each point represents a valid gate-pump loop whose centre is displaced
while the complete trajectory remains within the allowed control
window. For each varied centre, the corresponding modulation amplitude
is adjusted so that neither the gate nor the pump trajectory leaves its
supported range.

Together with the geometric control-centre scan in the main manuscript,
the calculated trends identify \(V_G^{(0)}=3.5~\mathrm{V}\) and
\(I_{\mathrm{pump}}^{(0)}=1.3\) as the working centre of the
full-range candidate loop. The geometric cumulants are largest near
this central full-range trajectory. Displacing either centre reduces
the admissible modulation amplitude in at least one control direction
and consequently reduces the effective area enclosed in the
\((V_G(t),I_{\mathrm{pump}}(t))\) plane.

The dynamic cumulants in Fig.~\ref{fig:dynamic} provide the
corresponding orientation-even background. Both
\(C_1^{\mathrm{dyn}}\) and \(C_2^{\mathrm{dyn}}\) generally increase
as either \(V_G^{(0)}\) or \(I_{\mathrm{pump}}^{(0)}\) is increased.
The pump-centre scan exhibits an approximately monotonic increase,
whereas the gate-centre scan increases toward the higher-gate region
and decreases slightly at the largest \(V_G^{(0)}\). The working centre
is therefore selected primarily from the maximum geometric response
within the supported control window, while the present scan records the
dynamic background associated with the same family of valid loops.

\FloatBarrier

\subsection{Loop-geometry and enclosed-area comparison
}
\label{sec:si_loop_area_test}

After fixing the working centre, the full-range candidate gate-pump
loop is
\(V_G(t)=3.5~\mathrm{V}
+3.0~\mathrm{V}\sin(2\pi t/\tau)\) and
\(I_{\mathrm{pump}}(t)=1.3
+1.2\sin(2\pi t/\tau+\pi/2)\). It spans
\(0.5\leq V_G(t)\leq6.5~\mathrm{V}\) and
\(0.1\leq I_{\mathrm{pump}}(t)\leq2.5\), thereby using the complete
experimentally supported gate and pump ranges while remaining inside
the Hamiltonian input surface.

As an additional loop-geometry test, the full-range candidate is
compared with the nested reduced-area loop
\(V_G^{\mathrm{red}}(t)=3.0~\mathrm{V}
+1.5~\mathrm{V}\sin(2\pi t/\tau)\) and
\(I_{\mathrm{pump}}^{\mathrm{red}}(t)=1.0
+0.7\sin(2\pi t/\tau+\pi/2)\), which spans
\(1.5\leq V_G^{\mathrm{red}}(t)\leq4.5~\mathrm{V}\) and
\(0.3\leq I_{\mathrm{pump}}^{\mathrm{red}}(t)\leq1.7\).
This reduced loop lies entirely within the control region of the
full-range candidate and is therefore treated as a nested member of the
loop-geometry assessment rather than as an independent loop outside the
supported control range. Both loops are evaluated at
\(F_z^{(0)}=+0.17~\mathrm{V\,nm^{-1}}\) and
\(\tau=5~\mathrm{ns}\).

The full-range candidate yields a \(4.3\times\) larger
\(C_1^{\mathrm{geo}}\) and a \(6.5\times\) larger
\(C_2^{\mathrm{geo}}\) than the nested reduced-area loop. The dynamic
cumulants are also larger for the full-range trajectory because the
cycle samples a broader portion of the driven rate landscape. However,
the stronger increase of the geometric sector establishes that
restricting the trajectory to the smaller nested region substantially
reduces the integrated geometric response. This comparison is therefore
interpreted together with the control-centre and loop-geometry scans and
not as the sole criterion used to select the baseline loop.

\begin{table}[!htbp]
\centering
\small
\caption{Cumulants for the full-range candidate and nested reduced-area
loops at \(\tau=5~\mathrm{ns}\).
}
\label{tab:loops}
\begin{tabular}{lcccc}
\toprule
Loop
&
\(C_1^{\mathrm{dyn}}\)
&
\(C_1^{\mathrm{geo}}\)
&
\(C_2^{\mathrm{dyn}}\)
&
\(C_2^{\mathrm{geo}}\)
\\
&
(meV/cycle)
&
(meV/cycle)
&
(meV\(^2\)/cycle)
&
(meV\(^2\)/cycle)
\\
\midrule

Full-range candidate \((0.5\text{-}6.5~\mathrm{V})\)
&
108.76
&
0.2145
&
\(4.978\times10^{3}\)
&
11.18
\\

Nested reduced-area \((1.5\text{-}4.5~\mathrm{V})\)
&
78.42
&
0.0501
&
\(2.809\times10^{3}\)
&
1.71
\\
\bottomrule
\end{tabular}
\end{table}

The control-centre and loop-geometry assessments identify
\(V_G^{(0)}=3.5~\mathrm{V}\),
\(I_{\mathrm{pump}}^{(0)}=1.3\),
\(A_G=3.0~\mathrm{V}\), and
\(A_I=1.2\) as the working loop-geometry quantities. This central
full-range trajectory gives the largest geometric response among the
tested range-constrained control-centre variations while keeping
\(V_G(t)\) inside the experimentally anchored Hamiltonian surface. The
nested reduced-area comparison further establishes that restricting the
enclosed control-space region substantially lowers both geometric
cumulants. On the basis of these combined assessments, the full-range
candidate is designated as the baseline driving loop for the subsequent
convergence, robustness, sensitivity, and enhancement calculations. At
\(F_z^{(0)}=+0.17~\mathrm{V\,nm^{-1}}\) and
\(\tau=5~\mathrm{ns}\), its baseline geometric cumulants are
\(C_1^{\mathrm{geo}}=0.2145~\mathrm{meV/cycle}\) and
\(C_2^{\mathrm{geo}}=11.18~\mathrm{meV^2/cycle}\).

\FloatBarrier


\clearpage

\section{Numerical initialization, convergence, and detectability}
\label{sec:si_numerical_detectability}

This section verifies the numerical conventions used to extract the
finite-cycle cumulants of the working loop. We first specify the
finite-time tilted-Liouvillian propagation used to construct the
orientation-resolved generating functions. We then compare different
initial population states and establish the periodic limit cycle as the
repeated-cycle initialization. The cycle period is classified relative
to the slowest instantaneous population-relaxation time, and the
finite-period orientation-even and orientation-odd cumulants are tested
against explicit time-extensive and plateau criteria. Finally, the
counting-field differentiation step and the intrinsic statistical
requirements of the CW-CCW measurement are assessed.

\subsection{Finite-time tilted-Liouvillian propagation}
\label{sec:si_finite_time_propagation}

For a uniform traversal of one cycle, the physical time interval
\(0\leq t\leq\tau\) is divided into \(N_t\) equal steps,
\begin{equation}
\Delta t=\frac{\tau}{N_t}.
\label{eq:si_uniform_time_step}
\end{equation}
Here, \(t\) is the running physical time within one cycle, \(\tau\) is
the total cycle period, and \(\Delta t\) is the numerical time step.
The \(k\)th interval is represented by its midpoint
\begin{equation}
t_k
=
\left(k-\frac{1}{2}\right)\Delta t,
\qquad
k=1,\ldots,N_t.
\label{eq:si_midpoint_time}
\end{equation}
For fixed \(N_t\), changing \(\tau\) leaves the sampled reduced-time
points \(t_k/\tau\) unchanged and therefore samples the same points on
the gate-pump loop. It changes only the physical traversal speed and
the interval duration \(\Delta t\). A larger \(\tau\) consequently
corresponds to slower traversal of the same loop.

At each midpoint \(t_k\), the interpolated experimentally anchored
parameter surface fixes
\(H_T[F_z^{(0)},V_G(t_k)]\). Diagonalization gives the instantaneous
dressed energies, gaps, and state-character fractions, from which the
rates \(a_n(t_k)\), \(b_n(t_k)\), \(d(t_k)\), and \(u(t_k)\) are
constructed. The tilted Liouvillian
\(\mathcal L_{s,\eta}(t_k)\), with
\(\eta\in\{\mathrm{CW},\mathrm{CCW}\}\), is then evaluated at the same
midpoint and treated as constant within that interval.

The resulting discrete one-cycle propagator is
\begin{equation}
\mathcal U_{s,\eta}(\tau)
=
\prod_{k=N_t}^{1}
\exp\!\left[
\mathcal L_{s,\eta}(t_k)\Delta t
\right],
\label{eq:si_discrete_propagator}
\end{equation}
where later time steps multiply from the left. Equation
\eqref{eq:si_discrete_propagator} is the midpoint-product
implementation of the time-ordered one-cycle propagator. The
orientation-resolved generating function is
\begin{equation}
G_{\eta}(s,\tau)
=
\ln\!\left[
\mathbf 1^{T}
\mathcal U_{s,\eta}(\tau)
\mathbf p_{\mathrm{LC},\eta}(0)
\right],
\label{eq:si_finite_time_generating_function}
\end{equation}
where \(\mathbf p_{\mathrm{LC},\eta}(0)\) is the periodic-limit-cycle
initial state defined in the following subsection. The cumulants are
obtained by differentiating
Eq.~\eqref{eq:si_finite_time_generating_function} with respect to the
counting field and then forming the CW-CCW half-sum and
half-difference.

\subsection{Periodic limit-cycle initialization and
preparation-transient check}
\label{si:init_check}

The one-cycle generating function depends not only on the tilted
propagator but also on the population vector from which the cycle is
started. We distinguish three initialization choices. A bare-ground
start uses \(\mathbf p_g=(1,0,0,0)^T\). A frozen instantaneous steady
state uses the normalized null vector of the ordinary Liouvillian at the
initial control point. The periodic limit-cycle state is instead defined
as the fixed point of the complete ordinary one-cycle propagator for
each orientation,
\begin{equation}
\mathcal U_{0,\eta}(\tau)\,
\mathbf p_{\mathrm{LC},\eta}(0)
=
\mathbf p_{\mathrm{LC},\eta}(0),
\qquad
\mathbf 1^T\mathbf p_{\mathrm{LC},\eta}(0)=1,
\qquad
\eta\in\{\mathrm{CW},\mathrm{CCW}\}.
\label{eq:si_periodic_initial_state}
\end{equation}
Here, \(\mathcal U_{0,\eta}(\tau)\) is the ordinary one-cycle propagator
evaluated at counting field \(s=0\) for the orientation
\(\eta\), where \(\eta=\mathrm{CW}\) or
\(\eta=\mathrm{CCW}\). It propagates the population vector through one
complete cycle of duration \(\tau\).
The vector \(\mathbf p_{\mathrm{LC},\eta}(0)\) is the population
distribution at the beginning of the cycle in the periodic repeated-cycle
regime. The fixed-point condition
\(\mathcal U_{0,\eta}(\tau)\mathbf p_{\mathrm{LC},\eta}(0)
=\mathbf p_{\mathrm{LC},\eta}(0)\) means that, after one complete
cycle, the populations return to the same values from which that cycle
started. The row vector \(\mathbf 1^T=(1,1,1,1)\) sums the four
populations, and the condition
\(\mathbf 1^T\mathbf p_{\mathrm{LC},\eta}(0)=1\) imposes probability
normalization.

Numerically, \(\mathbf p_{\mathrm{LC},\eta}(0)\) is obtained as the
right eigenvector of \(\mathcal U_{0,\eta}(\tau)\) associated with
eigenvalue unity. The eigenvalue \(1\) expresses the fact that this
population vector is unchanged after one full cycle. The corresponding
eigenvector is normalized so that its components sum to unity. A
candidate eigenvector is rejected if any population is negative beyond
the numerical tolerance used in the propagation. Negative values are not
clipped. The same \(s=0\) periodic-limit-cycle vector is used as the
initial state for all five counting-field nodes entering the
finite-difference stencil.

The periodic state represents the repeated-cycle operating condition:
after one complete cycle, the population returns to its initial value,
so the following cycle is statistically equivalent to the previous one.
By contrast, a one-cycle calculation initialized in the bare ground
state contains a preparation transient.

Under the loop-reversal convention, the sign of the
gate-pump phase lag is reversed while the gate-phase origin is retained.
The CW and CCW calculations therefore begin at different values of the
pump control, and their bare-ground preparation transients are not
identical.
Consequently, a bare-ground start introduces unequal preparation biases
into the orientation half-difference. These one-cycle transients are
removed by using the corresponding periodic limit-cycle state for every
reported cumulant.

The initialization results at the baseline uniform operating point,
\(\tau=5000~\mathrm{ps}\), are collected in
Table~\ref{tab:si_init_check_5000}.
The frozen instantaneous steady state and the periodic limit cycle agree
at this slow period, whereas the bare-ground start produces a substantial
one-cycle bias.

\begin{table}[!htbp]
\centering
\scriptsize
\setlength{\tabcolsep}{4pt}
\renewcommand{\arraystretch}{1.10}
\caption{Initialization check for the working full range uniform loop at
\(\tau=5000~\mathrm{ps}\), using the canonical piecewise-linear
interpolation and \(N_t=800\).
At this period, the orientation-even and orientation-odd cumulants have
reached their slow-driving plateaus and are therefore reported using the
shorter dynamic and geometric labels.
}
\label{tab:si_init_check_5000}
\begin{tabular}{p{4.2cm}cccc}
\toprule
\textbf{Initialization}
&
\(C_1^{\mathrm{dyn}}\)
&
\(C_1^{\mathrm{geo}}\)
&
\(C_2^{\mathrm{dyn}}\)
&
\(C_2^{\mathrm{geo}}\)
\\
&
(meV/cycle)
&
(meV/cycle)
&
(meV\(^2\)/cycle)
&
(meV\(^2\)/cycle)
\\
\midrule
Bare ground state
& 108.700834 & 0.169359 & 4976.546740 & 9.564115 \\
Frozen instantaneous steady state
& 108.757665 & 0.214531 & 4978.586022 & 11.180100 \\
Periodic limit cycle
& 108.757669 & 0.214531 & 4978.586169 & 11.180098 \\
\bottomrule
\end{tabular}
\end{table}

Relative to the periodic limit cycle, the bare-ground start lowers
\(C_1^{\mathrm{geo}}\) by approximately \(21.1\%\) and
\(C_2^{\mathrm{geo}}\) by approximately \(14.5\%\). The dynamic
cumulants change much less because they are dominated by the extensive
time-integrated background. This contrast explains why the
initialization error is particularly important for the orientation-odd
signal obtained by subtracting two much larger orientation-resolved
cumulants.

The finite-period differences between the frozen instantaneous
steady state and the periodic limit cycle are listed in
Table~\ref{tab:si_inst_periodic_tau}.
At \(200~\mathrm{ps}\), the difference in
\(C_2^{\mathrm{odd}}\) is approximately
\(2.44\times10^{-2}~\mathrm{meV^2/cycle}\). It decreases to
\(4.58\times10^{-3}~\mathrm{meV^2/cycle}\) at
\(500~\mathrm{ps}\) and to approximately
\(2\times10^{-6}~\mathrm{meV^2/cycle}\) at
\(5000~\mathrm{ps}\). Thus, the frozen instantaneous state becomes an
excellent approximation in the deeply slow-driving regime, but the
periodic state remains the unambiguous repeated-cycle initialization
used for every reported value.

\begin{table}[!htbp]
\centering
\scriptsize
\setlength{\tabcolsep}{3.5pt}
\renewcommand{\arraystretch}{1.08}
\caption{
Finite-time comparison of frozen-instantaneous-steady-state and
periodic-limit-cycle initialization using \(N_t=800\). The
orientation-even and orientation-odd labels are retained because the
shorter-period values need not yet be identified with the fully
converged adiabatic dynamic and geometric contributions.
}
\label{tab:si_inst_periodic_tau}
\begin{tabular}{r l c c c c}
\toprule
\(\tau\) (ps)
&
\textbf{Initialization}
&
\(C_1^{\mathrm{even}}\)
&
\(C_1^{\mathrm{odd}}\)
&
\(C_2^{\mathrm{even}}\)
&
\(C_2^{\mathrm{odd}}\)
\\
&
&
(meV/cycle)
&
(meV/cycle)
&
(meV\(^2\)/cycle)
&
(meV\(^2\)/cycle)
\\
\midrule
200
& Frozen instantaneous
& 4.351769 & 0.208981 & 199.240323 & 10.795043 \\
200
& Periodic limit cycle
& 4.352614 & 0.208302 & 199.270483 & 10.770683 \\
\addlinespace
500
& Frozen instantaneous
& 10.877049 & 0.213363 & 497.963861 & 11.097774 \\
500
& Periodic limit cycle
& 10.877197 & 0.213233 & 497.969033 & 11.093195 \\
\addlinespace
5000
& Frozen instantaneous
& 108.757665 & 0.214531 & 4978.586022 & 11.180100 \\
5000
& Periodic limit cycle
& 108.757669 & 0.214531 & 4978.586169 & 11.180098 \\
\bottomrule
\end{tabular}
\end{table}

Finally, direct comparison of the \(N_t=400\) and \(N_t=800\)
calculations at the baseline \(5000~\mathrm{ps}\) operating point
reconciles the discretization used for the main scans with the refined
initialization check. Doubling the number of time steps changes
\(C_1^{\mathrm{dyn}}\), \(C_1^{\mathrm{geo}}\),
\(C_2^{\mathrm{dyn}}\), and \(C_2^{\mathrm{geo}}\) by approximately
\(0.00179\%\), \(0.00291\%\), \(0.00185\%\), and \(0.00269\%\),
respectively. The fifth-digit differences therefore reflect
time-discretization convergence rather than a change in the physical
model or initialization convention.

\FloatBarrier

\subsection{Cycle-period convergence and Liouvillian time-scale
classification}
\label{sec:si_tau_convergence}

The main manuscript examines the finite-time approach to the
slow-driving plateau over
\(\tau=200\text{-}5000~\mathrm{ps}\). Here, we provide the complete
four-cumulant values for this period scan, classify the tested periods
relative to the slowest instantaneous population-relaxation time, and
extend the orientation-odd calculation to \(10\), \(20\), and
\(50~\mathrm{ns}\). The extended calculation determines whether the
baseline \(5~\mathrm{ns}\) result remains on the same orientation-odd
plateau at still slower traversal speeds.

The instantaneous ordinary Liouvillian
\(\mathcal L_{s=0}(t)\) has one zero eigenvalue associated with
probability conservation. Its instantaneous population-relaxation gap
is defined as
\(\Delta_L(t)=\min_{m\neq0}
[-\operatorname{Re}\lambda_m(0,t)]\), where the minimum is taken over
the nonstationary eigenmodes of the ordinary Liouvillian. Along the
baseline gate-pump loop,
\begin{equation}
\Delta_L^{\min}
=
\min_{0\leq t\leq\tau}\Delta_L(t)
\simeq
0.160~\mathrm{ps^{-1}},
\qquad
\tau_{\mathrm{rel}}^{\max}
=
\frac{1}{\Delta_L^{\min}}
\simeq
6.25~\mathrm{ps}.
\label{eq:si_liouvillian_gap}
\end{equation}
Here, \(\Delta_L^{\min}\) is the smallest population-relaxation gap
encountered during one complete loop, while
\(\tau_{\mathrm{rel}}^{\max}\) is the corresponding longest
instantaneous population-relaxation time.

A convenient dimensionless time-scale diagnostic is
\begin{equation}
\epsilon_{\mathrm{ad}}(\tau)
=
\frac{2\pi/\tau}{\Delta_L^{\min}}.
\label{eq:si_adiabaticity_indicator}
\end{equation}
The numerator \(2\pi/\tau\) is the angular frequency of the uniform
traversal, whereas \(\Delta_L^{\min}\) is the slowest population
relaxation rate encountered along the loop. A smaller value of
\(\epsilon_{\mathrm{ad}}\) therefore corresponds to slower traversal
relative to the population-relaxation scale.

The ratio in Eq.~\eqref{eq:si_adiabaticity_indicator} is not, by itself,
a strict open-system adiabatic condition. A complete condition also
contains derivatives of the instantaneous left and right eigenvectors
and their spectral separation from the remaining Liouvillian modes.
We therefore use \(\epsilon_{\mathrm{ad}}\) only as a time-scale
diagnostic and take the observed orientation-odd plateau as the primary
numerical evidence supporting the Berry-Sinitsyn interpretation.

At an arbitrary finite period, the CW-CCW half-sum and half-difference
are denoted by \(C_n^{\mathrm{even}}(\tau)\) and
\(C_n^{\mathrm{odd}}(\tau)\), respectively. These are the exact
finite-period orientation-even and orientation-odd cumulants and are not
identified automatically with the dynamic and geometric contributions.

We use two numerical conditions for this identification. First, the
orientation-odd cumulant must agree with its long-period reference value
within the tolerance limit,
\begin{equation}
\varepsilon_n^{\mathrm{odd}}(\tau)
=
\frac{
\left|
C_n^{\mathrm{odd}}(\tau)
-
C_n^{\mathrm{odd}}(\tau_{\mathrm{ref}})
\right|
}{
\left|
C_n^{\mathrm{odd}}(\tau_{\mathrm{ref}})
\right|
}
<
\varepsilon_{\mathrm{tol}}.
\label{eq:si_odd_plateau_criterion}
\end{equation}
Here, \(\tau_{\mathrm{ref}}\) is a cycle period on the numerically
converged long-period plateau. Second, the orientation-even cumulant
must display time-extensive scaling,
\begin{equation}
\varepsilon_n^{\mathrm{even}}(\tau)
=
\left|
\frac{
C_n^{\mathrm{even}}(\tau)/\tau
}{
C_n^{\mathrm{even}}(\tau_{\mathrm{ref}})
/
\tau_{\mathrm{ref}}
}
-1
\right|
<
\varepsilon_{\mathrm{tol}}.
\label{eq:si_even_extensive_criterion}
\end{equation}
We use \(\varepsilon_{\mathrm{tol}}=0.01\). When both
Eqs.~\eqref{eq:si_odd_plateau_criterion} and
\eqref{eq:si_even_extensive_criterion} are satisfied, we identify
\begin{equation}
C_n^{\mathrm{odd}}(\tau)
\simeq
C_n^{\mathrm{geo}},
\qquad
C_n^{\mathrm{even}}(\tau)
\simeq
C_n^{\mathrm{dyn}}(\tau).
\label{eq:si_finite_time_identification}
\end{equation}
Outside this validated regime, the exact quantities retain the labels
\(C_n^{\mathrm{odd}}(\tau)\) and
\(C_n^{\mathrm{even}}(\tau)\). The Liouvillian time-scale diagnostic in
Eq.~\eqref{eq:si_adiabaticity_indicator} supports this classification
but does not replace either numerical criterion.

\begin{table*}[!htbp]
\centering
\scriptsize
\setlength{\tabcolsep}{3.5pt}
\renewcommand{\arraystretch}{1.12}
\caption{Merged four-cumulant period scan and Liouvillian time-scale
classification for the uniform traversal of the working full range loop with
periodic limit cycle initialization.
The orientation-even cumulants grow approximately in proportion to the
cycle period, whereas the orientation-odd cumulants approach finite
per-cycle plateaus. The time-scale estimates use
\(\Delta_L^{\min}=0.160~\mathrm{ps^{-1}}\) and
\(\tau_{\mathrm{rel}}^{\max}=6.25~\mathrm{ps}\). The dimensionless
diagnostic is
\(\epsilon_{\mathrm{ad}}=(2\pi/\tau)/\Delta_L^{\min}\).
}
\label{tab:si_merged_period_scan}
\begin{tabular}{
r
c
c
c
c
c
c
p{0.22\textwidth}
}
\toprule
\(\tau\)
&
\shortstack{\(C_1^{\mathrm{even}}\)\\
(meV/cycle)}
&
\shortstack{\(C_1^{\mathrm{odd}}\)\\
(meV/cycle)}
&
\shortstack{\(C_2^{\mathrm{even}}\)\\
(meV\(^{2}\)/cycle)}
&
\shortstack{\(C_2^{\mathrm{odd}}\)\\
(meV\(^{2}\)/cycle)}
&
\shortstack{\(\tau/\tau_{\mathrm{rel}}^{\max}\)}
&
\shortstack{\(\epsilon_{\mathrm{ad}}\)\\
\(=(2\pi/\tau)/\Delta_L^{\min}\)}
&
\multicolumn{1}{c}{\textbf{Interpretation}}
\\
\midrule
\(200~\mathrm{ps}\)
&
4.352614
&
0.208302
&
199.270483
&
10.770683
&
32
&
0.196
&
Shortest-period point in the main scan. The largest finite speed
corrections occur in the orientation-odd sector.
\\
\(500~\mathrm{ps}\)
&
10.877197
&
0.213233
&
497.969033
&
11.093195
&
80
&
0.079
&
Slow traversal. Both orientation-odd cumulants are already close to
their long-period values.
\\
\(1000~\mathrm{ps}\)
&
21.751998
&
0.214179
&
995.777041
&
11.156491
&
160
&
0.039
&
Well inside the slow driving population regime. Only small residual
period dependence remains.
\\
\(2000~\mathrm{ps}\)
&
43.502717
&
0.214446
&
1991.444089
&
11.174425
&
320
&
0.020
&
Deep slow driving. The orientation-odd cumulants are within
approximately \(0.05\%\) of their long period plateau values.
\\
\(5000~\mathrm{ps}\)
&
108.755724
&
0.214525
&
4978.493917
&
11.179798
&
800
&
0.0079
&
Conservative baseline period. The orientation-odd cumulants have
reached the numerically stable plateau.
\\
\bottomrule
\end{tabular}
\end{table*}

The values in Table~\ref{tab:si_merged_period_scan} display the
different period scaling of the two orientation sectors.
The orientation-even first and second cumulants grow approximately
linearly with \(\tau\), consistent with accumulation of the ordinary
dynamic mean and variance over a longer cycle. In contrast, the
orientation-odd cumulants rapidly approach finite per-cycle values.

At \(\tau=200~\mathrm{ps}\), \(C_1^{\mathrm{odd}}\) is approximately
\(2.9\%\) below its long-period value, while
\(C_2^{\mathrm{odd}}\) is approximately \(3.7\%\) below its
long-period value. The \(200~\mathrm{ps}\) results should therefore be
described as finite-time orientation-odd estimators rather than as fully
converged geometric cumulants. At \(\tau=1000~\mathrm{ps}\), the
corresponding deviations have already decreased to approximately
\(0.17\%\) and \(0.21\%\), respectively. By
\(\tau=2000\text{-}5000~\mathrm{ps}\), both orientation-odd
cumulants are effectively plateaued.

The longer-period calculation is summarized separately because only the
orientation-odd cumulants are required for the extended plateau check.

\begin{table}[!htbp]
\centering
\scriptsize
\setlength{\tabcolsep}{6pt}
\renewcommand{\arraystretch}{1.10}
\caption{Extended orientation-odd plateau check. The \(5~\mathrm{ns}\) row is repeated as the baseline point, followed
by the \(10\), \(20\), and \(50~\mathrm{ns}\) calculations. The
displayed cumulants are rounded, while the quoted relative plateau
variation is evaluated using the unrounded numerical values.
}
\label{tab:si_extended_tau_plateau}
\begin{tabular}{rcccc}
\toprule
\(\tau\) (ps)
&
\shortstack{\(C_1^{\mathrm{odd}}\)\\
(meV/cycle)}
&
\shortstack{\(C_2^{\mathrm{odd}}\)\\
(meV\(^{2}\)/cycle)}
&
\(\tau/\tau_{\mathrm{rel}}^{\max}\)
&
\(\epsilon_{\mathrm{ad}}\)
\\
\midrule
5000  & 0.21453 & 11.180 & 800  & 0.0079 \\
10000 & 0.21453 & 11.180 & 1600 & 0.0039 \\
20000 & 0.21454 & 11.180 & 3200 & 0.0020 \\
50000 & 0.21454 & 11.180 & 8000 & 0.00079 \\
\bottomrule
\end{tabular}
\end{table}

\begin{figure}[!htbp]
\centering
\includegraphics[
width=4in,
height=3in,
keepaspectratio
]{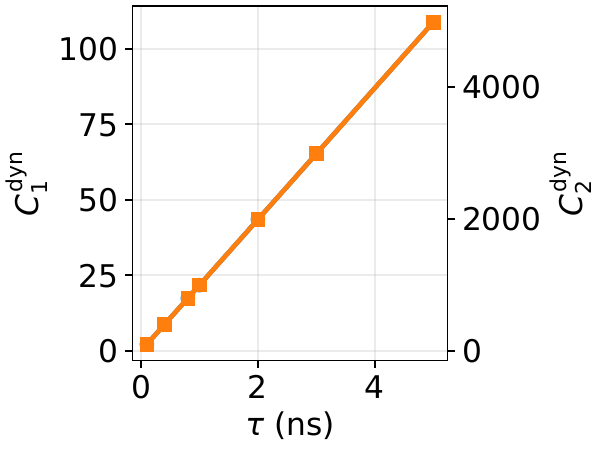}
\caption{(Calculated.)
Additional dense uniform-period scan of the orientation-even dynamic
cumulants over \(0.1\text{-}5~\mathrm{ns}\). This diagnostic includes
intermediate periods beyond the five points listed in
Table~\ref{tab:si_merged_period_scan}. Both dynamic cumulants increase
approximately linearly with the cycle duration because the ordinary
counted-energy mean and variance continue to accumulate throughout the
longer cycle.
}
\label{fig:si_report_period_dynamic}
\end{figure}

\begin{figure}[!htbp]
\centering
\includegraphics[
width=\columnwidth,
keepaspectratio
]{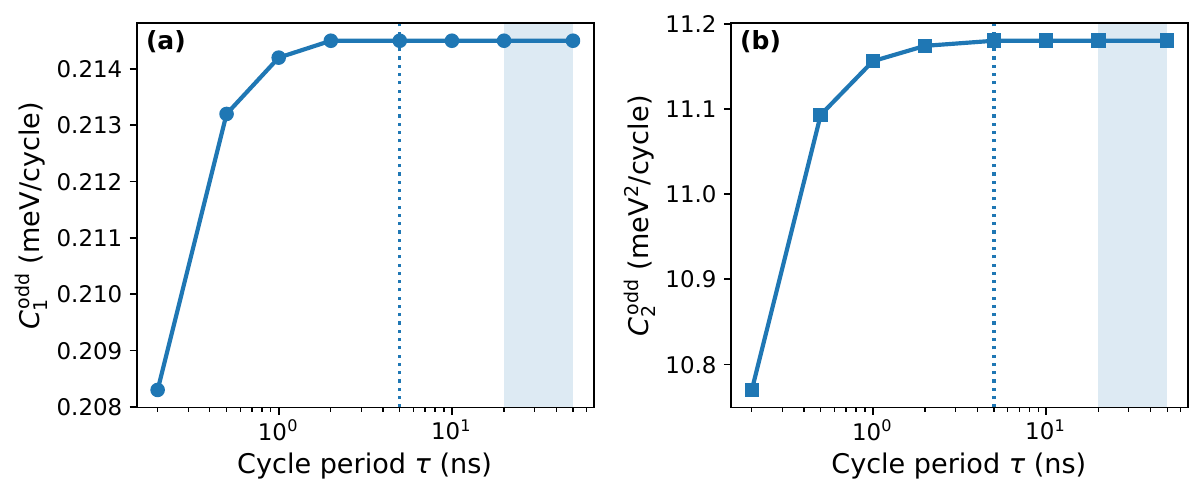}
\caption{Calculated convergence of the orientation-odd cumulants with
cycle period.
Plot (a) First orientation-odd cumulant,and 
(b) Second orientation-odd cumulant.
The dotted vertical line marks the \(5~\mathrm{ns}\) baseline period
used in the main manuscript, while the shaded region marks the
\(20\text{-}50~\mathrm{ns}\) extended long-period window. Both
quantities remain on the same numerical plateau throughout the
extended-period calculation.
}
\label{fig:si_tau_convergence}
\end{figure}

Using the unrounded numerical results, the changes in the two
orientation-odd cumulants between \(5\) and \(50~\mathrm{ns}\) are only
of order \(0.005\%\). The working uniform protocol at
\(\tau=5000~\mathrm{ps}\) therefore lies on a stable long-period
orientation-odd plateau rather than merely near a local finite-time
saturation point.

The combined calculations cover
\(200~\mathrm{ps}\leq\tau\leq50~\mathrm{ns}\), corresponding to
approximately \(32\) to \(8000\) times the longest instantaneous
population-relaxation time per cycle and to
\(0.196\geq\epsilon_{\mathrm{ad}}\geq7.9\times10^{-4}\). This interval
provides a broad validation within the slow-driving sector of the
population Liouvillian. It does not sample the sudden or strongly
nonadiabatic regime near
\(\tau\sim\tau_{\mathrm{rel}}^{\max}\), and no claim is made here about
coherence-resolving dynamics in that regime.

\FloatBarrier

\subsection{Counting-field-step convergence}
\label{si:delta_convergence}
\label{sec:si_counting_field_convergence}

We next verify the counting-field step \(\delta\), which is the second
numerical discretization entering the cumulant evaluation. The
orientation-resolved finite-time cumulants \(C_n^\eta(\tau)\), with
\(\eta\in\{\mathrm{CW},\mathrm{CCW}\}\), are obtained from the
loop-propagated generating function \(G_\eta(s,\tau)\) using the same
fourth-order five-point symmetric finite-difference stencils. For
compactness, \(G_\eta(m\delta)\) denotes
\(G_\eta(m\delta,\tau)\) at the fixed loop period. The first cumulant is
evaluated as
\(C_1^\eta(\tau)\approx
[-G_\eta(2\delta)+8G_\eta(\delta)-8G_\eta(-\delta)
+G_\eta(-2\delta)]/(12\delta)\), whereas the second cumulant is
evaluated as
\(C_2^\eta(\tau)\approx
[-G_\eta(2\delta)+16G_\eta(\delta)-30G_\eta(0)
+16G_\eta(-\delta)-G_\eta(-2\delta)]/(12\delta^2)\).

These stencils are evaluated at fixed time discretization \(N_t=400\)
for the baseline loop. Their fourth-order accuracy makes the result only
weakly dependent on \(\delta\): varying \(\delta\) over the values in
Table~\ref{tab:delta_convergence} leaves both geometric cumulants
unchanged to the displayed precision. The five nodes are
\(s=0,\pm\delta,\pm2\delta\), so the largest tilt excursion satisfies
\(2\delta\Omega_{12}\lesssim0.12\ll1\). The tilted factors therefore
remain in their smooth regime, while the step is sufficiently large to
avoid floating-point cancellation in the second difference. The
selected value \(\delta=10^{-3}~\mathrm{meV^{-1}}\) is consequently a
numerical differentiation choice rather than a physical parameter.

\begin{table}[!htbp]
\centering
\small
\caption{
Counting-field-step convergence for the five-point finite-difference
evaluation of the geometric cumulants of the baseline loop at fixed
time discretization \(N_t=400\).
}
\label{tab:delta_convergence}
\begin{tabular}{ccc}
\toprule
\(\delta\) (\(\mathrm{meV^{-1}}\))
&
\(C_1^{\mathrm{geo}}\) (\(\mathrm{meV/cycle}\))
&
\(C_2^{\mathrm{geo}}\) (\(\mathrm{meV^2/cycle}\))
\\
\midrule
\(5\times10^{-4}\) & 0.2145 & 11.180 \\
\(1\times10^{-3}\) & 0.2145 & 11.180 \\
\(2\times10^{-3}\) & 0.2145 & 11.180 \\
\bottomrule
\end{tabular}
\end{table}

\FloatBarrier
\subsection{Intrinsic SNR versus cycle period}
\label{si:snr_tau}

We next estimate the intrinsic full counting statistics signal to noise
ratio associated with the difference between the CW and CCW
measurements. This quantity characterizes the statistics of the
transferred energy itself and does not include detector noise,
imperfect optical transduction, cycle-switching overhead, or other
experimental technical noise.

Let \(N\) the number of statistically independent cycles recorded
per orientation; thus, \(N=1\) corresponds to one CW cycle and one CCW
cycle, constituting a single CW-CCW pair. let
\(\overline{Q}_{\eta,N}
=N^{-1}\sum_{k=1}^{N}Q_{\eta,k}\), with
\(\eta\in\{\mathrm{CW},\mathrm{CCW}\}\), denote the sample-averaged
transferred energy per cycle. Using
\(C_1^{\mathrm{odd}}
=(C_{1,\mathrm{CW}}-C_{1,\mathrm{CCW}})/2\) and
\(C_2^{\mathrm{even}}
=(C_{2,\mathrm{CW}}+C_{2,\mathrm{CCW}})/2\), the mean CW-CCW
difference is
\(\langle\overline{Q}_{\mathrm{CW},N}
-\overline{Q}_{\mathrm{CCW},N}\rangle
=2C_1^{\mathrm{odd}}\). For statistically independent measurements in
the two orientations, its variance is
\(\operatorname{Var}(
\overline{Q}_{\mathrm{CW},N}
-\overline{Q}_{\mathrm{CCW},N})
=2C_2^{\mathrm{even}}/N\). The corresponding intrinsic
signal to noise ratio is therefore
\begin{equation}
\operatorname{SNR}_{N}(\tau)
=
\sqrt{2N}\,
\frac{
\left|C_1^{\mathrm{odd}}(\tau)\right|
}{
\sqrt{C_2^{\mathrm{even}}(\tau)}
}.
\label{eq:si_snr_tau}
\end{equation}
Here,  the factor \(\sqrt{2N}\) follows from the
\(2C_1^{\mathrm{odd}}\) mean difference and the
\(2C_2^{\mathrm{even}}/N\) variance of that difference.

At an arbitrary finite period, the quantities entering
Eq.~\eqref{eq:si_snr_tau} retain the labels
\(C_1^{\mathrm{odd}}(\tau)\) and
\(C_2^{\mathrm{even}}(\tau)\). On the numerically validated
slow-driving plateau, they are identified with
\(C_1^{\mathrm{geo}}\) and \(C_2^{\mathrm{dyn}}\), respectively. For a
fixed number \(N\) of cycles in each orientation,
\(C_1^{\mathrm{odd}}\) approaches a finite per-cycle plateau, while
\(C_2^{\mathrm{even}}\) grows approximately linearly with \(\tau\).
Equation~\eqref{eq:si_snr_tau} therefore gives
\(\operatorname{SNR}_{N}\propto\tau^{-1/2}\) within the slow-driving
regime.

Equation~\eqref{eq:si_snr_tau} compares protocols at a fixed number of
recorded cycles. Protocols with different cycle periods are instead
compared at the same total acquisition time \(T\). Within the intrinsic
FCS estimate, \(T=2N\tau\) for equal numbers of CW and CCW cycles and
includes only the durations of the recorded driving cycles.
Initialization, orientation switching, detector dead time, readout, and
other experimental overheads are excluded.  Eliminating \(N\) in favour
of \(T\) in Eq.~\eqref{eq:si_snr_tau} shows that the signal to noise
ratio at fixed \(T\) is \(\sqrt{T}\) multiplied by the
protocol-dependent factor \(\mathcal{D}_{T}(\tau)\). This factor is
the time normalized detectability, that is, the signal to noise ratio per square root of total acquisition time,
\begin{equation}
\mathcal{D}_{T}(\tau)
=
\frac{
\left|C_1^{\mathrm{odd}}(\tau)\right|
}{
\sqrt{\tau C_2^{\mathrm{even}}(\tau)}
},
\label{eq:si_detectability_measure}
\end{equation}

Since, \(C_1^{\mathrm{geo}}\) and \(C_2^{\mathrm{dyn}}\) grows linearly with \(\tau\),
\(\mathcal{D}_{T}\) scales as \(\tau^{-1}\). Reducing the cycle period
therefore provides both a smaller per cycle dynamic variance and a
larger number of completed loops within the same acquisition time.

Inverting Eq.~\eqref{eq:si_snr_tau}, the required number of
statistically independent cycles per orientation for a prescribed
target \(\operatorname{SNR}_{\star}\) is
\(N_{\star}(\tau)
=\operatorname{SNR}_{\star}^{\,2}C_2^{\mathrm{even}}(\tau)/
\{2[C_1^{\mathrm{odd}}(\tau)]^2\}\). The physical cycle count is
\(\lceil N_{\star}\rceil\). For the representative target
\(\operatorname{SNR}_{\star}=10\), we denote the corresponding estimate
by \(N_{10}\). The total number of recorded cycles is \(2N_{10}\), and
the associated acquisition time within the intrinsic FCS estimate is
\(2N_{10}\tau\). The values reported below are rounded to three
significant figures.

\begin{table}[!htbp]
\centering
\scriptsize
\setlength{\tabcolsep}{5pt}
\renewcommand{\arraystretch}{1.10}
\caption{Intrinsic detectability budget for the uniform traversal of the
baseline loop. The entries are evaluated using the finite-period
quantities \(C_1^{\mathrm{odd}}(\tau)\) and
\(C_2^{\mathrm{even}}(\tau)\). Here, \(\operatorname{SNR}_{N=1}\)
corresponds to one CW cycle and one CCW cycle, \(N_{10}\) is the
estimated number of statistically independent cycles per orientation
required for a target intrinsic \(\operatorname{SNR}=10\), and the last
column gives the corresponding acquisition time \(2N_{10}\tau\). The
estimates exclude cycle correlations, detector inefficiency,
orientation switching, optical-transduction uncertainty, readout time,
and other experimental overheads.
}
\label{tab:si_merged_snr_budget}
\begin{tabular}{rccc}
\toprule
\(\tau\) (ps)
&
\(\operatorname{SNR}_{N=1}\)
&
\(N_{10}\) (cycles/orientation)
&
Acquisition time
\\
\midrule
200
&
0.02087
&
\(2.30\times10^{5}\)
&
\(91.9~\mu\mathrm{s}\)
\\
500
&
0.01351
&
\(5.48\times10^{5}\)
&
\(0.548~\mathrm{ms}\)
\\
1000
&
0.00960
&
\(1.09\times10^{6}\)
&
\(2.17~\mathrm{ms}\)
\\
2000
&
0.00680
&
\(2.17\times10^{6}\)
&
\(8.66~\mathrm{ms}\)
\\
5000
&
0.00430
&
\(5.41\times10^{6}\)
&
\(54.1~\mathrm{ms}\)
\\
\bottomrule
\end{tabular}
\end{table}

The values in Table~\ref{tab:si_merged_snr_budget} quantify the
intrinsic statistical advantage of the shorter periods. At
\(\tau=5000~\mathrm{ps}\), approximately
\(5.41\times10^{6}\) independent cycles are required in each orientation
to reach \(\operatorname{SNR}=10\), corresponding to an acquisition
time of approximately \(54.1~\mathrm{ms}\) within the intrinsic FCS
estimate. Reducing the
period to \(1000~\mathrm{ps}\) lowers the estimate to approximately
\(1.09\times10^{6}\) cycles per orientation and
\(2.17~\mathrm{ms}\). At \(500~\mathrm{ps}\), the corresponding
estimates are \(5.48\times10^{5}\) cycles per orientation and
\(0.548~\mathrm{ms}\).

The shortest tested period, \(\tau=200~\mathrm{ps}\), gives the smallest
cycle-count and acquisition-time requirements because it combines
a smaller orientation-even variance with a larger number of completed
loops per unit time. Its orientation-odd first cumulant, however,
remains approximately \(2.9\%\) below the long-period plateau. It should
therefore be interpreted as a statistically favourable finite-time
orientation-odd estimator rather than as the fully converged geometric
cumulant. The \(500\text{-}1000~\mathrm{ps}\) range provides a more
balanced compromise between retention of the geometric signal and
intrinsic statistical detectability.

The values in Table~\ref{tab:si_merged_snr_budget} are intrinsic FCS
estimates based on statistically independent cycles. Actual experimental
acquisition times can be longer because of detector inefficiency,
detector dead time, orientation switching, background subtraction,
inter-cycle correlations, readout time, and uncertainty in the optical
transduction of the counted phonon signal.

\subsection{Optical proxy cumulants for experimental detection}
\label{sec:si_optical_proxy}

The cumulants analyzed in the main text and in the preceding subsection
quantify the energy transferred to the phonon bath through the retained
\(\Omega_{12}(t)\) transition channel. Direct cycle-resolved measurement
of the energy transferred through a microscopic phonon channel may,
however, be difficult experimentally.

As a general experimental possibility, a phonon-assisted relaxation
process may be monitored through a spectrally resolved optical feature
associated with the same exciton-phonon dynamics. Phonon replicas of
dark excitons and related phonon-coupled emission features have been
observed in monolayer and bilayer WSe$_2$, showing that phonon-assisted
excitonic processes can produce spectrally resolvable optical
signatures~\cite{Li2019DarkExcitonReplica,Liu2019ChiralPhononReplica,
Ripin2023TunablePhononic}. These observations motivate the general
detection procedure outlined below, but they do not establish a
universal quantitative relation between phonon energy transfer and an
optical signal.

We therefore introduce a general optical proxy that may be used in a
driven system where a phonon-assisted optical feature can be spectrally
isolated and a control loop can be implemented in both CW and CCW
orientations. The proxy is not a microscopic phonon-counting equation
and is not evaluated numerically for the present model. Instead, it
defines a possible experimental procedure for constructing an optical
statistic that can be compared between the two loop orientations.

For cycle \(k\) and loop orientation
\(\eta\in\{\mathrm{CW},\mathrm{CCW}\}\), we define the detected optical
energy within a selected spectral window as
\begin{equation}
Q_\eta^{(k)}
=
\sum_\nu E_\nu N_{\nu,\eta}^{(k)},
\label{eq:optical_proxy_energy}
\end{equation}
where \(E_\nu\) is a calibrated photon energy associated with spectral
bin \(\nu\), and \(N_{\nu,\eta}^{(k)}\) is the background-corrected
number of detected photons in that bin during cycle \(k\) for
orientation \(\eta\). Thus, \(Q_\eta^{(k)}\) is an experimentally
constructed optical-energy statistic rather than the microscopic phonon
energy transferred during the cycle.

A spectral window used in this construction would first have to be
identified experimentally around a phonon-assisted optical feature
associated with a selected phonon-mediated relaxation process in the
system under study. The relevant transition energy and spectral position
would have to be established independently for that system. The proposed
proxy does not assume a universal equality between a phonon energy and a
detected photon energy. The photon yield, spectral weight, linewidth,
branching ratio, and detection efficiency of the selected feature would
also have to be determined experimentally.

The first and second optical-proxy cumulants are then obtained by
averaging over repeated cycles,
\begin{equation}
C_1^{\eta,\mathrm{exp}}
=
\left\langle Q_\eta^{(k)}\right\rangle_k,
\label{eq:optical_proxy_first}
\end{equation}
and
\begin{equation}
C_2^{\eta,\mathrm{exp}}
=
\left\langle
\left(Q_\eta^{(k)}\right)^2
\right\rangle_k
-
\left\langle Q_\eta^{(k)}\right\rangle_k^2,
\label{eq:optical_proxy_second}
\end{equation}
respectively. The loop-reversal half-difference defines the
orientation-odd optical-proxy cumulant,
\begin{equation}
C_n^{\mathrm{odd,exp}}
=
\frac{
C_n^{\mathrm{CW,exp}}
-
C_n^{\mathrm{CCW,exp}}
}{2},
\qquad n=1,2,
\label{eq:optical_proxy_odd}
\end{equation}
while the corresponding half-sum defines the orientation-even
optical-proxy cumulant,
\begin{equation}
C_n^{\mathrm{even,exp}}
=
\frac{
C_n^{\mathrm{CW,exp}}
+
C_n^{\mathrm{CCW,exp}}
}{2},
\qquad n=1,2.
\label{eq:optical_proxy_even}
\end{equation}

After reversal-dependent experimental systematics have been excluded,
\(C_n^{\mathrm{odd,exp}}\) may serve as a candidate experimental
estimator of a geometric contribution to the optical statistic. Such an
interpretation would require the same type of finite-period validation
used for the phonon cumulants; namely, an orientation-odd optical
cumulant should approach a period-independent plateau, whereas an
orientation-even optical cumulant should display the corresponding
time-extensive scaling.

The proposed method does not supply a conversion factor between phonon
energy cumulants and measured optical-proxy cumulants. Any such
conversion would depend on the material, the selected phonon-assisted
optical feature, and the experimental detection setup. We therefore do
not assign a numerical optical-transduction factor or predict an
absolute optical signal. Establishing such a relation would require
calibration of the spectral branching ratio, collection efficiency,
detector efficiency, instrumental response, background contribution, and
the fraction of relaxation events contributing to the selected spectral
window. Detector noise would also have to be characterized, especially
when extracting a second cumulant.

A possible readout would rely on synchronized accumulation over repeated
cycles and a CW-CCW half-difference to suppress a larger
orientation-even optical background. One possible implementation would
proceed in three stages. First, reflectance, photoluminescence,
photoluminescence-excitation, or related spectroscopy could be used to
identify a phonon-assisted optical feature associated with a selected
relaxation process. Second, time-resolved photoluminescence or pump-probe
spectroscopy could be used to test this association and constrain a
relevant relaxation timescale. Third, an optical signal could be
recorded over repeated phase-locked CW and CCW driving cycles and
separated into cycle-resolved spectral bins. The first and second
orientation-odd proxy cumulants could then be extracted using
Eq.~\eqref{eq:optical_proxy_odd}.

\FloatBarrier

\FloatBarrier

\clearpage
\section{Baseline physical robustness and model dependence}
\label{sec:si_baseline_robustness}

The phonon bath temperature, phonon window, competing channel,
and branch choice checks are grouped here because they test distinct
physical assumptions underlying the same uniform
\(\tau=5~\mathrm{ns}\) baseline calculation.

\subsection{Phonon bath temperature}
\label{sec:si_cold_temperature}

\begin{figure}[!htbp]
\centering
\includegraphics[
width=0.92\textwidth,
height=0.40\textheight,
keepaspectratio
]{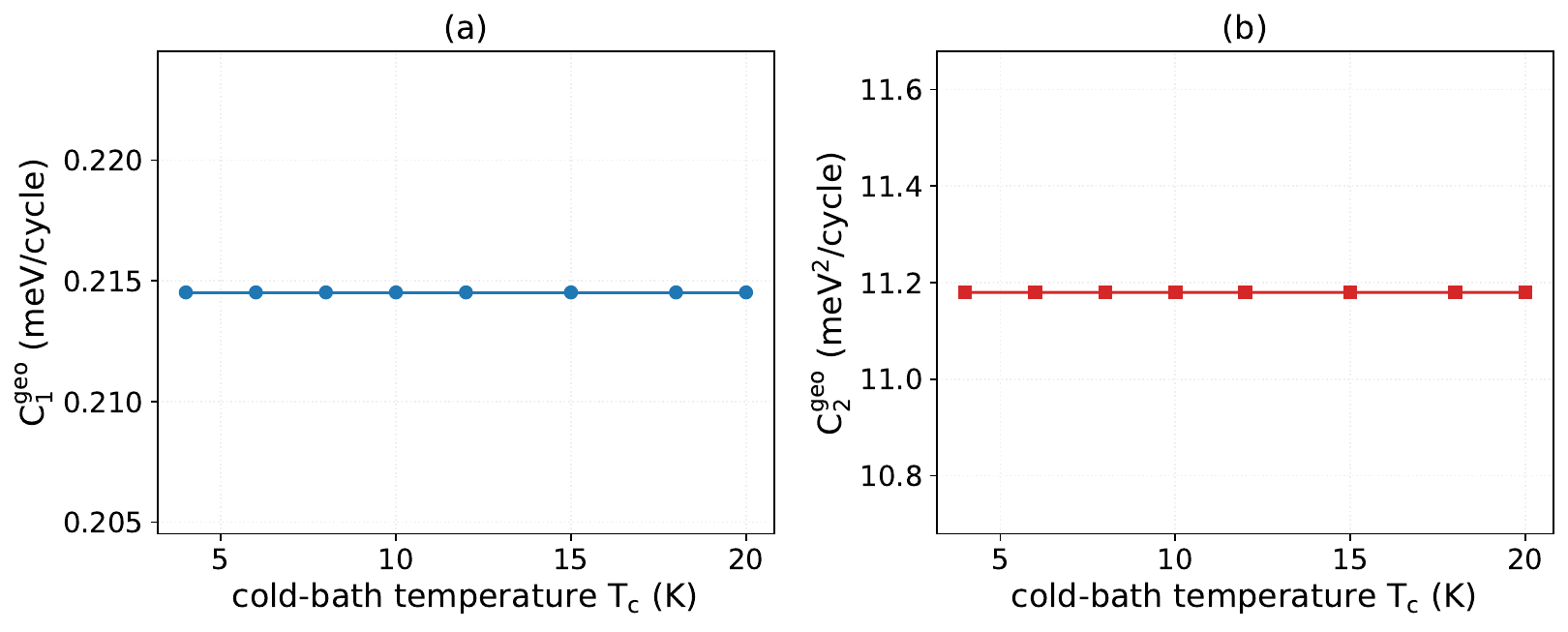}
\caption{
Phonon bath temperature dependence of the geometric cumulants. Panels~(a) and (b) correspond to the geometric first and second
cumulants, respectively, for the uniform baseline loop at
\(\tau=5~\mathrm{ns}\). No variation is resolved at the displayed
numerical precision over \(T_c=4\text{-}20~\mathrm{K}\).
}
\label{fig:si_temperature}
\end{figure}

The Bose occupation of a phonon bath phonon mode of energy \(\Omega\) is
\(
n_c(\Omega)
=
\frac{1}
{\exp[\Omega/(k_BT_c)]-1}.
\)
Here, \(k_B\) is the Boltzmann constant and \(T_c\) is the
phonon bath temperature. Along the uniform \(5~\mathrm{ns}\) baseline loop, the counted
dressed-state gap spans
\(
\Omega_{12}=31.03\text{-}58.21~\mathrm{meV}.
\)
Over the tested temperature interval
\(T_c=4\text{-}20~\mathrm{K}\), the corresponding thermal scale is
\(
k_BT_c
=
0.345\text{-}1.72~\mathrm{meV}.
\)
The largest Bose occupation occurs at the smallest counted gap and the
highest temperature, \(n_c^{\max}
=
n_c(31.03~\mathrm{meV},20~\mathrm{K})
\simeq
1.5\times10^{-8}.
\)
Thus,
\(
n_c[\Omega_{12}(t)]\ll1
\)
throughout the loop, and upward phonon bath absorption is negligible. As summarized in Fig.~\ref{fig:si_temperature}, both geometric
cumulants remain unchanged at the displayed numerical precision over
the investigated temperature range at \(\tau=5~\mathrm{ns}\).
\(
C_1^{\mathrm{geo}}
\simeq
0.2145~\mathrm{meV/cycle},
\qquad
C_2^{\mathrm{geo}}
\simeq
11.180~\mathrm{meV^2/cycle}.
\) This weak temperature dependence is consistent with a response
governed primarily by the loop-driven modulation of the dressed-state
populations and the counted phonon emission rate rather than by
thermal absorption from the phonon bath.

\FloatBarrier

\subsection{Phonon window and competing relaxation channel}
\label{sec:si_cutoff_competing}

After diagonalizing the experimentally anchored Hamiltonian input
\(H_T[F_z^{(0)},V_G(t)]\) along the gate-voltage trajectory of the
uniform \(5~\mathrm{ns}\) baseline loop,
the instantaneous dressed-state gaps are defined by
\(
\Omega_{ij}(t)
=
\epsilon_j(t)-\epsilon_i(t),
\qquad
j>i.
\)
Their numerical ranges are
\(
\Omega_{12}(t)
\in
[31.03,58.21]~\mathrm{meV},
\)
\(
\Omega_{01}(t)
\in
[67.64,82.82]~\mathrm{meV},
\)
and
\(
\Omega_{02}(t)
\in
[104.99,141.04]~\mathrm{meV}.
\)
Only \(\Omega_{12}(t)\) remains completely inside the phonon window
\(
\Omega_{\mathrm{cut}}=60~\mathrm{meV}.
\)
The counting field is therefore attached to the counted phonon
relaxation
\(
\lvert\psi_2(t)\rangle
\rightarrow
\lvert\psi_1(t)\rangle.
\). The
\(\lvert\psi_1(t)\rangle\rightarrow\lvert\psi_0(t)\rangle\) and
\(\lvert\psi_2(t)\rangle\rightarrow\lvert\psi_0(t)\rangle\)
relaxations associated with \(\Omega_{01}(t)\) and
\(\Omega_{02}(t)\), respectively, are therefore not included as
counted phonon channels.

No quantitative two-phonon rate is propagated or inferred in the
present model. A microscopic second-order calculation would require
branch- and momentum-resolved exciton-phonon matrix elements, virtual
intermediate-state energy denominators and lifetimes, and a properly
normalized two-phonon phase-space integral. Because these quantities are
not available for the present WSe$_2$/WS$_2$/hBN structure, the
discussion of the higher-gap pathways is restricted to their kinematic
energy requirements rather than to a quantitative rate estimate
\cite{Paradisanos2021,Brem2020,Lee2020TwoPhonon}.

For an illustrative equal-energy partition satisfying
\(
\hbar\omega_{\mathbf q}
+
\hbar\omega_{\mathbf q'}
=
\Omega_{ij}, where
\qquad
\hbar\omega_{\mathbf q}
=
\hbar\omega_{\mathbf q'}
=
\frac{\Omega_{ij}}{2},
\)
relaxation across \(\Omega_{01}\) would require individual phonon
energies of approximately
\(34\text{-}41~\mathrm{meV}\), whereas relaxation across
\(\Omega_{02}\) would require approximately
\(52\text{-}71~\mathrm{meV}\) per phonon. These estimates describe only
the energy combinations required for such processes. They neither
establish that the corresponding pathways are absent nor imply that
their rates are negligibly small. The counted \(\Omega_{12}\)phonon channel should therefore
be interpreted as the leading phonon relaxation pathway retained in
the model, while higher-order relaxation remains an unresolved
microscopic correction. Under the model criterion that a transition is included at a given
point on the loop when
\(\Omega_{ij}(t)\leq\Omega_{\mathrm{cut}}\),
the set of counted phonon transitions remains unchanged throughout
the interval
\(
58.21~\mathrm{meV}
\leq
\Omega_{\mathrm{cut}}
<
67.64~\mathrm{meV}.
\)
A cutoff below \(58.21~\mathrm{meV}\) would exclude part of the
\(\Omega_{12}\) trajectory, whereas a cutoff at or above
\(67.64~\mathrm{meV}\) would begin to admit part of the
\(\Omega_{01}\) branch. This interval is therefore a
model-selection-stability window and should not be interpreted as a
first-principles boundary of the phonon density of states.

\begin{figure}[!htbp]
\centering
\safeincludegraphics[
width=0.60\textwidth,
height=0.33\textheight,
keepaspectratio
]{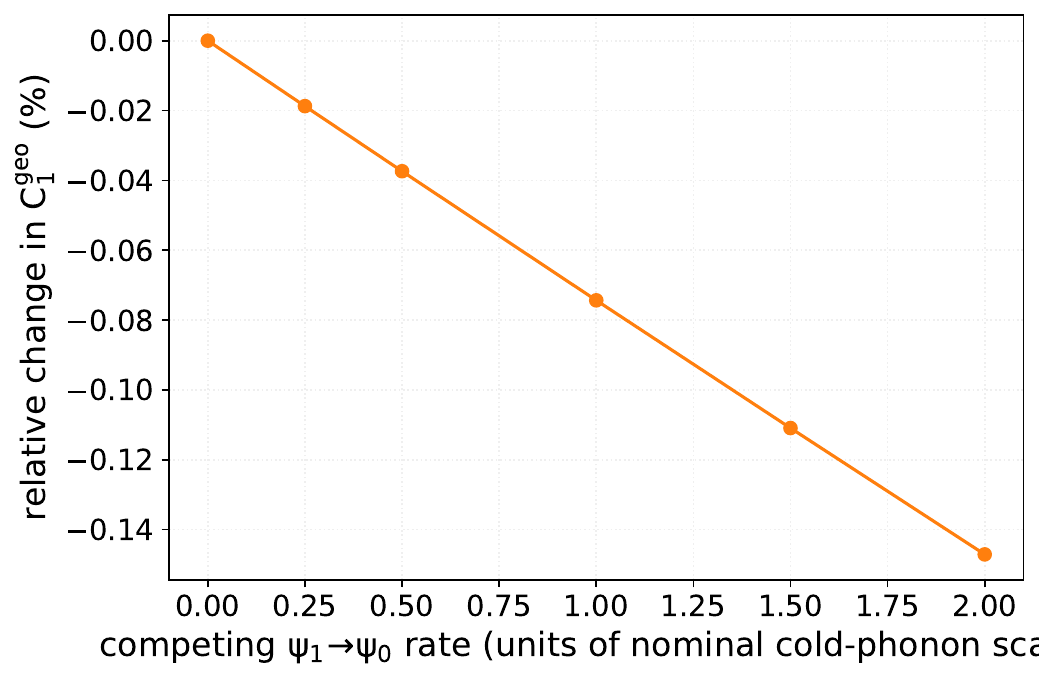}
\caption{Sensitivity to a competing uncounted relaxation pathway.
Relative change of the counted \(\Omega_{12}\) geometric first
cumulant after adding an effective uncounted
\(\lvert\psi_1\rangle\rightarrow\lvert\psi_0\rangle\) relaxation
channel. The horizontal axis is
\(\gamma_{10}^{\mathrm{add}}/\gamma_{10}^{\mathrm{ref}}\), and the
relative change is measured from the result at
\(\gamma_{10}^{\mathrm{add}}=0\). The added pathway redistributes the
dressed-state populations but does not carry the counting field.}

\label{fig:si_competing_channel}
\end{figure}

To test the sensitivity of the counted response to relaxation outside
the retained single-channel model, we add an effective uncounted
relaxation pathway
\(
\lvert\psi_1(t)\rangle
\rightarrow
\lvert\psi_0(t)\rangle
\)
with rate
\(\gamma_{10}^{\mathrm{add}}\).
The horizontal-axis variable in Fig.~\ref{fig:si_competing_channel} is
the dimensionless ratio
\(
\frac{\gamma_{10}^{\mathrm{add}}}
{\gamma_{10}^{\mathrm{ref}}},
\)where \(\gamma_{10}^{\mathrm{ref}}\) is the nominal phonon
rate used only to normalize the added channel in this robustness test. It does not represent a separately propagated baseline
\(\lvert\psi_1\rangle\rightarrow\lvert\psi_0\rangle\) transition.

The added pathway does not modify the Hamiltonian gap
\(\Omega_{12}(t)\) and does not carry the counting field. It influences
\(C_1^{\mathrm{geo}}\) only indirectly by redistributing the
instantaneous dressed-state populations during the driven cycle.
Accordingly, this construction is a sensitivity test rather than a
microscopic calculation of a specific omitted multiphonon process.

The relative change plotted in
Fig.~\ref{fig:si_competing_channel} is evaluated with respect to the
calculation without the added channel,
\(\delta_{\mathrm{add}} C_1^{\mathrm{geo}}
=
100
\frac{
C_1^{\mathrm{geo}}(\gamma_{10}^{\mathrm{add}})
-
C_1^{\mathrm{geo}}(0)
}{
C_1^{\mathrm{geo}}(0)
}.
\)
The relative change remains small over the tested interval. Even at
\(
\gamma_{10}^{\mathrm{add}}
=
2\gamma_{10}^{\mathrm{ref}},
\)
the counted geometric first cumulant changes by only approximately
\(-0.15\%\).

Thus, the counted-channel selection is unchanged for any cutoff
inside the stated stability window, while the calculated
\(C_1^{\mathrm{geo}}\) remains stable against the tested effective
uncounted relaxation pathway.

\FloatBarrier

\subsection{Branch choice sensitivity}
\label{sec:si_branch_choice}

Across the doping transitions
\(\nu=0\!\rightarrow\!1\),
at approximately \(V_G=2\text{-}3~\mathrm{V}\), and
\(\nu=1\!\rightarrow\!2\),
at approximately \(V_G=4\text{-}5~\mathrm{V}\), the experimental fit
resolves two \(X_1\)-like resonances simultaneously. These branches are
listed as
\(\varepsilon_1^{\mathrm{dom}}(V_G)\) and
\(\varepsilon_1^{\mathrm{sub}}(V_G)\) in
Table~\ref{tab:params}.

\begin{figure}[!htbp]
\centering
\includegraphics[
width=0.75\textwidth,
keepaspectratio
]{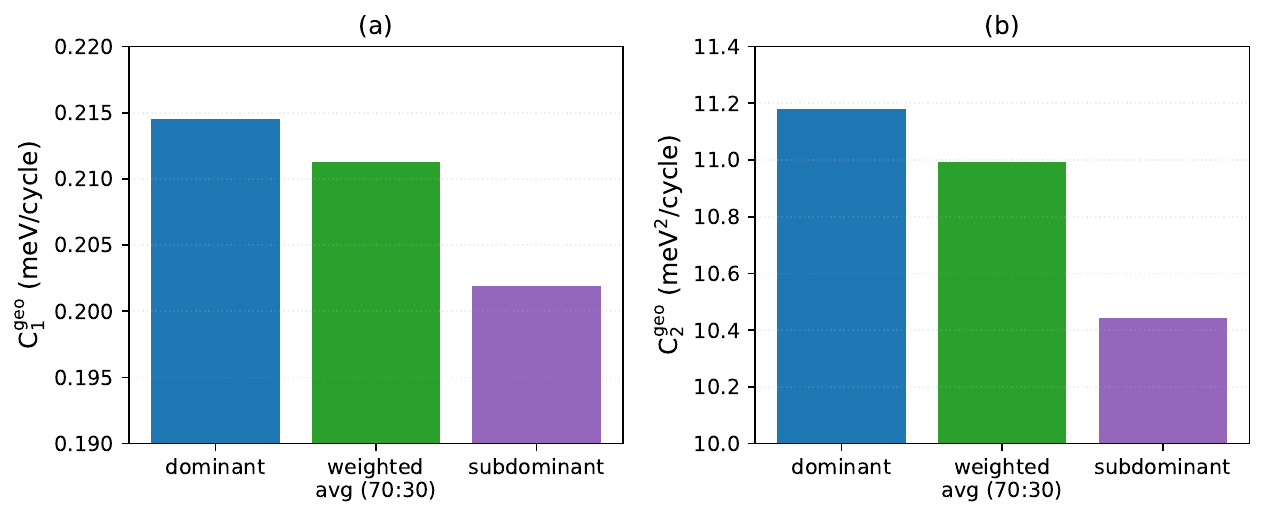}
\caption{Sensitivity to the \(E_1(V_G)\) branch prescription. Panels~(a) and (b) correspond to the geometric first and second
cumulants, respectively, obtained using the dominant,
\(70{:}30\) weighted-average, and subdominant prescriptions at
\(\tau=5~\mathrm{ns}\). All values remain positive, and the dominant
and weighted-average results agree within a few percent. The vertical
axes are expanded to resolve the relative differences.}

\label{fig:si_branch_choice}
\end{figure}

The experimentally anchored Hamiltonian requires one input
\(E_1(V_G)\) at each gate voltage. The dominant-branch prescription is
\(
E_1^{\mathrm{dom}}(V_G)
=
\varepsilon_1^{\mathrm{dom}}(V_G).
\)
We compare this prescription with an illustrative
\(70{:}30\) weighted average,
\(
E_1^{70{:}30}(V_G)
=
0.7\,
\varepsilon_1^{\mathrm{dom}}(V_G)
+
0.3\,
\varepsilon_1^{\mathrm{sub}}(V_G),
\)
and with the subdominant-branch prescription
\(
E_1^{\mathrm{sub}}(V_G)
=
\varepsilon_1^{\mathrm{sub}}(V_G).
\)
These alternative prescriptions are applied in the gate-voltage
regions where both branches are resolved. Outside the two-branch regions, the single experimentally resolved input
is retained unchanged for all three prescriptions. The relative first-cumulant change is defined with respect to the
dominant branch result as
\(
\delta C_1^{\mathrm{geo}}
=
100
\frac{
C_1^{\mathrm{geo}}
-
C_{1,\mathrm{dom}}^{\mathrm{geo}}
}{
C_{1,\mathrm{dom}}^{\mathrm{geo}}
}.
\)
Table~\ref{tab:branch} lists the resulting geometric cumulants,
and Fig.~\ref{fig:si_branch_choice} compares them graphically, at the
baseline period \(\tau=5~\mathrm{ns}\).

\begin{table}[!htbp]
\centering
\small
\setlength{\tabcolsep}{5pt}
\renewcommand{\arraystretch}{1.08}

\caption{Sensitivity of the geometric cumulants to the prescription used
for the Hamiltonian input \(E_1(V_G)\). The percentage change is measured relative to the
dominant-branch prescription. All values are evaluated at the baseline
period \(\tau=5~\mathrm{ns}\).
}
\label{tab:branch}

\begin{tabular}{lccc}
\toprule
\(E_1(V_G)\) prescription
&
\shortstack{\(C_1^{\mathrm{geo}}\)\\
(meV/cycle)}
&
\shortstack{\(\delta C_1^{\mathrm{geo}}\)\\
(\%)}
&
\shortstack{\(C_2^{\mathrm{geo}}\)\\
(meV\(^{2}\)/cycle)}
\\
\midrule

Dominant branch
&
0.2145
&
\(0.0\)
&
11.18
\\

Weighted average (\(70{:}30\))
&
0.2112
&
\(-1.5\)
&
10.99
\\

Subdominant branch
&
0.2019
&
\(-5.9\)
&
10.44
\\

\bottomrule
\end{tabular}
\end{table}

The \(70{:}30\) weighted prescription changes
\(C_1^{\mathrm{geo}}\) by only \(-1.5\%\).
An additional weighting scan from \(80{:}20\) to \(50{:}50\) gives
changes below approximately \(3\%\).
Using the subdominant branch alone produces the larger reduction of
\(-5.9\%\), while preserving the positive sign of the first geometric
cumulant. The second geometric cumulant follows the same ordering.

The dominant-branch and illustrative weighted-average
prescriptions therefore differ only slightly, whereas the
subdominant-only test produces a moderate quantitative change. In all
three cases, the sign of \(C_1^{\mathrm{geo}}\) and the qualitative
geometric-pumping conclusion are unchanged. These results support the
use of the dominant branch as the Hamiltonian input without
claiming complete quantitative independence from the branch-choice
prescription.

\FloatBarrier
\clearpage



\section{Uniform protocol signal enhancement and detectability diagnostics}
\label{sec:si_enhanced_signal}

The one-parameter sensitivity scans identify which physical directions
increase the geometric cumulants. This section combines the favourable parameter variations cumulatively for
the uniform (5~ns) baseline and evaluates their performance
using the absolute geometric signal, the geometric-to dynamic mean
contrast, and the variance-limited intrinsic SNR.
 The geometric first cumulant
\(C_1^{\mathrm{geo}}\)
gives the orientation-odd phonon energy transferred per cycle and
therefore measures the absolute size of the desired signal. For the cumulative comparison below, the multiplicative signal gain
is measured relative to the uniform
\(\tau=5~\mathrm{ns}\) baseline as
\(\lvert C_1^{\mathrm{geo}}\rvert/
 \lvert C_1^{\mathrm{geo,base}}\rvert\). The dynamic first cumulant
\(C_1^{\mathrm{dyn}}\)
gives the orientation-even mean energy background accumulated during the
same cycle. The ratio
\(\NEW{\lvert C_1^{\mathrm{geo}}\rvert/C_1^{\mathrm{dyn}}}\)
therefore measures the relative mean-background contrast. \NEW{This contrast is useful for assessing the required CW-CCW
common-mode rejection, but it is not a statistical SNR. The intrinsic
statistical noise is set by the dynamic second cumulant
\(C_2^{\mathrm{dyn}}\), as derived in
Sec.~\ref{si:snr_tau}.}

We consider a cumulative enhancement sequence while keeping the
experimentally supported Hamiltonian range fixed.
Starting from the uniform \(5~\mathrm{ns}\) baseline, the pump-control
range is widened to
\(I_{\mathrm{pump}}=0.1\text{-}4.9\), the optical-injection prefactor
\(\Gamma_{\mathrm{pump}}^{0}\) is doubled, both phonon prefactors
are doubled, and both radiative-loss prefactors are reduced by one half. \NEW{Only the widening of \(I_{\mathrm{pump}}(t)\) is a direct change
of the externally imposed pump protocol. Doubling
\(\Gamma_{\mathrm{pump}}^{0}\) represents a conditional increase of the
effective optical-injection rate scale associated with the coupling to
the external photon terminal, whereas the phonon and radiative
changes represent conditional modifications of the corresponding
environmental rate scales. None of these three rate-prefactor changes
modifies the experimentally anchored exciton Hamiltonian \(H_T\).} In the \(\gamma_c\times2\) step, both
\(\gamma_c^{\mathrm{intra}}\) and
\(\gamma_c^{\mathrm{inter}}\) are doubled. In the
\(\gamma_{\mathrm{rad}}\times0.5\) step, both
\(\gamma_{\mathrm{rad}}^{\mathrm{intra}}\) and
\(\gamma_{\mathrm{rad}}^{\mathrm{inter}}\) are reduced by one half. \NEW{The pump-range expansion and the change of
\(\Gamma_{\mathrm{pump}}^{0}\) both increase population injection into
the dressed manifold, but they are conceptually distinct: the former
changes the imposed control waveform, while the latter changes the
effective system-optical-environment injection scale. Increasing
\(\gamma_c\) strengthens phonon-assisted relaxation through the counted
channel, whereas reducing \(\gamma_{\mathrm{rad}}\) suppresses
radiative loss and leaves more excited-state population available for
phonon-mediated transfer.}

The gate trajectory is not extended beyond
\(0.5\leq V_G(t)\leq6.5~\mathrm{V}\), because this would require
extrapolation outside the digitized \NEW{experimentally supported}
Hamiltonian surface used to construct \(H_T\).
Increasing \(F_z\) beyond the selected operating point is also not
considered within the retained channel selection. At
\(F_z=+0.20~\mathrm{V\,nm^{-1}}\), the maximum counted gap reaches
\(\Omega_{12}^{\max}=64.6~\mathrm{meV}\), above the \NEW{\(60~\mathrm{meV}\) phonon window}.
The operating-field and counted-channel selection are detailed in
Sec.~\ref{sec:si_field_counted_selection}.

Lower fields reduce the geometric signal, giving
\(\NEW{C_1^{\mathrm{geo}}}
=0.147~\mathrm{meV/cycle}\) at
\(F_z=+0.10~\mathrm{V\,nm^{-1}}\) and
\(\NEW{C_1^{\mathrm{geo}}}
=0.186~\mathrm{meV/cycle}\) at
\(F_z=+0.14~\mathrm{V\,nm^{-1}}\).
The phase lag \(\Delta\phi=\pi/2\) also gives the largest response among
the tested loop orientations:
\(\Delta\phi=\pi/4\) gives \(0.64\times\) the baseline geometric
response, while \(\Delta\phi=3\pi/4\) gives \(0.71\times\).

\NEW{Table~\ref{tab:si_enhanced_stack} summarizes the cumulative
uniform-\(5~\mathrm{ns}\) sequence. The percentage contrast is reported
as \(100\,\lvert C_1^{\mathrm{geo}}\rvert/C_1^{\mathrm{dyn}}\).}

\begin{table}[!htbp]
\centering
\scriptsize
\setlength{\tabcolsep}{5pt}
\renewcommand{\arraystretch}{1.15}
\caption{Cumulative enhancement at the uniform
\(\tau=5~\mathrm{ns}\) baseline.
Each row adds one favourable change to the preceding configuration.
The \(\gamma_c\times2\) step doubles both phonon prefactors,
whereas the \(\gamma_{\mathrm{rad}}\times0.5\) step reduces both
radiative-loss prefactors by one half.
\NEW{The gain is measured relative to
\(\lvert C_1^{\mathrm{geo,base}}\rvert=0.2145~
\mathrm{meV/cycle}\), and the final column is the orientation-odd
mean-signal magnitude relative to the orientation-even dynamic mean.}}
\label{tab:si_enhanced_stack}
\begin{tabular}{p{2.0cm} c c c c}
\toprule
\textbf{Configuration} &
\(\boldsymbol{C_1^{\mathrm{geo}}}\) &
\textbf{Gain} &
\(\boldsymbol{C_1^{\mathrm{dyn}}}\) &
\(\boldsymbol{100\,|C_1^{\mathrm{geo}}|/C_1^{\mathrm{dyn}}}\) \\
&
(meV/cycle) &
&
(meV/cycle) &
(\%) \\
\midrule
\NEW{Baseline}
& 0.2145 & \(1.00\times\) & 108.8 & 0.197 \\

\(I_{\mathrm{pump}}=0.1\text{-}4.9\)
& 0.2763 & \(1.29\times\) & 154.8 & 0.178 \\

\(+\,\Gamma_{\mathrm{pump}}^{0}\times2\)
& 0.2958 & \(1.38\times\) & 206.7 & 0.143 \\

\(+\,\gamma_c\times2\)
& 0.5524 & \(2.58\times\) & 402.1 & 0.137 \\

\(+\,\gamma_{\mathrm{rad}}\times0.5\)
& 0.8896 & \(4.15\times\) & 468.9 & 0.190 \\
\bottomrule
\end{tabular}
\end{table}

\NEW{The cumulative sequence increases the absolute geometric first
cumulant from \(0.2145~\mathrm{meV/cycle}\) at the baseline to
\(0.8896~\mathrm{meV/cycle}\) after all four changes, corresponding to
a \(4.15\times\) enhancement.}

\NEW{The pump-range expansion gives}
\(\NEW{C_1^{\mathrm{geo}}}
=0.2763~\mathrm{meV/cycle}\), or a \(1.29\times\) gain.

\NEW{Because this change increases the pump-coordinate excursion, it
also changes the closed path and enclosed area in the
\((V_G,I_{\mathrm{pump}})\) control plane. It is therefore a
different driving loop rather than a reparametrization of the baseline
loop.}
Doubling \(\Gamma_{\mathrm{pump}}^{0}\) raises the signal further to
\(0.2958~\mathrm{meV/cycle}\), corresponding to a
\(1.38\times\) gain.

\NEW{Within this cumulative stack, increasing the effective
optical-injection scale therefore produces only a moderate additional
increase of the geometric signal.}

The larger gains occur when the relaxation environment is modified.
Doubling the phonon scale gives
\(\NEW{C_1^{\mathrm{geo}}}
=0.5524~\mathrm{meV/cycle}\), corresponding to a
\(2.58\times\) enhancement.
This is expected because the counted terminal is the phonon
bath environment. Increasing \(\gamma_c\) increases the rate at which
population in \(\lvert\psi_2\rangle\) relaxes to
\(\lvert\psi_1\rangle\) while transferring energy into the counted
terminal.

Halving the radiative-loss scale gives the full cumulative value
\(\NEW{C_1^{\mathrm{geo}}}
=0.8896~\mathrm{meV/cycle}\).
Weaker radiative recombination prevents the excited population from
being removed too rapidly to the ground state and leaves more population
available for phonon-mediated transfer during the control cycle.

The dynamic background also increases during the cumulative sequence.
It changes from
\(\NEW{C_1^{\mathrm{dyn}}}
=108.8~\mathrm{meV/cycle}\) at the baseline to
\(154.8~\mathrm{meV/cycle}\) after widening the pump range,
\(206.7~\mathrm{meV/cycle}\) after doubling
\(\Gamma_{\mathrm{pump}}^{0}\),
\(402.1~\mathrm{meV/cycle}\) after doubling \(\gamma_c\), and
\(468.9~\mathrm{meV/cycle}\) after all four changes.

Consequently, the mean-background contrast does not improve
substantially. It is \(0.197\%\) at the baseline and \(0.190\%\) after
all four changes.

\NEW{The cumulative sequence therefore increases the absolute
orientation-odd signal but does not reduce the required cancellation of
the much larger orientation-even dynamic mean.}

The calculated increase in
\(\left|C_1^{\mathrm{geo}}\right|\) indicates the relative enhancement
that would be inherited by an experimentally calibrated optical proxy
under a common transduction relation. The construction and experimental
limitations of such an optical proxy are discussed in
Sec.~\ref{sec:si_optical_proxy}.

\NEW{The corresponding dynamic second cumulants and intrinsic
single-pair SNR values are reported in the enhanced-signal subsection
of the main manuscript. The single-pair SNR increases from
\(0.004299\) at the uniform baseline to \(0.008743\) after all four
changes, a factor of \(2.03\). Under the statistically independent-cycle
assumption of Sec.~\ref{si:snr_tau}, the number of cycles required for
a fixed target SNR therefore decreases by approximately
\((2.03)^2=4.14\).}

\NEW{Among the four changes, widening \(I_{\mathrm{pump}}(t)\) is the
directly imposed protocol modification. The changes of
\(\Gamma_{\mathrm{pump}}^{0}\), \(\gamma_c\), and
\(\gamma_{\mathrm{rad}}\) are conditional optical-injection, phonon,
and radiative engineering scenarios. Accordingly, the
\(4.15\times\) absolute-signal gain should not be presented as a
protocol-only prediction.}

\NEW{This cumulative uniform-baseline test must also be distinguished
from the same-loop current-shaped optimization in
Sec.~\ref{sec:si_current_shaped}. The pump-range expansion used here
changes the path and enclosed area, whereas the current-shaped protocol
preserves the established loop and changes only its time
parametrization. The latter is therefore the appropriate comparison for
isolating the effect of time redistribution on statistical
detectability.}

\FloatBarrier

\clearpage
\section{Independent Berry-Sinitsyn first-cumulant cross-check}
\label{sec:si_geometric_checks}
\label{si:curvature_crosscheck}

The finite-time loop-reversal calculation is cross-checked independently
through the first-cumulant Berry-Sinitsyn curvature. The curvature map,
its loop integral, and its interpolation sensitivity are kept together
in this section so that the geometric validation is presented once and
is not repeated in the subsequent control-optimization analysis.

The four reported cumulants are obtained from finite-time CW and CCW
propagation. As an independent check of the geometric interpretation, we
evaluate the frozen generator Berry Sinitsyn curvature for the first
cumulant over the \((V_G,I_{\mathrm{pump}})\) control plane
\cite{Sinitsyn2007,Ren2010,Yuge2012,Paulino2024}. At each frozen
\((V_G,I_{\mathrm{pump}})\) point, the tilted generator has dominant
biorthogonal modes \(\langle l_0|\) and \(|r_0\rangle\). The
gauge-invariant mode-sum expression is
\begin{equation}
B_s^{V_G,I_{\mathrm{pump}}}
=
\sum_{m\neq0}
\frac{
\langle l_0|\partial_{V_G}\mathcal L_s|r_m\rangle
\langle l_m|\partial_{I_{\mathrm{pump}}}\mathcal L_s|r_0\rangle
-
\langle l_0|\partial_{I_{\mathrm{pump}}}\mathcal L_s|r_m\rangle
\langle l_m|\partial_{V_G}\mathcal L_s|r_0\rangle
}{
(\lambda_0-\lambda_m)^2
},
\label{eq:si_bsn_mode_sum}
\end{equation}
with \(\langle l_i|r_j\rangle=\delta_{ij}\). The first-cumulant
curvature density is
\begin{equation}
B_1^{V_G,I_{\mathrm{pump}}}
=
\left.
\frac{\partial
B_s^{V_G,I_{\mathrm{pump}}}}
{\partial s}
\right|_{s=0}.
\label{eq:si_first_cumulant_curvature}
\end{equation}
The left eigenvectors are constructed from the inverse right-eigenvector
matrix. The largest eigenvector-matrix condition number across the map
is below \(3.85\), and the residual imaginary part is zero at the
displayed precision.

\begin{figure}[!htbp]
\centering
\safeincludegraphics[width=0.95\textwidth]
{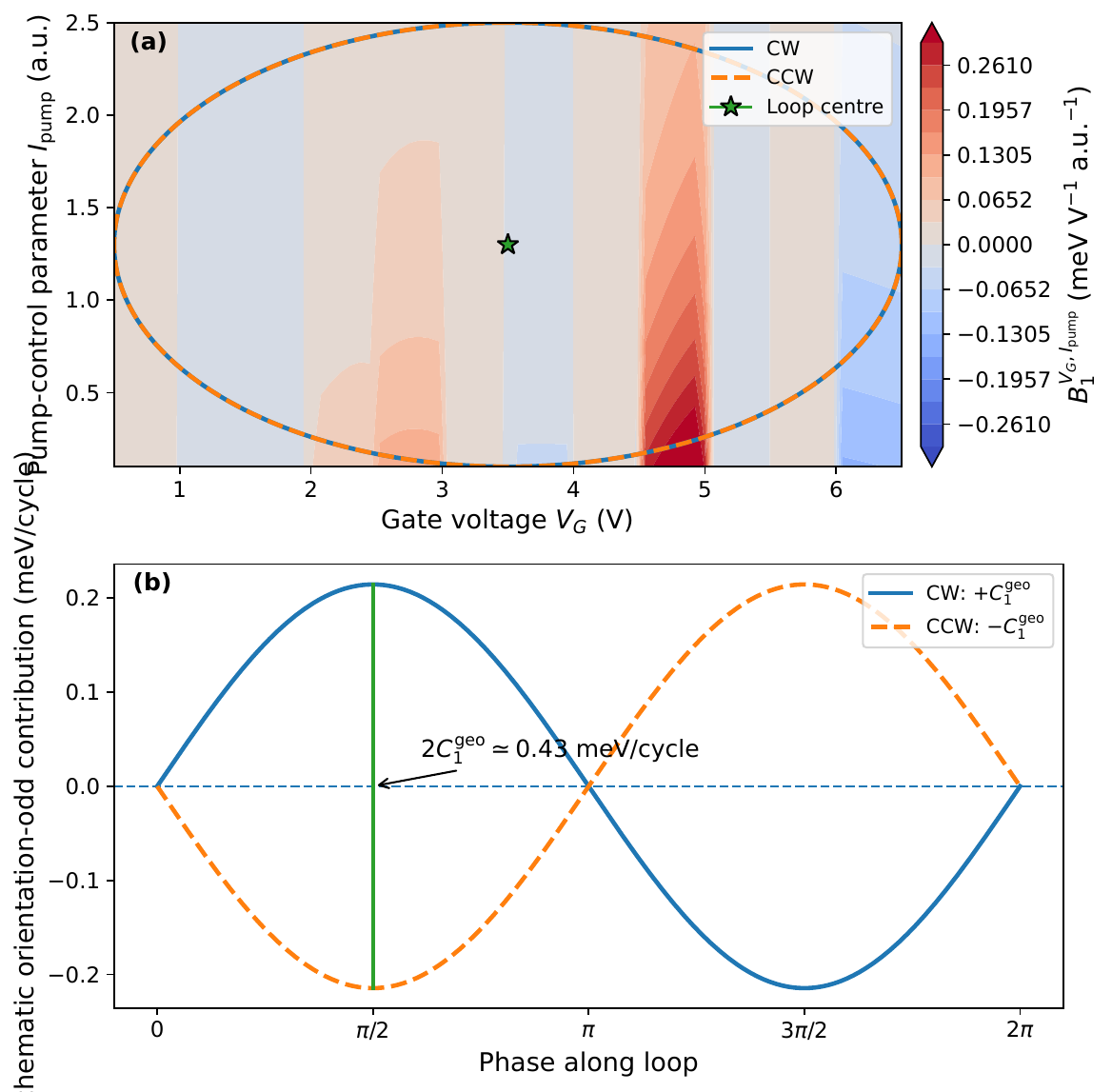}
\caption{Independent first-cumulant Berry-Sinitsyn curvature map and
the loop-reversal signature. The curvature is concentrated mainly
toward the higher-\(V_G\) side of the working loop. The full CW-CCW
separation on the slow-driving plateau is
\(2C_1^{\mathrm{geo}}\simeq0.43~\mathrm{meV/cycle}\). The curvature
map is not used as an input to the finite-time cumulant engine.}
\label{fig:si_curvature_loop}
\end{figure}

\begin{figure}[!htbp]
\centering
\begin{minipage}[t]{0.32\textwidth}
\centering
\safeincludegraphics[width=\linewidth]{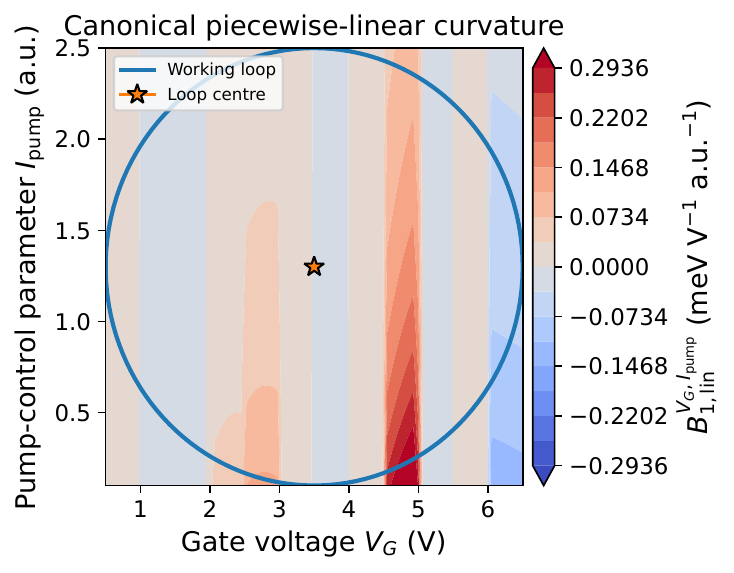}
\textbf{(a)}
\end{minipage}\hfill
\begin{minipage}[t]{0.32\textwidth}
\centering
\safeincludegraphics[width=\linewidth]{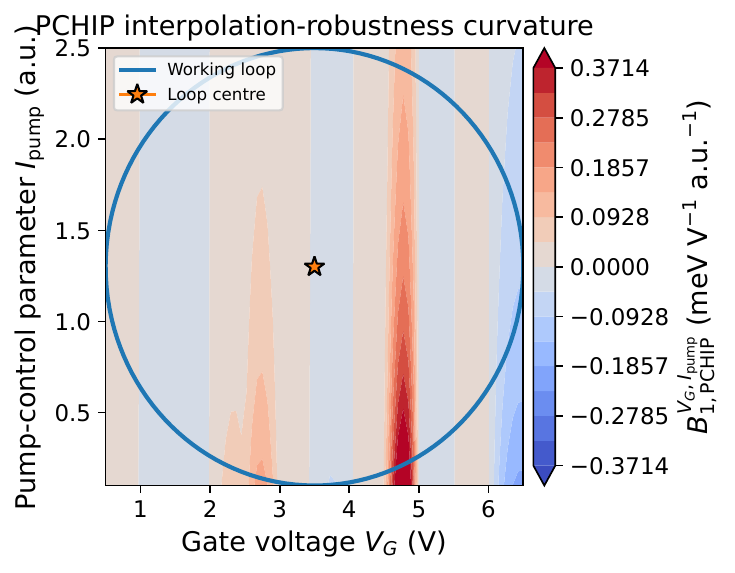}
\textbf{(b)}
\end{minipage}\hfill
\begin{minipage}[t]{0.32\textwidth}
\centering
\safeincludegraphics[width=\linewidth]{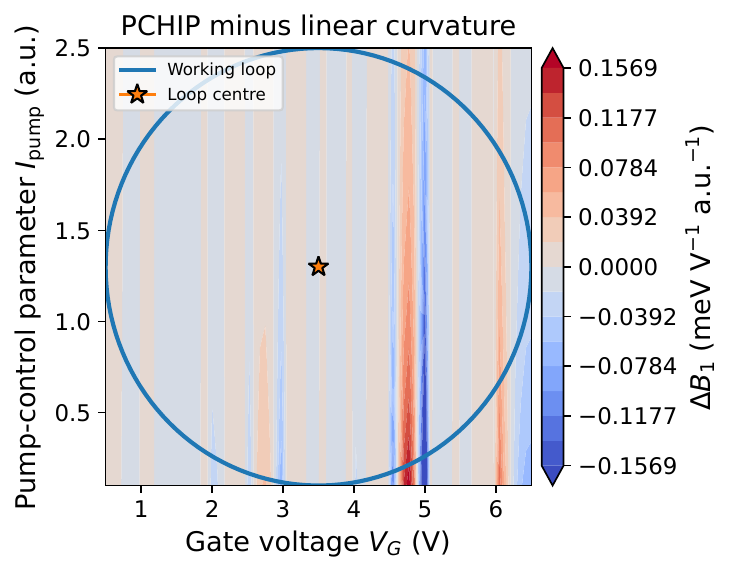}
\textbf{(c)}
\end{minipage}
\caption{Interpolation robustness of the first-cumulant
Berry-Sinitsyn curvature. (a) Canonical piecewise-linear
interpolation. (b) Shape-preserving PCHIP interpolation used only as a
robustness check. (c) Pointwise PCHIP-minus-linear difference. The two
maps have correlation coefficient \(0.9292\) and retain the same broad
higher-\(V_G\) structure.}
\label{fig:si_curvature_interpolation}
\end{figure}



For the canonical piecewise-linear interpolation model, the independent Berry-Sinitsyn curvature-flux calculation gives
\(
C_{1,\mathrm{curv}}^{\mathrm{geo}}
\simeq 0.2247~\mathrm{meV/cycle},
\)
whereas the converged CW-CCW loop-reversal propagation gives
\(
C_{1,\mathrm{prop}}^{\mathrm{geo}}
\simeq 0.2145~\mathrm{meV/cycle}.
\)
The two values differ by approximately \(4.8\%\). Because the propagated orientation-odd cumulant remains unchanged over the \(5\)-\(50~\mathrm{ns}\) period range, this difference cannot be attributed primarily to finite-speed effects. It instead reflects the numerical sensitivity of the curvature calculation to the finite control-space grid, the finite-difference derivatives used to construct the curvature, and the nonsmooth derivatives of the piecewise-linear Hamiltonian surface at the interpolation nodes.

As an interpolation-sensitivity test, both calculations were repeated using the same PCHIP-interpolated Hamiltonian inputs. The PCHIP curvature-flux calculation gives
\(
C_{1,\mathrm{curv}}^{\mathrm{geo}}
\simeq 0.2175~\mathrm{meV/cycle},
\)
while the independent PCHIP finite-time loop-reversal propagation gives
\(
C_{1,\mathrm{prop}}^{\mathrm{geo}}
\simeq 0.2188~\mathrm{meV/cycle}.
\)
Their difference is then reduced to approximately (0.6$\%$). Relative to the canonical linear-interpolation results, PCHIP changes the propagated ($C_1^{\mathrm{geo}}$) and ($C_2^{\mathrm{geo}}$) by approximately (2.0\%) and (3.5\%), respectively. We therefore retain piecewise-linear interpolation as the explicitly defined canonical model and use PCHIP only to estimate interpolation sensitivity. The close agreement between the independent curvature-flux and loop-reversal calculations, particularly for the smooth PCHIP surface, confirms that the converged orientation-odd first cumulant has the expected Berry-Sinitsyn geometric origin.

\FloatBarrier
\clearpage

\section{Alternative loop-shape comparison}
\label{sec:si_alternative_protocols}

\NEW{Alternative loop shapes are examined to determine whether
concentrating the enclosed control-space area around the
first-cumulant Berry-Sinitsyn curvature hotspot improves either the
absolute geometric signal or its intrinsic statistical detectability.
This exploratory calculation is retained as a control for the
full range baseline loop and is kept separate from the same loop
NUFM traversal of Sec.~\ref{sec:si_current_shaped}, which
preserves the established path and changes only its time
parametrization.}

\NEW{Candidate ellipses are screened using the PCHIP-interpolated
first-cumulant curvature surface employed in the interpolation-sensitivity
check of Sec.~\ref{si:curvature_crosscheck}, together with the frozen
dynamic first-cumulant-rate map}
\(
\NEW{
j_1(V_G,I_{\mathrm{pump}})
=
\left.
\frac{\partial\lambda_0
(s;V_G,I_{\mathrm{pump}})}
{\partial s}
\right|_{s=0},
}
\)
\NEW{where \(\lambda_0\) is the dominant eigenvalue of the frozen tilted
Liouvillian. The curvature map identifies regions that can contribute
to the orientation-odd geometric signal, whereas
\(j_1(V_G,I_{\mathrm{pump}})\), in
\(\mathrm{meV\,ps^{-1}}\), identifies regions with a large local
orientation-even mean-energy-transfer rate. The label ``ranked
ellipse'' denotes the highest-ranked alternative from this combined
screening, while the narrow hotspot ellipse is a deliberately localized
test around the visible curvature maximum.}

\NEW{All candidate trajectories remain within the experimentally
supported control window,
\(0.5\leq V_G\leq6.5~\mathrm{V}\) and
\(0.1\leq I_{\mathrm{pump}}\leq2.5\).
For the PCHIP interpolation used only in this exploratory comparison,
the counted dressed-state gap \(\Omega_{12}(t)\) remains inside the\(60~\mathrm{meV}\) phonon window throughout every tested
loop, with a largest value of \(58.40~\mathrm{meV}\). The corresponding
maximum for the canonical piecewise-linear Hamiltonian surface is
\(58.21~\mathrm{meV}\). The two values therefore refer to different
interpolation models and are not inconsistent.}

\begin{figure}[!htbp]
\centering
\begin{minipage}[t]{0.49\textwidth}
\centering
\safeincludegraphics[
width=\linewidth,
keepaspectratio
]{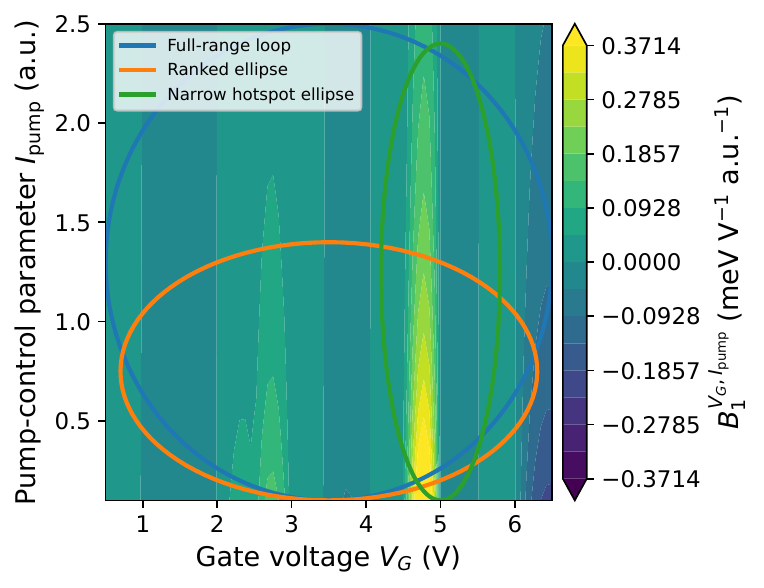}
\textbf{(a)}
\end{minipage}
\hfill
\begin{minipage}[t]{0.49\textwidth}
\centering
\safeincludegraphics[
width=\linewidth,
keepaspectratio
]{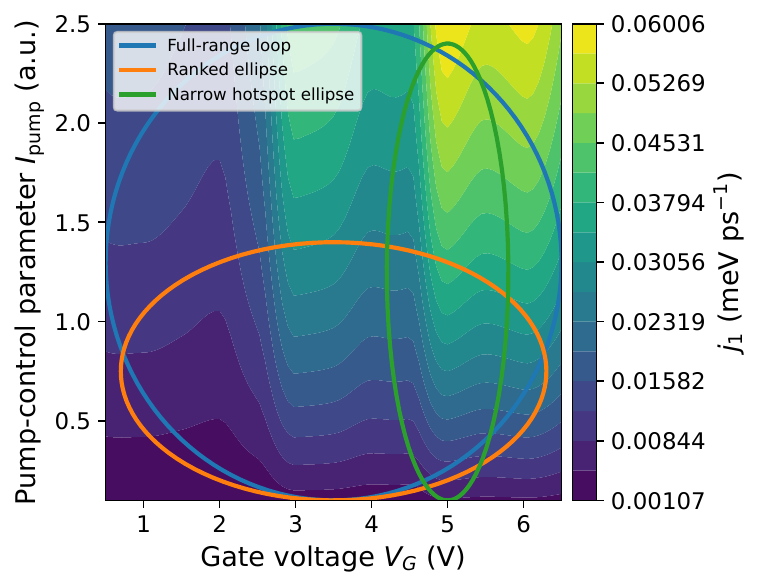}
\textbf{(b)}
\end{minipage}
\caption{\NEW{Control-space comparison of alternative loop shapes on the
PCHIP interpolation-sensitivity surface. Panel~(a) gives the first-cumulant Berry-Sinitsyn curvature density
together with the full-range baseline loop, the ranked alternative
ellipse, and the narrow curvature-hotspot ellipse. Panel~(b) gives the
frozen dynamic first-cumulant-rate map for the same trajectories.
Concentrating a loop around a local curvature maximum does not
necessarily improve the combined signal and detectability because the
new loop may enclose less net curvature flux or traverse regions with a
larger dynamic mean-energy-transfer rate.}
}
\label{fig:si_loop_shape_maps}
\end{figure}

\NEW{Figure~\ref{fig:si_loop_shape_maps} compares the three trajectories
on the PCHIP curvature and frozen-current maps.}

\NEW{For consistency with Sec.~\ref{si:snr_tau} and the main
manuscript, we compare the loops using the mean-background contrast}
\(
\NEW{
100\,
\frac{|C_1^{\mathrm{geo}}|}
{C_1^{\mathrm{dyn}}},
}
\)
\NEW{and the intrinsic single-pair SNR}
\(
\NEW{
\operatorname{SNR}_{N=1}
=
\sqrt{2}\,
\frac{|C_1^{\mathrm{geo}}|}
{\sqrt{C_2^{\mathrm{dyn}}}}.
}
\)
\NEW{Here, \(N=1\) denotes one statistically independent CW cycle and
one statistically independent CCW cycle. Because all three loops are
evaluated at the same period, \(\tau=1~\mathrm{ns}\), their ordering
according to \(\operatorname{SNR}_{N=1}\) is also their ordering for a
fixed total acquisition time. No additional detectability symbol is
introduced for this equal-period comparison.}

\NEW{The corresponding finite-time PCHIP cumulants, contrasts, and
single-pair SNR values are listed in
Table~\ref{tab:si_loop_shape_scan}.}

\begin{table}[!htbp]
\centering
\scriptsize
\setlength{\tabcolsep}{4pt}
\renewcommand{\arraystretch}{1.10}
\caption{\NEW{PCHIP-based finite-time comparison of the full-range
baseline loop and two alternative elliptical loops at
\(\tau=1~\mathrm{ns}\).
All three trajectories are evaluated with the same PCHIP interpolation
and finite-time FCS conventions. These exploratory
interpolation-sensitivity values do not replace the canonical
piecewise-linear results used in the principal calculations.}
}
\label{tab:si_loop_shape_scan}
\begin{tabular}{lccccc}
\toprule
\textbf{Loop}
&
\shortstack{\(C_1^{\mathrm{geo}}\)\\
(\(\mathrm{meV/cycle}\))}
&
\shortstack{\(C_1^{\mathrm{dyn}}\)\\
(\(\mathrm{meV/cycle}\))}
&
\shortstack{\(C_2^{\mathrm{dyn}}\)\\
(\(\mathrm{meV^2/cycle}\))}
&
\shortstack{\(\NEW{100|C_1^{\mathrm{geo}}|/
C_1^{\mathrm{dyn}}}\)\\(\%)}
&
\shortstack{\(\operatorname{SNR}_{N=1}\)}
\\
\midrule
Full-range baseline
& 0.21843 & 21.840 & 1009.87 & 1.000
& \NEW{0.009721}\\
Ranked ellipse
& 0.16438 & 15.289 & 728.94 & 1.075
& \NEW{0.008610}\\
Narrow hotspot ellipse
& 0.16468 & 29.611 & 1435.61 & 0.556
& \NEW{0.006147}\\
\bottomrule
\end{tabular}
\end{table}

The ranked ellipse increases the mean-background contrast by only
\(7.5\%\), while retaining approximately \(75.3\%\) of the geometric
first cumulant of the full-range baseline loop. Its reduction in
\(C_2^{\mathrm{dyn}}\) is insufficient to compensate for the loss of
geometric signal, and
\(\NEW{\operatorname{SNR}_{N=1}}\)
decreases by approximately \(11.4\%\).

The narrow hotspot ellipse gives a geometric first cumulant similar to
that of the ranked ellipse but samples a substantially larger dynamic
mean and variance. It consequently retains only \(55.6\%\) of the
full-range mean-background contrast and \(63.2\%\) of the corresponding \NEW{intrinsic single-pair SNR}.

\NEW{The local maximum of the Berry-Sinitsyn curvature density is
therefore not sufficient by itself to identify an improved driving
loop. The geometric first cumulant depends on the oriented curvature
flux through the complete enclosed area. The required rejection of the
orientation-even mean depends on \(C_1^{\mathrm{dyn}}\), whereas the
intrinsic statistical noise entering the CW-CCW SNR is set by
\(C_2^{\mathrm{dyn}}\). Under the tested constraints, neither
alternative ellipse improves both the absolute geometric first
cumulant and its intrinsic detectability. The full-range baseline loop
is therefore retained for the subsequent same-loop current-shaped
protocol of Sec.~\ref{sec:si_current_shaped}.}

\FloatBarrier
\clearpage


\FloatBarrier
\section{\NEW{Nonuniform frequency modulated traversal of the same loop}}
\label{sec:si_current_shaped}

\NEW{This section gives the complete construction and validation of the
nonuniform frequency modulated (\(\mathrm{NUFM}\)) protocol used in the
main manuscript. The established oriented path in the
\((V_G,I_{\mathrm{pump}})\) control plane is preserved while only its
time parametrization is changed. We first define the constrained angular
speed, then examine the parameter grid and select the representative
operating point, evaluate its cumulant response, and finally test its
period dependence, numerical convergence, waveform feasibility, and
rate-model robustness.}

\subsection{Protocol definition and constrained angular speed law}
\label{sec:si_current_shaped_traversal}

The uniform protocol used throughout the main manuscript is
\begin{align}
V_G(t)
&=
V_G^{(0)}
+
A_G\sin(\omega t+\phi_G),
\\
I_{\mathrm{pump}}(t)
&=
I_{\mathrm{pump}}^{(0)}
+
A_I\sin(\omega t+\phi_I),
\end{align}
where
\(
\omega=2\pi/\tau
\)
and
\(
\Delta\phi=\phi_I-\phi_G
\).
For the established baseline loop, the time origin may be chosen such
that
\(
\phi_G=0
\)
and
\(
\phi_I=\Delta\phi=\pi/2
\).
This choice fixes only the phase origin and does not alter the relative
phase, loop shape, enclosed area, or orientation.

\NEW{For the \(\mathrm{NUFM}\) traversal, the uniformly advancing phase
\(\omega t\) is replaced by a common accumulated loop coordinate
\(\theta(t)\),}
\begin{align}
V_G^{\mathrm{NUFM}}(t)
&=
V_G^{(0)}
+
A_G\sin[\theta(t)+\phi_G],
\\
I_{\mathrm{pump}}^{\mathrm{NUFM}}(t)
&=
I_{\mathrm{pump}}^{(0)}
+
A_I\sin[\theta(t)+\phi_I].
\end{align}
The boundary conditions
\(
\theta(0)=0
\)
and
\(
\theta(\tau)=2\pi
\)
enforce one complete traversal in the prescribed cycle period. Because
the same \(\theta(t)\) is used for both controls, the relative phase
remains fixed at \(\Delta\phi\). Eliminating \(\theta\) therefore gives
the same closed path, enclosed area, and orientation as the corresponding
uniform traversal. The instantaneous angular frequency is
\(
\dot{\theta}(t)
\),
which is equal to \(\omega\) for the uniform protocol and becomes
position dependent for the \(\mathrm{NUFM}\) protocol. Its cycle average
remains
\(
2\pi/\tau=\omega
\).

Let \(\theta\in[0,2\pi]\) denote position along the fixed loop. The
dominant eigenvalue of the instantaneous tilted population Liouvillian
\(\mathcal{L}_s(\theta)\) is denoted by
\(\lambda_0(s,\theta)\). The frozen rate of the first dynamic cumulant is
defined as
\(
j_1(\theta)
=
\left.
\partial_s\lambda_0(s,\theta)
\right|_{s=0}
\).
It represents the instantaneous mean counted-energy current associated
with the frozen system parameters at that loop position. At \(s=0\), the probability-conserving Liouvillian has the stationary
eigenvalue
\(
\lambda_0(0,\theta)=0
\).
The ordinary population-relaxation gap is defined from the remaining
eigenvalues as
\(
\Delta_L(\theta)
=
\min_{m\neq0}
[-\operatorname{Re}\lambda_m(0,\theta)]
\),
with corresponding relaxation time
\(
\tau_{\mathrm{rel}}(\theta)=1/\Delta_L(\theta)
\). \NEW{Because \(\Delta_L(\theta)\) is obtained from the complete
population Liouvillian, it contains the combined effects of optical
injection, phonon relaxation, and radiative loss. It is therefore a
population-relaxation gap of the full open-system generator rather than
a phonon-only relaxation rate.}

The imposed angular speed is
\begin{equation}
\begin{aligned}
\dot{\theta}(\theta)
=
\operatorname{clip}\Bigg[
&
K
\left(
\frac{j_1(\theta)}
     {j_{1,*}}
\right)^{\alpha}
\left(
\frac{\Delta_L(\theta)}
     {\Delta_{L,*}}
\right)^{\beta},
\\[-1mm]
&
\frac{\omega}{r},
\min\!\left(
r\omega,
\epsilon_{\max}\Delta_L(\theta)
\right)
\Bigg],
\end{aligned}
\label{eq:si_gapaware_speed}
\end{equation}
where
\(
\operatorname{clip}(x,a,b)
=
\min[\max(x,a),b]
\).
The normalization constants are
\(
j_{1,*}=\max_\theta j_1(\theta)
\)
and
\(
\Delta_{L,*}=\max_\theta\Delta_L(\theta)
\),
so both profile factors in Eq.~\eqref{eq:si_gapaware_speed} are
dimensionless. The exponent \(\alpha\) controls the strength of the current-based
redistribution. Increasing \(\alpha\) increases the proposed angular
speed where \(j_1(\theta)\) is large and consequently reduces the local
residence time in those regions. The exponent \(\beta\) introduces a
direct dependence of the unconstrained profile on the local relaxation
gap. The bounds
\(
\omega/r\leq\dot{\theta}\leq r\omega
\)
prevent excessively slow or fast motion relative to the cycle-averaged
angular frequency. The additional local condition
\(
\dot{\theta}(\theta)
\leq
\epsilon_{\max}\Delta_L(\theta)
\)
limits the phase advance relative to the slowest instantaneous
population-relaxation rate. Equivalently,
\(
\dot{\theta}(\theta)/\Delta_L(\theta)
\leq
\epsilon_{\max}
\).
For \(r=5\), the nominal speed bounds permit
\(
\omega/5\leq\dot{\theta}\leq5\omega
\),
corresponding to a maximum-to-minimum allowed speed ratio of \(25\)
before the local relaxation-gap constraint is applied. The normalization factor \(K\) is determined from the prescribed-period
condition
\(
\int_0^{2\pi}d\theta/\dot{\theta}(\theta)=\tau
\).
Because clipping may be active over only part of the loop, \(K\) is
obtained numerically from the root of
\(
F(K)
=
\int_0^{2\pi}
d\theta/\dot{\theta}(\theta;K)-\tau
\). When
\(
\alpha=\beta=0
\),
the unconstrained profile is constant. Provided that the uniform speed
lies inside the allowed interval at every loop position, the period
constraint gives
\(
K=\omega
\)
and therefore
\(
\dot{\theta}(\theta)=\omega
\),
recovering the uniform traversal exactly. \NEW{The values of
\((\alpha,\beta,r,\epsilon_{\max})\) used for the main calculations are
selected from the complete protocol scan discussed in
Sec.~\ref{sec:si_complete_parameter_grid}.} For a loop discretized into \(N_t\) equal phase intervals,
\(
\Delta\theta=2\pi/N_t
\),
the uniform protocol assigns the same dwell time to every interval,
\(
\Delta t_k^{\mathrm{uniform}}=\tau/N_t
\).
The \(\mathrm{NUFM}\) traversal instead assigns
\(
\Delta t_k^{\mathrm{NUFM}}
=
\Delta\theta/\dot{\theta}(\theta_k)
\). \NEW{The finite-time tilted-Liouvillian propagation, periodic limit-cycle
initialization, counting-field differentiation, and CW-CCW
loop-reversal procedure are otherwise identical to those used for the
uniform calculations. The \(\mathrm{NUFM}\) protocol therefore changes
the physical dwell time assigned to each position on the loop rather
than introducing a different counting procedure.} The effect on the dynamic first cumulant follows directly from the
dwell-time weighting,
\(
C_1^{\mathrm{dyn}}
\simeq
\int_0^{2\pi}
\frac{j_1(\theta)}
{\dot{\theta}(\theta)}
\,d\theta .
\)
The protocol therefore reduces the contribution from regions with large
\(j_1(\theta)\) by traversing them more rapidly and redistributes the
remaining cycle time to regions with smaller frozen current. The second
dynamic cumulant is modified by the same dwell redistribution but also
depends on the population-relaxation history and temporal correlations
and is therefore evaluated from the complete tilted propagator. \NEW{By contrast, the adiabatic geometric contribution is a line
integral over the oriented parameter-space path and contains no explicit
\(1/\dot{\theta}\) weighting. Changing the traversal schedule while
preserving the same oriented loop therefore leaves the adiabatic
geometric contribution unchanged. The small protocol-dependent
differences observed at finite period are retained as finite-speed
corrections rather than interpreted as changes of the geometric path.} For the CW and CCW calculations, the same physical dwell profile is
traversed in opposite order. Corresponding loop positions therefore
receive the same dwell times. The maximum numerical mismatch between
corresponding CW and CCW dwell times is
\(
5\times10^{-12}~\mathrm{ps}
\),
confirming that the orientation-odd subtraction is not biased by unequal
local exposure times.

\FloatBarrier
\subsection{Parameter grid and selection of the balanced NUFM protocol}
\label{sec:si_complete_parameter_grid}

\NEW{The protocol parameters were first examined through a complete
factorial scan at
\(\tau=1~\mathrm{ns}\) and \(N_t=400\). The grid contains all
\(5^4=625\) combinations}
\(
\alpha\in\{0,1,2,4,6\},
\qquad
\beta\in\{0,0.5,1,1.5,2\},
\)
\(
r\in\{2,3,5,8,10\},
\qquad
\epsilon_{\max}
\in
\{0.04,0.06,0.08,0.10,0.12\}.
\)
All \(625\) reported combinations are feasible. Exactly \(25\) rows,
corresponding to \(\alpha=\beta=0\) for the \(25\) tested
\((r,\epsilon_{\max})\) pairs, reproduce the uniform protocol to
numerical precision because the raw profile is constant and clipping is
inactive. \NEW{The overall suppression hierarchy is dominated by the current
exponent \(\alpha\) and the permitted speed range \(r\). The response
begins to saturate for \(\alpha\simeq4\) to \(6\), whereas direct gap
weighting through \(\beta\) produces a comparatively weaker change once
the current-based shaping becomes strong.}

\begin{figure}[!htbp]
\centering
\safeincludegraphics[width=4in,height=3in]
{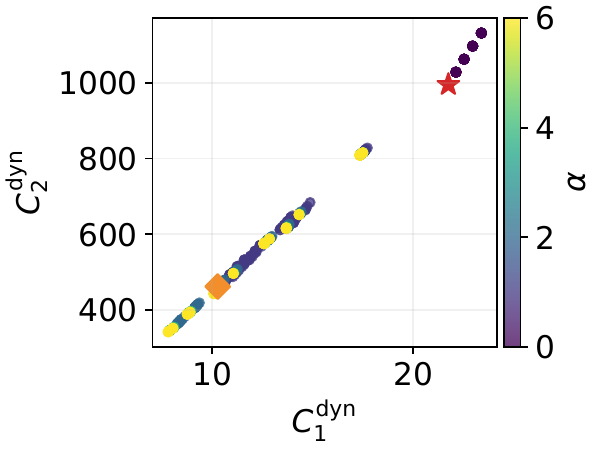}
\caption{\NEW{Complete \(625\)-combination NUFM parameter grid at
\(\tau=1~\mathrm{ns}\) and \(N_t=400\).
The dynamic first and second cumulants are shown for every feasible
combination. Colour encodes \(\alpha\). The star denotes the uniform
traversal and the diamond denotes the selected
\((\alpha,\beta,r,\epsilon_{\max})=(6,0,5,0.08)\) protocol.}}
\label{fig:si_new_625_grid}
\end{figure}

\subsubsection{Role of the current and gap exponents}
\label{sec:si_profile_exponent_audit}

To isolate the roles of \(\alpha\) and \(\beta\), the \(25\)
combinations
\(
\alpha\in\{0,1,2,4,6\},
\qquad
\beta\in\{0,0.5,1,1.5,2\}
\)
were evaluated at fixed
\(
r=5
\),
\(
\epsilon_{\max}=0.08
\),
\(
\tau=1000~\mathrm{ps}
\),
and
\(
N_t=400
\). \NEW{Under the matched same-point reversal convention used for this
audit, the uniform \(1~\mathrm{ns}\) protocol gives
\(C_1^{\mathrm{dyn}}=21.751998~\mathrm{meV/cycle}\),
\(C_2^{\mathrm{dyn}}=995.777041~\mathrm{meV^2/cycle}\),
\(C_1^{\mathrm{odd}}=0.214179~\mathrm{meV/cycle}\), and
\(C_2^{\mathrm{odd}}=11.123571~\mathrm{meV^2/cycle}\).} The matched same-point value
\(C_2^{\mathrm{odd}}=11.123571~\mathrm{meV^2/cycle}\)
differs slightly from the direct phase-lag-reversal value
\(11.156724~\mathrm{meV^2/cycle}\). This small difference is associated
with the one-cycle reversal-boundary convention and does not represent
a change in the Hamiltonian, loop, or bulk Berry-Sinitsyn contribution.
The first orientation-odd cumulant is insensitive to this convention at
the displayed precision.

\begin{figure}[!htbp]
\centering

\begin{minipage}[t]{0.48\textwidth}
\centering
\safeincludegraphics[width=\linewidth]
{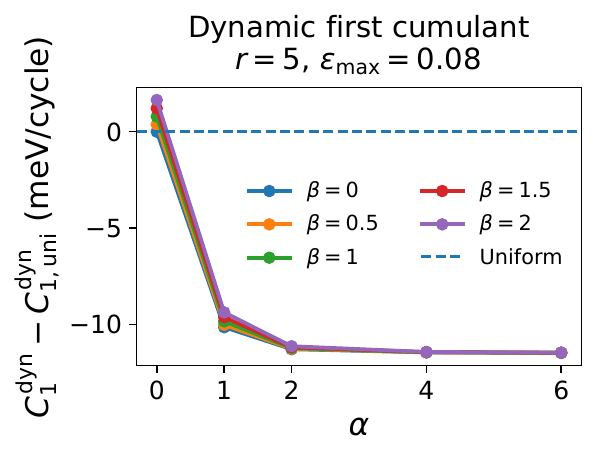}
\end{minipage}
\hfill
\begin{minipage}[t]{0.48\textwidth}
\centering
\safeincludegraphics[width=\linewidth]
{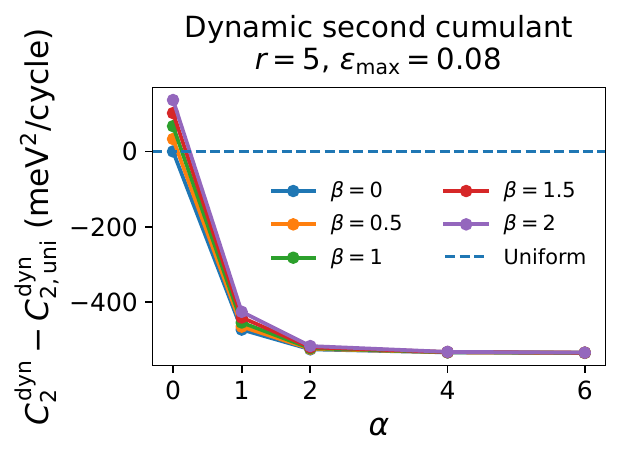}
\end{minipage}

\vspace{2mm}

\begin{minipage}[t]{0.48\textwidth}
\centering
\safeincludegraphics[width=\linewidth]
{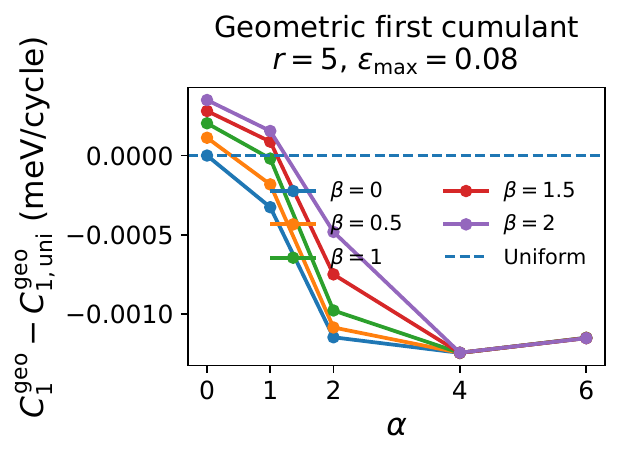}
\end{minipage}
\hfill
\begin{minipage}[t]{0.48\textwidth}
\centering
\safeincludegraphics[width=\linewidth]
{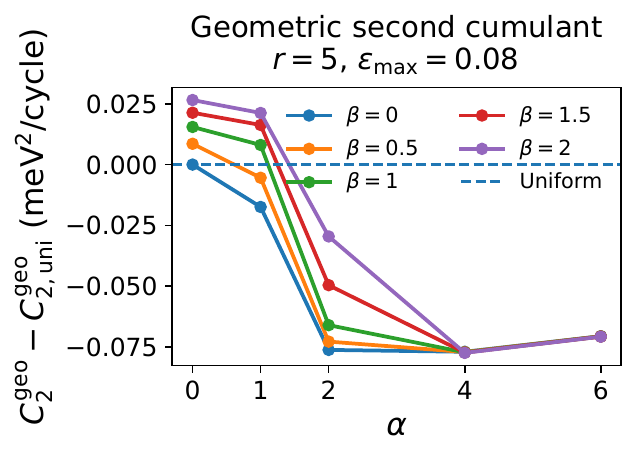}
\end{minipage}

\caption{
Dependence of the four cumulants on the current exponent \(\alpha\) and
gap exponent \(\beta\) at fixed
\(r=5\) and \(\epsilon_{\max}=0.08\).
The plotted quantities are signed differences from the independently
evaluated uniform \(1~\mathrm{ns}\) comparison.
Top left:
\(\NEW{C_1^{\mathrm{dyn}}(\mathrm{NUFM})
-C_1^{\mathrm{dyn}}(\mathrm{uniform})}\).
Top right:
\(\NEW{C_2^{\mathrm{dyn}}(\mathrm{NUFM})
-C_2^{\mathrm{dyn}}(\mathrm{uniform})}\).
Bottom left:
\(\NEW{C_1^{\mathrm{odd}}(\mathrm{NUFM})
-C_1^{\mathrm{odd}}(\mathrm{uniform})}\).
Bottom right:
\(\NEW{C_2^{\mathrm{odd}}(\mathrm{NUFM})
-C_2^{\mathrm{odd}}(\mathrm{uniform})}\).
Each curve corresponds to one value of \(\beta\), while the horizontal
axis gives \(\alpha\). The dashed horizontal line denotes zero
difference from the uniform result.
}
\label{fig:si_alpha_beta_uniform_comparison}
\end{figure}

Increasing \(\alpha\) from \(0\) to \(1\) produces the largest initial
reduction of both dynamic cumulants, with additional suppression up to
approximately \(\alpha=4\) to \(6\). In contrast, varying \(\beta\)
mainly produces a smaller separation between the curves, and this
dependence becomes weak once the current-based shaping is strong.
Gap weighting alone does not necessarily suppress the dynamic
background; for \(\alpha=0\) and sufficiently large \(\beta\), both
dynamic cumulants can exceed their uniform values.

Across the \(25\) tested \((\alpha,\beta)\) combinations,
\(
10.274030
\leq
C_1^{\mathrm{dyn}}
\leq
23.392809~\mathrm{meV/cycle},
\)
and
\(
461.556568
\leq
C_2^{\mathrm{dyn}}
\leq
1132.164847~\mathrm{meV^2/cycle}.
\)
The maximum reductions relative to the uniform comparison are
approximately \(52.77\%\) for \(C_1^{\mathrm{dyn}}\) and
\(53.65\%\) for \(C_2^{\mathrm{dyn}}\).

Over the same grid,
\(
0.212932
\leq
C_1^{\mathrm{odd}}
\leq
0.214529~\mathrm{meV/cycle},
\)
and
\(
11.046208
\leq
C_2^{\mathrm{odd}}
\leq
11.150097~\mathrm{meV^2/cycle}.
\)
The largest relative deviations from the uniform matched-reversal
values remain below approximately \(0.59\%\) for
\(C_1^{\mathrm{odd}}\) and \(0.70\%\) for
\(C_2^{\mathrm{odd}}\). Their substantially weaker dependence on the
profile parameters is consistent with preservation of the same oriented
loop, while the residual spread is retained as a finite-period
correction.

\subsubsection{Constrained-threshold comparison}

The reduction of the dynamic first cumulant follows from the
dwell-weighted frozen-current expression
\(
C_1^{\mathrm{dyn}}
\simeq
\int_0^{2\pi}
j_1(\theta)
\frac{d\theta}{\dot{\theta}(\theta)}.
\)
The strongly shaped profiles therefore approach the threshold structure
expected from minimizing this approximate functional under the imposed
pointwise speed bounds.

A direct construction of the constrained threshold schedule at
\(\tau=1000~\mathrm{ps}\) and \(N_t=800\) gives
\(
C_1^{\mathrm{dyn}}
=
10.2527~\mathrm{meV/cycle}
\),
with \(99.875\%\) of phase cells at a dwell bound, compared with
\(
C_1^{\mathrm{dyn}}
=
10.2741~\mathrm{meV/cycle}
\)
for the selected \(\alpha=6\) schedule. The selected schedule is
therefore within approximately \(0.21\%\) of the constrained threshold
minimum of the dwell-weighted frozen-current approximation.

\NEW{This comparison does not establish a global optimum of the exact
finite-time FCS functional. In particular, the second dynamic cumulant
cannot be minimized from the same local-current expression because it
also contains temporal correlations generated by the driven stochastic
dynamics. Its reduction is therefore obtained from the full
tilted-Liouvillian propagation.}

\begin{table}[!htbp]
\centering
\scriptsize
\setlength{\tabcolsep}{5pt}
\renewcommand{\arraystretch}{1.10}
\caption{\NEW{Comparison of the uniform, selected NUFM, and constrained
threshold schedules at \(\tau=1~\mathrm{ns}\).
All rows use the same \(N_t=800\) refinement and the matched same-point
reversal convention. Cumulant units are \(\mathrm{meV/cycle}\) for the
first cumulants and \(\mathrm{meV^2/cycle}\) for the second cumulants.}}
\label{tab:si_new_threshold}
\begin{tabular}{lrrrr}
\toprule
Protocol
&
\(C_1^{\rm dyn}\)
&
\(C_1^{\rm odd}\)
&
\(C_2^{\rm dyn}\)
&
\(C_2^{\rm odd}\)
\\
\midrule

Uniform
& 21.752370
& 0.214184
& 995.795
& 11.123784
\\

Selected NUFM
& 10.274144
& 0.213036
& 461.561
& 11.053405
\\

Threshold schedule
& 10.252711
& 0.213396
& 460.484
& 11.073720
\\

\bottomrule
\end{tabular}
\end{table}

\noindent\textbf{\NEW{Selection of the balanced operating protocol.}}

\NEW{The complete parameter grid contains more aggressive settings that
produce smaller dynamic cumulants. In particular,
\((\alpha,\beta,r,\epsilon_{\max})=(6,0,10,0.12)\) gives
\(C_1^{\mathrm{dyn}}=7.8074~\mathrm{meV/cycle}\) and
\(C_2^{\mathrm{dyn}}=341.55~\mathrm{meV^2/cycle}\), compared with
\(10.2740~\mathrm{meV/cycle}\) and
\(461.56~\mathrm{meV^2/cycle}\) for
\((6,0,5,0.08)\).}

The stronger suppression at
\((6,0,10,0.12)\) is obtained with a larger allowed speed range and a
weaker local relaxation constraint. The value \(r=10\) permits a nominal
maximum-to-minimum speed ratio of \(r^2=100\), compared with \(25\) for
\(r=5\), while \(\epsilon_{\max}=0.12\) permits a \(50\%\) larger value
of \(\dot{\theta}/\Delta_L\) than \(\epsilon_{\max}=0.08\).
The corresponding dwell-time range is approximately
\(0.437\) to \(25~\mathrm{ps}\) at \(N_t=400\), compared with
\(0.656\) to \(12.5~\mathrm{ps}\) for the selected protocol.

The more aggressive setting also produces a larger finite-period change
in the orientation-odd quantities. Under the same matched-reversal
convention, the retention of \(C_1^{\mathrm{odd}}\) decreases from
approximately \(99.46\%\) for \((6,0,5,0.08)\) to \(98.74\%\), while
the corresponding \(C_2^{\mathrm{odd}}\) retention decreases from
approximately \(99.37\%\) to \(98.45\%\).

\NEW{We therefore use
\((\alpha,\beta,r,\epsilon_{\max})=(6,0,5,0.08)\) as the balanced
operating point in the main manuscript. It combines substantial
suppression of the dynamic background with high geometric retention and
moderate imposed speed bounds and is not claimed to be the absolute
minimum of the discrete parameter grid.}

For this selected protocol, \(\beta=0\), so the unconstrained speed
profile is determined by
\(
\dot{\theta}_{\mathrm{raw}}(\theta)
=
K
\left[
\frac{j_1(\theta)}
     {j_{1,*}}
\right]^6 .
\)
The relaxation gap nevertheless remains active through the local upper
constraint
\(
\dot{\theta}(\theta)
\leq
0.08\,\Delta_L(\theta).
\)
Thus, \(j_1(\theta)\) determines where the protocol attempts to
redistribute the dwell time, while \(\Delta_L(\theta)\) limits the
permitted local angular speed.

\FloatBarrier
\subsection{Cumulant response and local-profile mechanism}
\label{sec:si_current_shaped_response}

\NEW{Having selected
\((\alpha,\beta,r,\epsilon_{\max})=(6,0,5,0.08)\), we now compare its
response at \(\tau=1~\mathrm{ns}\) with the uniform protocols. All rows
of Table~\ref{tab:si_gapaware_protocol} use \(N_t=400\), and the
independent convergence analysis in
Sec.~\ref{sec:si_protocol_convergence} verifies that this resolution is
sufficient.}

At the stated finite period, the directly calculated quantities are the
orientation-even and orientation-odd half-sum and half-difference.
Following the notation used in the main manuscript,
\(C_1^{\mathrm{odd}}\) is denoted by \(C_1^{\mathrm{geo}}\) after its
near-plateau behaviour has been established, while
\(C_1^{\mathrm{dyn}}\) and \(C_2^{\mathrm{dyn}}\) denote the
orientation-even components. The matched same-point
\(C_2^{\mathrm{odd}}\) is analyzed separately and is not required for
the SNR quantities reported here.

\begin{table}[!htbp]
\centering
\scriptsize
\setlength{\tabcolsep}{3.5pt}
\renewcommand{\arraystretch}{1.10}
\caption{\NEW{Canonical piecewise-linear results for the uniform and
NUFM traversals. All rows use \(N_t=400\). The contrast is
\(100|C_1^{\mathrm{geo}}|/C_1^{\mathrm{dyn}}\),
\(\operatorname{SNR}_{N=1}\) corresponds to one statistically
independent CW cycle and one statistically independent CCW cycle, and
\(\mathcal D_T
=
|C_1^{\mathrm{geo}}|/
\sqrt{\tau C_2^{\mathrm{dyn}}}\)
is the fixed-total-time detectability measure. The labels \(0.06\),
\(0.08\), and \(0.10\) denote the tested values of
\(\epsilon_{\max}\) at
\((\alpha,\beta,r)=(6,0,5)\) and \(\tau=1~\mathrm{ns}\).}}
\label{tab:si_gapaware_protocol}

\begin{tabular}{lcccccc}
\toprule
\textbf{Protocol}
&
\shortstack{\(C_1^{\mathrm{geo}}\)\\(meV/cycle)}
&
\shortstack{\(C_1^{\mathrm{dyn}}\)\\(meV/cycle)}
&
\shortstack{\(C_2^{\mathrm{dyn}}\)\\(meV\(^{2}\)/cycle)}
&
\shortstack{contrast\\(\%)}
&
\shortstack{\(\operatorname{SNR}_{N=1}\)}
&
\shortstack{\(\mathcal D_T\) (\(\mathrm{ps^{-1/2}}\))}
\\
\midrule

Uniform, \(5~\mathrm{ns}\)
& 0.214525
& 108.755724
& 4978.493917
& 0.197
& 0.004300
& 0.0000430
\\

Uniform, \(1~\mathrm{ns}\)
& 0.214179
& 21.751998
& 995.777041
& 0.985
& 0.009599
& 0.0002146
\\

NUFM, \(\epsilon_{\max}=0.06\)
& 0.213787
& 12.612953
& 576.468407
& 1.695
& 0.012592
& 0.0002816
\\

NUFM, \(\epsilon_{\max}=0.08\)
& 0.213027
& 10.274030
& 461.556568
& 2.073
& 0.014023
& 0.0003136
\\

NUFM, \(\epsilon_{\max}=0.10\)
& 0.212163
& 8.885957
& 393.336196
& 2.388
& 0.015129
& 0.0003383
\\

\bottomrule
\end{tabular}
\end{table}

\NEW{At \(1~\mathrm{ns}\), the selected NUFM protocol reduces
\(C_1^{\mathrm{dyn}}\) from \(21.752\) to
\(10.274~\mathrm{meV/cycle}\) and \(C_2^{\mathrm{dyn}}\) from
\(995.777\) to \(461.557~\mathrm{meV^2/cycle}\), corresponding to
reductions of \(52.8\%\) and \(53.6\%\), respectively.
\(C_1^{\mathrm{geo}}\), in contrast, changes only from
\(0.214179\) to \(0.213027~\mathrm{meV/cycle}\), retaining
\(99.46\%\) of the uniform \(1~\mathrm{ns}\) value. The corresponding
intrinsic gains relative to the uniform \(5~\mathrm{ns}\) baseline are
\(3.26\times\) per CW-CCW pair and \(7.29\times\) at fixed total
time.}

\begin{figure}[!htbp]
\centering
\includegraphics[width=0.95\textwidth]
{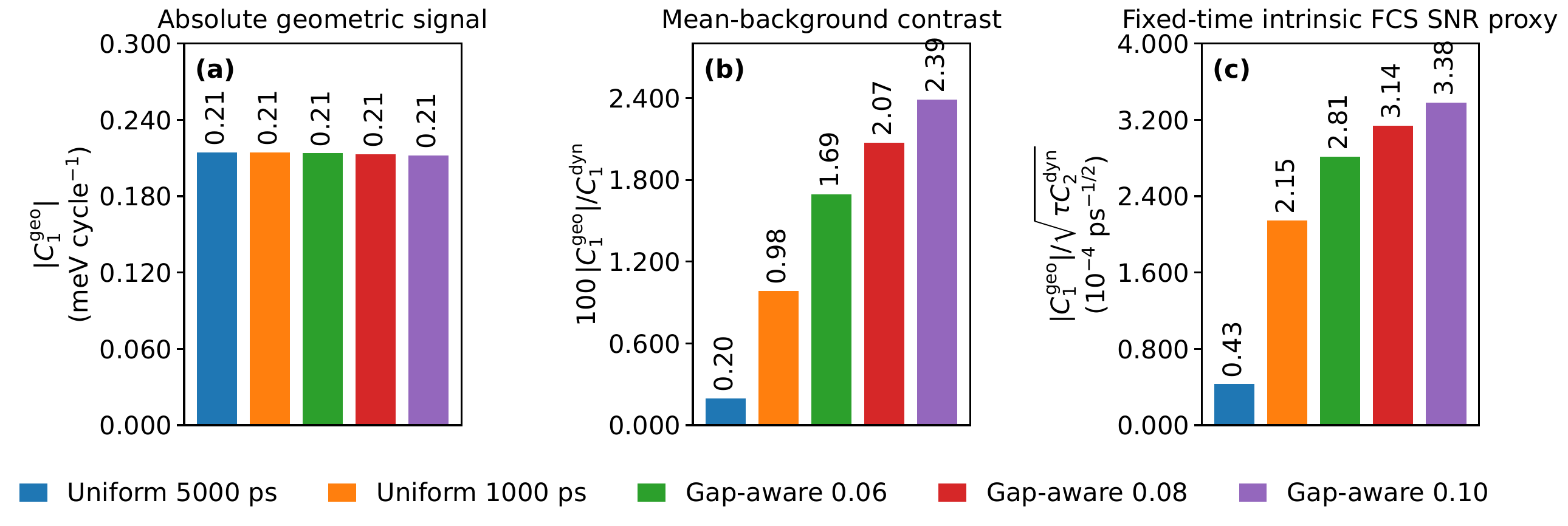}
\caption{\NEW{Performance comparison between the uniform and NUFM
traversals. Panel~(a) gives the absolute geometric first cumulant,
panel~(b) gives the mean-background contrast
\(100|C_1^{\mathrm{geo}}|/C_1^{\mathrm{dyn}}\), and panel~(c) gives
the fixed-total-time detectability
\(\mathcal D_T=
|C_1^{\mathrm{geo}}|/\sqrt{\tau C_2^{\mathrm{dyn}}}\).}}
\label{fig:si_gapaware_performance}
\end{figure}

\FloatBarrier

\begin{figure}[!htbp]
\centering
\safeincludegraphics[width=0.95\textwidth]
{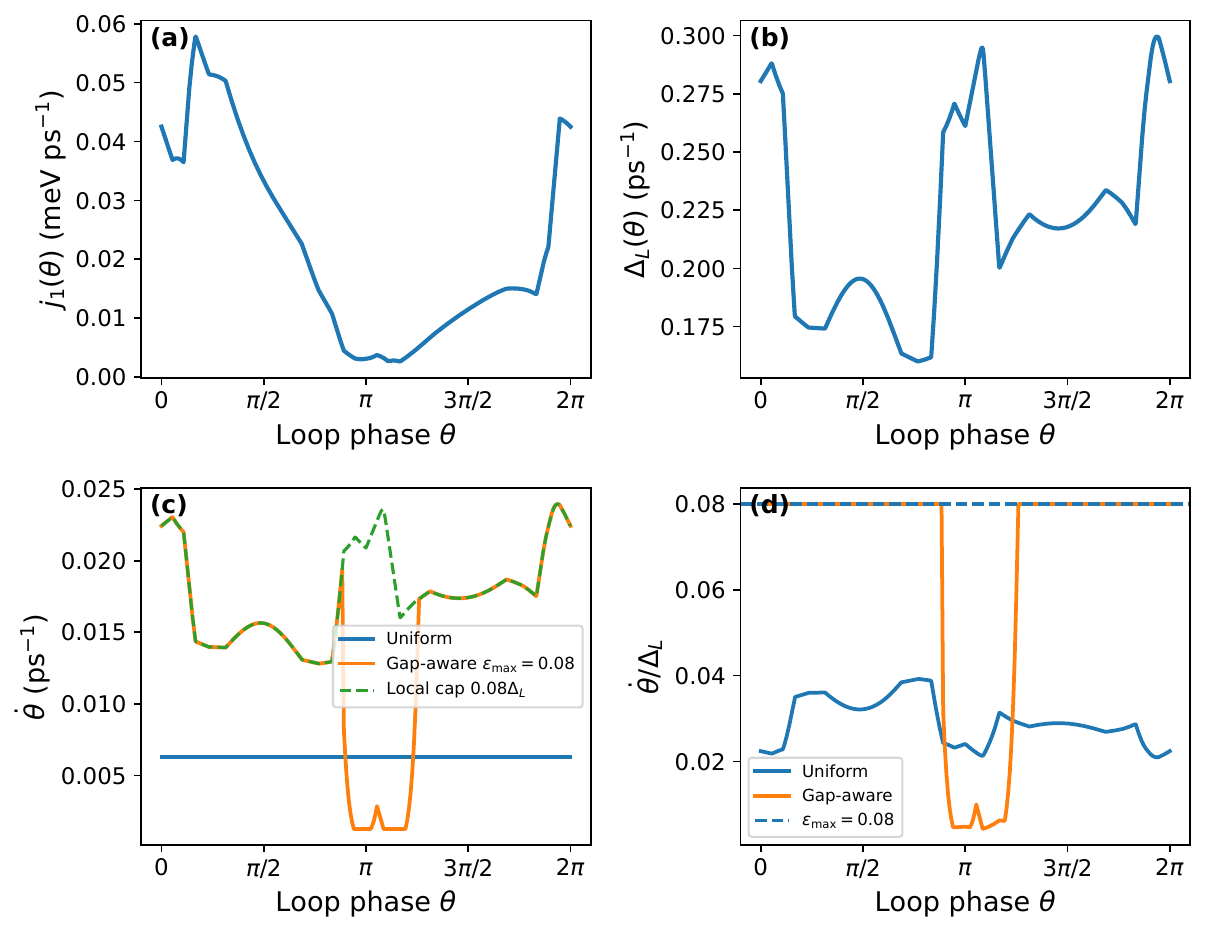}
\caption{\NEW{Local mechanism of the selected NUFM protocol at
\(\tau=1~\mathrm{ns}\).
Panels~(a) and (b) show the frozen dynamic first-cumulant rate
\(j_1(\theta)\) and ordinary-Liouvillian relaxation gap
\(\Delta_L(\theta)\), respectively. Panel~(c) compares the uniform
angular speed, selected NUFM speed, and the local upper bound
\(0.08\Delta_L(\theta)\). The pronounced speed minimum near
\(\theta\simeq\pi\) follows from the minimum of \(j_1(\theta)\)
amplified by the factor
\([j_1(\theta)/j_{1,*}]^6\), rather than from a closing of the
relaxation gap. Panel~(d) gives
\(\dot{\theta}(\theta)/\Delta_L(\theta)\), which remains below the
imposed value \(\epsilon_{\max}=0.08\).}}
\label{fig:si_gapaware_local}
\end{figure}

\NEW{The local origin of the dwell redistribution is shown in
Fig.~\ref{fig:si_gapaware_local}.}

\FloatBarrier
\subsection{Period-dependent cap activity and feasibility}
\label{sec:si_cap_activity}

For the selected parameter set, the relaxation-gap ceiling is active
only in the sufficiently fast-driving regime. A useful loop-wide
crossover follows from
\(
r\omega=\epsilon_{\max}\Delta_L^{\min}
\),
which gives
\(
\tau_{\mathrm{cap}}
=
\frac{2\pi r}
{\epsilon_{\max}\Delta_L^{\min}}
\simeq
2.45~\mathrm{ns}
\)
for
\(
(r,\epsilon_{\max})=(5,0.08)
\)
and
\(
\Delta_L^{\min}\simeq0.160~\mathrm{ps^{-1}}
\).

For
\(
\tau>\tau_{\mathrm{cap}}
\),
the inequality
\(
r\omega<\epsilon_{\max}\Delta_L(\theta)
\)
holds throughout the loop, so the local relaxation-gap cap is inactive
everywhere. Below this threshold, the cap may become active depending on
the raw NUFM profile. For the selected protocol it is active over part
of the loop at \(\tau=1~\mathrm{ns}\), whereas it is inactive at
\(5~\mathrm{ns}\) and longer periods.

A proposed set of local bounds is feasible only when the fastest allowed
traversal can complete the loop within the prescribed period. The
necessary condition is
\(
\int_0^{2\pi}
\frac{d\theta}
{\min[
r\omega,
\epsilon_{\max}\Delta_L(\theta)
]}
\leq\tau.
\)

A separate feasibility test at
\(
\epsilon_{\max}=0.02
\),
outside the reported \(625\)-point grid, fails this condition at
\(\tau=1~\mathrm{ns}\). Even propagation at the largest locally allowed
speed then requires more than the available cycle time. This failure
arises from the imposed physical bounds rather than from numerical
root finding.

\begin{figure}[!htbp]
\centering
\safeincludegraphics[width=4in,height=3in]
{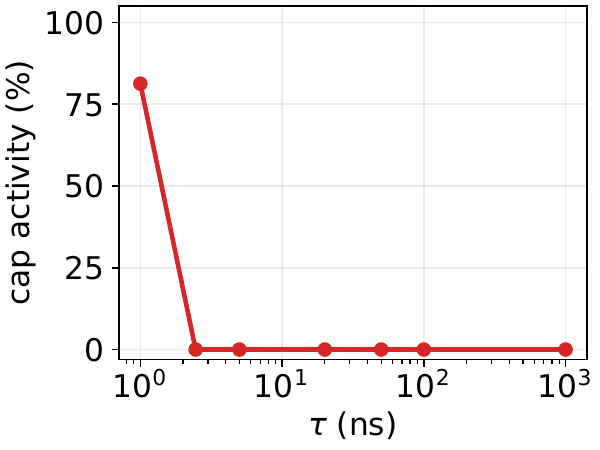}
\caption{Period dependence of local relaxation-gap-cap activity.
The plotted quantity is the fraction of sampled phase cells for which
the selected NUFM speed is pinned to
\(\epsilon_{\max}\Delta_L(\theta)\). The cap is active over
approximately \(81.3\%\) of the phase cells at
\(\tau=1~\mathrm{ns}\) and becomes inactive near the numerically
confirmed crossover
\(\tau\simeq2.454~\mathrm{ns}\). This percentage measures constraint
activity and is not a dynamic-cumulant reduction.}
\label{fig:si_new_cap_activity}
\end{figure}

\begin{figure}[!htbp]
\centering
\safeincludegraphics[width=4in,height=3in]
{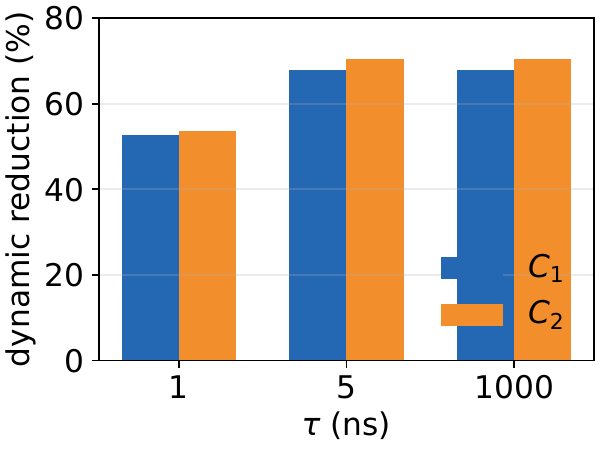}
\caption{Period dependence of dynamic-cumulant suppression.
The selected protocol is gap limited at \(1~\mathrm{ns}\), while the
local gap cap is inactive at \(5\) and \(1000~\mathrm{ns}\). The
approximately repeated reductions of \(67.9\%\) in
\(C_1^{\mathrm{dyn}}\) and \(70.5\%\) in
\(C_2^{\mathrm{dyn}}\) at the two cap-inactive periods are consistent
with rescaling of the same normalized dwell profile once the local
relaxation-gap ceiling is inactive.}
\label{fig:si_new_period_scaling}
\end{figure}

\FloatBarrier
\subsection{Protocol-specific time-step convergence}
\label{sec:si_protocol_convergence}

The discretization convergence must be verified for each prescribed
physical waveform. Refining \(N_t\) improves the ordered-product
representation of a given traversal but does not transform the uniform
protocol into the NUFM protocol. We therefore compare the uniform and selected NUFM traversals at
\(\tau=5~\mathrm{ns}\) and \(\tau=1~\mathrm{ns}\) for
\(
N_t=400,800,1000,2000,
\)
and \(5000\).

\begin{figure}[!htbp]
\centering
\includegraphics[
width=0.8\textwidth,
keepaspectratio
]{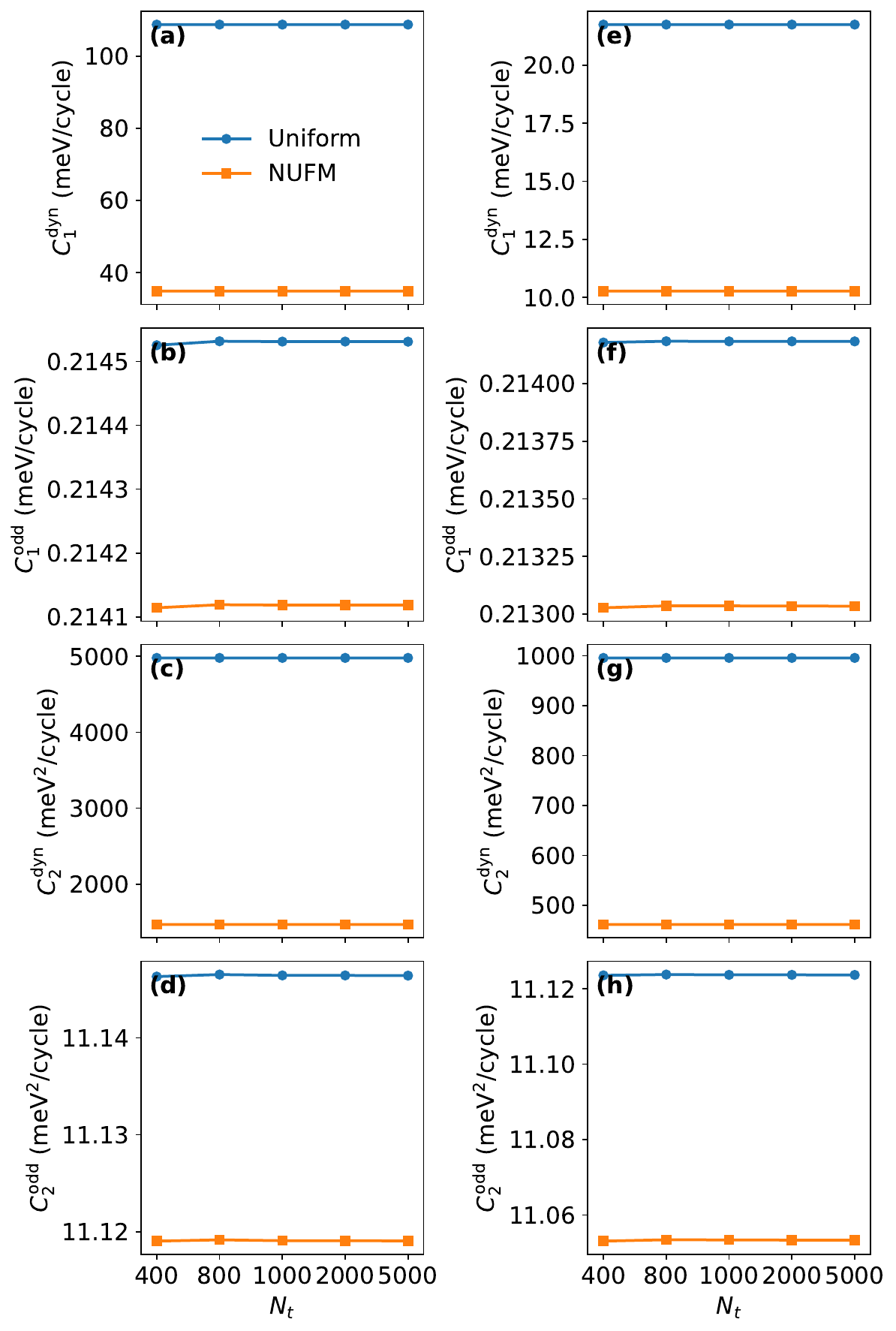}
\caption{Time-step convergence of the four cumulants for the uniform and
selected NUFM traversals. The left and right columns correspond to
\(\tau=5~\mathrm{ns}\) and \(1~\mathrm{ns}\), respectively.
Panels~(a,e) show \(C_1^{\mathrm{dyn}}\), panels~(b,f) show
\(C_1^{\mathrm{odd}}\), panels~(c,g) show
\(C_2^{\mathrm{dyn}}\), and panels~(d,h) show
\(C_2^{\mathrm{odd}}\).
The calculations use
\(N_t=400,800,1000,2000,\) and \(5000\).
All curves are already numerically stable at \(N_t=400\), and
refinement to \(N_t=5000\) changes every displayed cumulant by less
than \(0.004\%\). The same matched same-point reversal convention is
used for both protocols.
}
\label{fig:si_combined_protocol_convergence}
\end{figure}

\begin{table*}[!htbp]
\centering
\scriptsize
\setlength{\tabcolsep}{4pt}
\renewcommand{\arraystretch}{1.08}
\caption{
Values at the largest plotted resolution,
\(N_t=5000\), corresponding to
Fig.~\ref{fig:si_combined_protocol_convergence}.
The relative changes compare the selected NUFM traversal with the
uniform traversal at the same cycle period. The orientation-odd
quantities use the same matched same-point reversal convention for both
protocols.
}
\label{tab:si_merged_protocol_convergence}
\begin{tabular}{r l r r r r}
\toprule
\(\tau\)
&
Protocol
&
\shortstack{\(C_1^{\mathrm{dyn}}\)\\(meV/cycle)}
&
\shortstack{\(C_1^{\mathrm{odd}}\)\\(meV/cycle)}
&
\shortstack{\(C_2^{\mathrm{dyn}}\)\\(meV\(^2\)/cycle)}
&
\shortstack{\(C_2^{\mathrm{odd}}\)\\(meV\(^2\)/cycle)}
\\
\midrule

\(5~\mathrm{ns}\)
& Uniform
& 108.757262
& 0.214531
& 4978.589125
& 11.146417
\\

& NUFM
& 34.891515
& 0.214119
& 1469.011671
& 11.119058
\\

& Relative change
& \(-67.92\%\)
& \(-0.192\%\)
& \(-70.49\%\)
& \(-0.245\%\)
\\

\addlinespace

\(1~\mathrm{ns}\)
& Uniform
& 21.752287
& 0.214183
& 995.821473
& 11.123682
\\

& NUFM
& 10.274055
& 0.213034
& 461.555829
& 11.053291
\\

& Relative change
& \(-52.77\%\)
& \(-0.536\%\)
& \(-53.65\%\)
& \(-0.633\%\)
\\

\bottomrule
\end{tabular}
\end{table*}

Between \(N_t=400\) and \(N_t=5000\), the largest relative change
among the eight calculated curves is approximately \(0.0034\%\).
The \(N_t=400\) discretization used for the uniform period scan,
parameter-grid screening, selected NUFM calculation, and subsequent
rate-modification calculations is therefore numerically converged at
the precision required here. The separation between the uniform and NUFM dynamic cumulants persists
under refinement and therefore results from their different physical
dwell-time distributions. The much smaller differences in the
orientation-odd cumulants are retained as finite-period corrections
under the same matched reversal convention rather than interpreted as
changes of the underlying oriented loop.

\FloatBarrier
\subsection{Waveform feasibility and rate robustness of the selected NUFM protocol}
\label{sec:si_new_waveform_robustness}

For experimental waveform assessment, the physically relevant
quantities are the continuous maxima
\(
\max_\theta|\dot{\theta}|
\),
\(
\max_\theta|dV_G/dt|
\),
and
\(
\max_\theta|dI_{\mathrm{pump}}/dt|
\),
rather than the duration of an individual numerical phase cell, which
depends on \(N_t\). For the sinusoidal coordinates,
\(
\left|\frac{dV_G}{dt}\right|
=
|A_G\cos\theta|\dot{\theta},
\qquad
\left|\frac{dI_{\mathrm{pump}}}{dt}\right|
=
|A_I\cos(\theta+\Delta\phi)|\dot{\theta}.
\) Direct evaluation along the selected clipped waveform shows that in the
cap-inactive regime the fast portions of the trajectory include the
largest-slope regions of both controls. At
\(\tau=50~\mathrm{ns}\) with \(r=5\),
\(
\max|dV_G/dt|
=
1.885~\mathrm{V\,ns^{-1}},
\)
and
\(
\max|dI_{\mathrm{pump}}/dt|
=
0.754~\mathrm{ns^{-1}}.
\) At the gap-limited
\(\tau=1000~\mathrm{ps}\) operating point, the active local cap reduces
the maximum gate slew to approximately \(75\%\) of the value allowed by
the speed-range bound,
\(70.7\) compared with \(94.2~\mathrm{V\,ns^{-1}}\). At
\(\tau=2.454~\mathrm{ns}\simeq\tau_{\mathrm{cap}}\),
the directly calculated maximum ratio
\(
\dot{\theta}/\Delta_L
\)
is \(0.0800\). Experimental implementation would therefore require gate and pump
control bandwidths compatible with these slew rates. A quantitative
hardware bandwidth estimate would additionally require modelling of the
waveform smoothing, gate-line transfer function, and pump-modulator
response.

\begin{table}[!htbp]
\centering
\scriptsize
\setlength{\tabcolsep}{4pt}
\renewcommand{\arraystretch}{1.08}
\caption{Continuous-waveform maxima for the selected NUFM protocol.
Cap activity denotes the percentage of sampled phase cells pinned to
the local relaxation-gap ceiling. The remaining columns give the
maximum gate and pump-coordinate slew magnitudes evaluated directly
from the continuous clipped waveform.}
\label{tab:si_new_slew}
\begin{tabular}{rrrr}
\toprule
\(\tau\) (ns)
&
cap activity (\%)
&
\(\max|dV_G/dt|\) (V/ns)
&
\(\max|dI_{\rm pump}/dt|\) (ns\(^{-1}\))
\\
\midrule

1.000 & 81.3 & 70.726 & 20.939\\
2.454 & 0.0 & 38.400 & 15.360\\
5.000 & 0.0 & 18.850 & 7.540\\
20.000 & 0.0 & 4.712 & 1.885\\
50.000 & 0.0 & 1.885 & 0.754\\

\bottomrule
\end{tabular}
\end{table}

The phenomenological rate modifications considered in the main
manuscript also change \(j_1(\theta)\) and
\(\Delta_L(\theta)\). The NUFM waveform must therefore be reconstructed
self-consistently after a rate modification rather than obtaining the
new cumulants by rescaling their baseline values. The protocol-level suppression was further tested against variations of
the phenomenological rate prefactors and the interpolation procedure.
Scaling both phonon prefactors by \(0.5\) to \(2\) changes the normalized
frozen-current profile only weakly, with a profile correlation exceeding
\(0.9999\), and leaves the \(C_1^{\mathrm{dyn}}\) reduction between
\(52.3\%\) and \(53.7\%\). Repeating the complete construction with PCHIP interpolation gives a
\(52.7\%\) reduction of \(C_1^{\mathrm{dyn}}\) with \(99.45\%\)
geometric retention, essentially unchanged from the canonical
piecewise-linear interpolation. Normalization by \(j_{1,*}\) also reduces the sensitivity to the
optical-injection prefactor. Scaling
\(\Gamma_{\mathrm{pump}}^0\) by \(0.5\) to \(2\) changes the normalized
profile by up to approximately \(0.038\) in RMS, while the
\(C_1^{\mathrm{dyn}}\) reduction remains between \(50.8\%\) and
\(53.4\%\). The largest tested sensitivity arises from the radiative-loss
prefactors. Scaling both radiative prefactors by \(0.5\) to \(2\)
changes the \(C_1^{\mathrm{dyn}}\) reduction from \(22.5\%\) to
\(68.5\%\), while the geometric first-cumulant retention remains above
approximately \(99.3\%\). The tested variations therefore show that geometric retention is
stable across the examined rate modifications, whereas the quantitative
suppression of the dynamic background is more sensitive to the
radiative-loss scale. Among the phenomenological variations considered
here, the radiative prefactors constitute the dominant model sensitivity
of the NUFM performance gain.

\begin{table}[!htbp]
\centering
\scriptsize
\setlength{\tabcolsep}{5pt}
\renewcommand{\arraystretch}{1.08}
\caption{Robustness hierarchy for the selected
\(1~\mathrm{ns}\) NUFM protocol.
Each rate row compares the NUFM and uniform traversals after the same
phenomenological prefactor modification and reports the resulting
dynamic-cumulant reductions and geometric first-cumulant retention.}
\label{tab:si_new_robustness}
\begin{tabular}{lrrr}
\toprule
Variation
&
\(C_1^{\rm dyn}\) reduction
&
\(C_2^{\rm dyn}\) reduction
&
\(C_1^{\rm geo}\) retention
\\
\midrule

Phonon rates \(0.5\)-\(2\times\)
& 52.27-53.69\%
& 53.11-54.65\%
& 99.45-99.47\%
\\

Pump prefactor \(0.5\)-\(2\times\)
& 50.85-53.44\%
& 52.45-53.85\%
& 99.32-99.55\%
\\

Radiative rates \(0.5\)-\(2\times\)
& 22.53-68.46\%
& 21.78-70.31\%
& 99.35-100.04\%
\\

PCHIP interpolation
& 52.70\%
& 53.57\%
& 99.45\%
\\

\bottomrule
\end{tabular}
\end{table}

\FloatBarrier

\clearpage

\section{Conclusion}
\label{sec:si_conclusion}

The calculations establish a consistent hierarchy of model, numerical,
physical, and control checks. The experimentally anchored Hamiltonian
inputs and the imposed gate and pump protocol are specified first,
followed by the selection of the operating field and counted transition,
the construction of the rates from the microscopic model, and the scope
of the population model. The full range trajectory is then selected as
the baseline driving loop through comparisons of the control centre, loop
geometry, and enclosed area. Finite time propagation, periodic limit
cycle initialization, cycle period scaling, counting field convergence,
and intrinsic CW-CCW detectability are subsequently verified before the
physical robustness and model dependence checks are assessed. The
geometric interpretation of the first cumulant is independently checked
using the Berry-Sinitsyn curvature. Alternative loop shapes do not
improve the intrinsic detectability relative to the full range baseline,
whereas the selected NUFM traversal of the same loop strongly suppresses
the dynamic mean and variance while retaining the geometric first
cumulant. Its constrained parameter grid, period dependence, convergence,
waveform feasibility, and rate model robustness support its use as the
preferred protocol for improving detectability under the tested
conditions.

\bibliographystyle{unsrt}
\bibliography{Reference}